\documentclass[twocolumn,prd,showpacs,preprintnumbers,amsmath,amssymb,nofootinbib,eqsecnum]{revtex4-1}

\usepackage[varg]{txfonts}
\usepackage{graphicx}
\usepackage{amssymb}
\usepackage{psfrag}
\usepackage{placeins}
\usepackage{graphicx}
\usepackage{diagbox}
\usepackage{soul}
\usepackage[colorlinks=true,citecolor=blue]{hyperref}
\usepackage{xcolor}

\newcommand{\Lcdm}{\ensuremath{\Lambda}CDM~}

\newcommand{\rhob}{\ensuremath{\bar{\rho}}}
\newcommand{\Pb}{\ensuremath{\bar{P}}}
\newcommand{\grad}{\ensuremath{\vec{\nabla}}}

\newcommand{\DeltaGI}{\ensuremath{{\Delta}}}

\newcommand{\omegazero}{\ensuremath{\omega^{(0)}_g}}
\newcommand{\omegaeq}{\ensuremath{\omega^{\rm eq}_g}}

\newcommand{\dpar}{\ensuremath{b}}

\newcommand{\Rcal}{\ensuremath{{\cal R}}}
\newcommand{\Ocal}{\ensuremath{{\cal O}}}
\newcommand{\Lcal}{\ensuremath{{\cal L}}}
\newcommand{\varW}{{\it var-w}}
\newcommand{\constw}{{\it const-w}}
\newcommand{\varwc}{{\it var-wc}}
\newcommand{\varc}{{\it var-c}}
\newcommand{\varweqc}{{\it var-w=c}}
\newcommand{\cvis}{\ensuremath{c^2_{\rm vis}}}
\newcommand{\cvisi}{\ensuremath{c^2_{{\rm vis},i}}}
\newcommand{\cvisI}[1]{\ensuremath{c^2_{{\rm vis},#1}}}
\newcommand{\Dim}{D}

\DeclareMathOperator\erf{erf\!}

\definecolor{OliveGreen}{RGB}{60,128,49}
\definecolor{Gray}{RGB}{148,150,152}

\newcommand{\colorvarwc}[1]{\textcolor{blue}{#1}}
\newcommand{\colorvarc}[1]{\textcolor{black}{#1}}
\newcommand{\colorvarw}[1]{\textcolor{OliveGreen}{#1}}
\newcommand{\colorLCDM}[1]{\textcolor{Gray}{#1}}

\begin{document}

\title{Dark matter properties through cosmic history}

\author{St\'ephane Ili\'c}
\email{stephane.ilic@obspm.fr}
\affiliation{CEICO, Institute of Physics of the Czech Academy of Sciences, Na Slovance 2, Praha 8, Czech Republic}
\affiliation{Universit\'e PSL, Observatoire de Paris, Sorbonne Universit\'e, CNRS, LERMA, F-75014, Paris, France}
\affiliation{IRAP, Universit\'e de Toulouse, CNRS, CNES, UPS, Toulouse, France}
\author{Michael Kopp}
\email{michael.kopp@su.se}
\affiliation{CEICO, Institute of Physics of the Czech Academy of Sciences, Na Slovance 2, Praha 8, Czech Republic}
\affiliation{Nordita, KTH Royal Institute of Technology and Stockholm University, Hannes Alfv\'ens v\"ag 12, SE-106 91 Stockholm, Sweden}
\author{Constantinos Skordis}
\email{skordis@fzu.cz}
\affiliation{CEICO, Institute of Physics of the Czech Academy of Sciences, Na Slovance 2, Praha 8, Czech Republic}
\author{Daniel B. Thomas}
\email{daniel.thomas-2@manchester.ac.uk}
\affiliation{Jodrell Bank Centre for Astrophysics, School of Physics \& Astronomy, The University of Manchester, Manchester M13 9PL, United Kingdom}

\date{\today}

\begin{abstract}
    We perform the first test of dark matter (DM) stress-energy evolution through cosmic history, using cosmic microwave background measurements supplemented with baryon acoustic oscillation
 data and the Hubble Space Telescope key project data. We constrain the DM equation of state (EoS) in 8 redshift bins, and its sound speed and (shear) viscosity in 9 redshift bins,
finding no convincing evidence for non-$\Lambda$CDM values in any of the redshift bins. Despite this enlarged parameter space, the sound speed and viscosity are constrained relatively
well at late times (due to the inclusion of CMB lensing), whereas the EoS is most strongly constrained around recombination.
These results constrain for the first time the level of ``coldness'' required of DM across various cosmological epochs at both the
background and perturbative levels. We show that simultaneously allowing time dependence for both the
EoS and sound speed parameters shifts the posterior of the DM abundance before recombination to a
higher value, while keeping the present day DM abundance similar to the $\Lambda$CDM value.
 This shifts the posterior for the present day Hubble constant compared to $\Lambda$CDM, suggesting that DM with time-dependent parameters
is well-suited to explore possible solutions to persistent tensions within the $\Lambda$CDM model. 
We perform a detailed comparison with our previous study involving a vanishing sound speed and viscosity using the same datasets 
in order to explain the physical mechanism behind these shifts.
\end{abstract}


\maketitle


\section{Introduction}

Galactic and cosmological observations indicate that if gravitational laws are dictated by general relativity, a large fraction of
the nonrelativistic matter in our Universe is in the form of particles having negligible interaction with e\-lec\-tro\-mag\-ne\-ti\-sm, baryons and themselves,
 and with negligible initial velocity dispersion.
The existence of these particles has been demonstrated through their gravitational effects on the largest (galactic to cosmological) scales of the Universe.
Collectively, they are successfully modeled as cold dark matter (CDM), a crucial component of the \Lcdm concordance model.

Although a plethora of concrete DM models have been proposed \cite{Bertone:2004pz}, de\-di\-ca\-ted direct and indirect  astrophysical searches have yielded
no convincing evidence for a DM particle so far.
The strongest exclusion limits in the mass vs cross-section plane
using direct detection through nuclear recoil come from the Xenon1T experiment~\cite{Aprile:2019jmx,Aprile:2019xxb,Aprile:2018dbl,Aprile:2019dbj}.
Meanwhile, possible signals of DM annihilation resulting in the positron excess detected by the Alpha Magnetic Spectrometer (AMS) instrument~\cite{AMS02-positron_excess}
are in conflict with the Planck collaboration \cite{Aghanim:2018eyx} observations of the Cosmic Microwave Background (CMB) anisotropies,
the latter being sensitive to energy injection in the intergalactic medium through such annihilations. Indeed, the AMS positron excess may be
explained by conventional astrophysical mechanisms~\cite{Ahlers:2009ae,Mertsch:2014poa}.

This lack of nongravitational evidence necessitates further testing of the CDM paradigm.
Taking a more agnostic approach with this in mind,
we test possible departures from CDM using the phenomenologically motivated generalized dark matter (GDM) model~\cite{Hu1998a}.
GDM compactly parametrizes the DM properties encapsulated by pressure and viscosity using three parametric functions: the background equation
of state (EoS) $w(a)$ of DM, sound speed $c^2_s(a,k)$ and the viscosity $\cvis(a,k)$, where $a$ is the scale factor and $k$ the wave number of the linearized
GDM fluid fluctuations.

In \cite{Hu1998a} it was shown that the expansion history and consequently the CMB anisotropies angular power spectrum is particularly sensitive to these parameters.
Moreover, when $w$ is a constant, \cite{Hu1998a} uncovers a degeneracy between $w$ and $\omegazero$, the dimensionless DM density today.
An extensive investigation of the model was presented in \cite{KoppSkordisThomas2016} where its possible connection to more fundamental theories was established,
particularly to $K$-essence scalar fields, a rich internally coupled dark sector (e.g. dark matter coupled to dark radiation), thermodynamics and effective field theories.
Furthermore, \cite{KoppSkordisThomas2016} analysed
 an exact solution of the perturbed Einstein equations in a flat GDM-dominated universe uncovering a degeneracy between
a constant sound speed and constant viscosity. Specifically, the effective perturbative parameter relevant for the CMB is $c_s^2 + \frac{8}{15} \cvis$
and in order to break this degeneracy, different types of observations are necessary.

Constraints on constant GDM parameters were placed previously by \cite{Muller2005, CalabreseMigliaccioPaganoEtal2009, KumarXu2012, XuChang2013}
 using a variety of datasets. The latest
constraints on constant GDM parameters were reported in \cite{ThomasKoppSkordis2016} and \cite{KunzNesserisSawicki2016} using CMB data from the Planck satellite setting
a limit on constant $|w| \lesssim 10^{-3}$ and $c_s^2, \cvis \lesssim 10^{-6}$. Significant improvements on the perturbative
parameters $c_s^2$ and $\cvis$ were obtained in \cite{KunzNesserisSawicki2016} and \cite{ThomasKoppMarkovic2019} through  the
inclusion of the late-time clustering data. Using late-time clustering data, however, is prone to introducing systematic modeling errors
due to the nonlinearities inherent in the processing of these datasets. Thus, to test that the improvement in $c_s^2$ and $\cvis$ is robust,
 \cite{ThomasKoppMarkovic2019} designed a nonlinear extension of the GDM model based on the ``warm and fuzzy'' dark matter halo model, which incorporates certain nonlinear phenomena.
Joint constraints on the sum of neutrino masses and constant GDM parameters were obtained in \cite{KumarNunesYadav2019, ThomasKoppMarkovic2019}.
A time-varying equation of state $w$ was considered in \cite{KoppSkordisThomasEtal2018} by piecewise parametrizing $w(a)$ in $8$ redshift bins,
while both $c_s^2$ and $\cvis$ were assumed to be zero. There, the most general time-evolution of the DM equation of state was tested,
yet no evidence for DM properties beyond CDM was found.
Interestingly, while data allow  $w$ to be fairly larger than zero in the late universe,
between matter-radiation equality and CMB recombination $|w|$ is $\lesssim 10^{-3}$ and thus DM must behave very closely to CDM during that time~\cite{KoppSkordisThomasEtal2018}.

Although a wealth of more constraining data exists, by sticking to observables
pertaining to linear perturbations and Friedmann-Lema\^itre-Robertson-Walker (FLRW) background one reduces systematic uncertainties and modeling errors
(on the nonlinear scales) to a minimum. This ensures that any potential detection of nonzero GDM parameters can be convincingly interpreted as a detection
of DM properties. We refer however to \cite{KunzNesserisSawicki2016, TutusausLamineBlanchard2018, ThomasKoppMarkovic2019}
 for potential applications to nonlinear scales.

In this article, we present the most exhaustive parameter search to date, allowing all three GDM parametric functions $w(a)$, $c^2_s(a)$ and $\cvis(a)$
to have a sufficiently general time dependence. This time dependence is modeled  by binning $w(a)$ in 8 and
 $c^2_s$ and $\cvis$ in 9 scale factor bins, totalling 26 new parameters beyond CDM. We use the same datasets as our previous study which had a time-dependent $w(a)$ but
zero $c^2_s(a)$ and $\cvis(a)$~\cite{KoppSkordisThomasEtal2018}; this allows us to perform a detailed comparison of the effects of the new enlarged parameter space
 corresponding to $c^2_s(a)$ and $\cvis(a)$ with respect to \cite{KoppSkordisThomasEtal2018}.

The structure of the article is as follows. We give a brief summary of the GDM model, describe our binning strategy
and present the various models and submodels that we study in Sec.~\ref{sec:model}.  In Sec.~\ref{sec:methods} we present our methodology, including numerical solutions,
the datasets and sampling method used which allowed exploration of the very high-dimensional parameter space and a discussion of our choice of priors.
Our results are presented in Sec.~\ref{sec:results}, specifically constraints on the DM  EoS and abundance, constraints on the sound speed and viscosity, degeneracies
and a special submodel where all three functions are set to be equal. We discuss the physical aspects and implications of our results in Sec.~\ref{sec:discussion}, particularly,
 the tight constraint of the GDM comoving density perturbation in the early universe
and how some GDM models may alleviate the Hubble tension.
We conclude in Sec.~\ref{sec:conclusion}.

The reader may find useful the three appendices. In Appendix~\ref{app:sigma8w0Degeneracy} we derive an expression for the growth index
in a $\Lambda w$DM (i.e. GDM with constant EoS and zero sound speed and viscosity) and discuss the Integrated Sachs-Wolfe (ISW) effect.
We describe our publicly available suite of codes used here for sampling the parameter space and visualizing the results in  Appendix~\ref{app:ECLAIR}.
A complete list of constraints for various choices of datasets, parametrization choices and priors as well as correlation matrices can be found in Appendix~\ref{app:bigtables}.


\section{The model}
\label{sec:model}

\subsection{Evolution equations}
We consider a flat FLRW background with only scalar perturbations, see \cite{KoppSkordisThomas2016} for more details and notation.
The GDM background density $\rhob_g$ and pressure $\Pb_g$ evolve according to the conservation law
\begin{align} \label{GDMconservation}
\dot{\rhob}_g  = - 3 H (1+w) \rhob_g\,, \qquad \Pb_g=w \rhob_g\,,
\end{align}
where $H =\frac{\dot a}{a}$ is the Hubble parameter, satisfying the Friedmann equation, and the overdot denotes derivatives with respect to cosmic time $t$.
The parametric function $w(a)$ can be freely specified and contains with $w=0$ the CDM model  ($\rhob_g = \rhob_c$).
The GDM model has two further free parametric functions, the speed of sound, $c_s^2(a,k)$, and the (shear) viscosity, $\cvis(a,k)$, both of which are zero in the case of CDM.

The synchronous gauge metric perturbed around a flat FLRW background is given by
\begin{multline}
ds^2 = -\, dt^2  +a^2 \Bigg[ \Big(1+\frac{1}{3} h\Big)\, \delta_{ij} + (\grad_i \grad_j - \frac{1}{3} \gamma_{ij} \grad^2) \nu  \Bigg] dx^i dx^j \,,
\label{def_perturbed_FLRW_metric}
\end{multline}
where $\grad_i$ is the covariant derivative compatible with the Euclidean metric $\gamma_{ij}$ and only scalar modes (in this gauge $h$ and $\nu$) are considered.

Switching to $k$-space, the general GDM fluid equations for the density contrast $\delta_g$ and velocity perturbation $\theta_g$ are given by
\begin{subequations}  \label{GDMperts}
\begin{equation}
 \dot{\delta}_g = 3  H \left( w \delta_g  - \Pi_g\right) -  (1+w) \left[ \frac{k^2}{a}\theta_g+ \frac{1}{2} \dot{h} \right]
\label{fluid_delta_equation}
\end{equation}
\begin{equation}
    a \dot{\theta}_g = -(1 -3 c_{a}^2)     aH   \theta_g
+  \frac{\Pi_g}{1+w}
-  \frac{2}{3}  k^2 \Sigma_g \,.
\label{fluid_theta_equation}
\end{equation}
with $ c_a^2 = \dot{\Pb}_g/\dot{\rhob}_g$. While the above equations are generically valid for any conserved fluid, the following special choice of closure equations, defines the GDM model \cite{Hu1998a}
\begin{align}
  \Pi_g &= c_s^2 \delta_g+3 (1+w) ( c_s^2 - c_a^2 )  aH  \theta_g  \\
\dot{\Sigma}_g  &=  - 3 H   \Sigma_g+ \frac{4}{1+w} \cvis (\frac{\theta_g}{a}  - \frac{1}{2}\dot{\nu})\,.
\label{ShearGDMeom}
\end{align}
\end{subequations}
The first equation is a perturbative EoS for the pressure perturbation $\Pi_g \equiv (P_g- \Pb_g)/ \rhob_g $ and the second equation is an evolution equation for the scalar part $\Sigma_g$ of the traceless part of the GDM stress tensor $T_g^i{}_j$. We refer the interested reader to \cite{KoppSkordisThomas2016} for further discussions of the theoretical motivation,
physical interpretation and notation.

The EoS $w$ is expected to be uncorrelated with the two perturbative parameters $c_s^2$ and $\cvis$ during the era of matter domination,
as shown in \cite{ThomasKoppSkordis2016}. However, as we show below, these parameters become correlated during the era of radiation domination, when
 adiabatic initial conditions are considered.


\subsection{Smooth bin parametrization}
In order to constrain the three purely time-dependent GDM parametric functions $w(a)$, $c^2_s(a)$ and $\cvis(a)$ in a way that is sufficiently general but still feasible,
we restricted the variation of these functions to $N=9$ scale factor bins. As our goal is to explore the allowed behavior of dark matter with as few restrictions as possible, having fewer bins would unnecessarily restrict the phenomenological freedom of the model.

\begin{figure}[t!]
\begin{center}
\includegraphics[width=0.44 \textwidth]{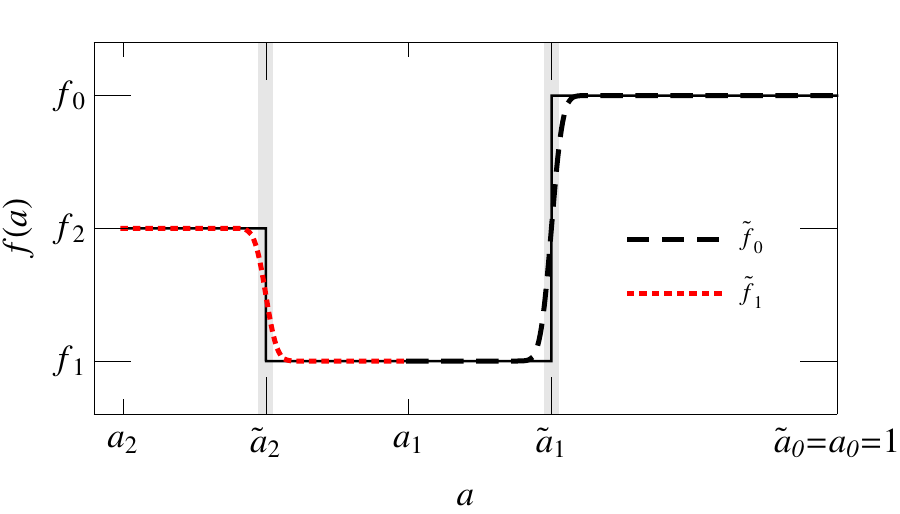}
\end{center}
\caption{ Dashed lines show the first two components of the sum in Eq.~\eqref{smoothbins} in an arbitrary case where $f_1<f_2<f_0$.
The black thin line shows the first three components of the sum in Eq.~\eqref{sharpbins}. The width of the grey bands corresponds to $\sigma_{\ln a}$.
}
\label{PixelExplanationSmall}
\end{figure}

The bin edges were chosen to be
\begin{align}
\tilde a_0 &=1 \notag\\
\tilde a_{1\leq i\leq N-1} &= 10^{- i \Delta_{\ln a}}~~\mathrm{with}~~\Delta_{\ln a}=0.5\\
\tilde a_{N} &= 0  \notag
\end{align}
so that $f(a)$ (here denoting any of $w$, $c^2_{s}$, $\cvis$) has piecewise constant values between them, that is,
\begin{equation}  \label{sharpbins}
f(a)= \sum_{i=0}^{N-1} f_i \Theta(a-\tilde a_{i+1}) \Theta(\tilde a_{i}-a)\,,
\end{equation}
with the $f_i$ coefficients comprising $N$ free parameters and $\Theta$ is the Heaviside step function.
In the case of the $w(a)$ function, the discontinuity at the bin edges $\tilde a_{1\leq i\leq N-1}$ implies $c^2_{a,i}(\tilde a_i)= \pm \infty$ for the
adiabatic sound speed.
In order to test whether our conclusions depend on this discontinuity, we regularized the transitions by a
lognormal smoothing of Eq.~\eqref{sharpbins} with width $\sigma_{\ln a}$ (and assuming $\sigma_{\ln a} \ll \Delta_{\ln a} $), leading to
\begin{align} \label{smoothbins}
f(a) &= \sum_{i=0}^{N-2} \tilde f_i(a)  \,\Theta(a- a_{i+1}) \Theta( a_{i}-a)\\
\tilde f_i(a) &=  \frac{f_i-f_{i+1} }{2} \erf\left( \frac{\ln (a/\tilde a_{i+1})}{\sigma_{\ln a}}\right)+\frac{ f_i+f_{i+1}}{2}
 \,, \notag
\end{align}
with corresponding logarithmic bin centers\footnote{For convenience we defined the bin centers for the first and last bin separately.
 Any definition of bin center is acceptable if it is several multiples of $\sigma_{\ln a}$ away from the transition times $\ln \tilde a_{1\leq i\leq N-2}$.}
\begin{align}
a_0&=1 \notag \\
a_{1\leq i\leq N-2 } &=\sqrt{ \tilde a_{i} \tilde a_{i+1}} \\
a_{N-1} &= 10^{-\Delta_{\ln a}} \tilde a_{N-1} \notag \,.
\end{align}
We set $\sigma_{\ln a}=0.1 \Delta_{\ln a}$, a sufficiently small choice in order to avoid introducing unwanted physical effects,
 but sufficiently wide to study potential differences to setting $\sigma_a=0$  corresponding to Eq.~\eqref{sharpbins}. Both choices produced the same constraints.
See Fig.\,\ref{PixelExplanationSmall} for a visual representation of Eq.~\eqref{smoothbins}.


\subsection{Definition of models and submodels}
We list here the different types of (sub)models that we used for our study of GDM.
\paragraph{Model ``\varwc'':}
The most general GDM model based on our parametrization
 with all 26 GDM parameters included is denoted by ``\varwc''. In addition to this model, we consider and study separately the three nested submodels below.
We note that as discussed in~\cite{KoppSkordisThomasEtal2018}, a degeneracy between $w$ and $\Lambda$ is present in the late universe.
With this in mind, the last two $w$-bins were merged by setting $w_0=w_1$ in this model, as well as its submodel \varW.

\paragraph{Submodel ``\varW'':}
The submodel obtained by setting $c_s^2=\cvis=0$ while keeping  the 8 $w_i$s free, is denoted by ``\varW''
and has been previously studied in~\cite{KoppSkordisThomasEtal2018}. It describes a GDM fluid that only modifies the background evolution of the Universe,
but maintains the geodesic motion of GDM fluid elements.
We include this model here for reference purposes, as the present paper is a direct generalization and logical
continuation of \cite{KoppSkordisThomasEtal2018}. The inclusion of this model allows us to check to what extent
the previously obtained \varW~constraints are recovered within the encompassing \varwc~model after marginalization
over the 18 additional $c^2_{s, i}$ and $\cvisi$ parameters.
The bins $w_0$ and $w_1$ were once again joined together.

\paragraph{Submodel ``\varc'':}
Using the same reasoning for studying the \varW~model,
 we also study the complementary submodel ``\varc'' defined by $w=0$ with all of the 18 $c^2_{s, i}$ and $\cvisi$ parameters left free.

\paragraph{Submodel ``\varweqc'':}
Finally, we also consider the submodel with the restriction  $w_i =c^2_{s, i} = \cvisi$.
This model is interesting due to its close relation with a number of well-motivated collisionless DM scenarios. Two examples are the case of warm DM
and the case of CDM when the effects of unresolved nonlinear small-scale physics is incorporated using
 the effective field theory of large-scale structure \cite[EFTofLSS,][]{BaumannNicolisSenatoreEtal2012,CarrascoHertzbergSenatore2012,CarrollLeichenauerPollack2013,ForemanSenatore2015}.
In this case we let $w_0$ and $w_1$ be mutually independent since late universe constraints on $w$ are driven by the $c_s^2$ and $\cvis$ functions,
as we explain in Sec.\,\ref{varweqc}.
\begin{table}[t!]
\begin{center}
\begin{tabular}{|l |l |l  |l  |}
\hline
 Model  & Additional & Restrictions & No.\,of additional  \\
   &  parameters & &  parameters \\
\hline
\hline
\varwc & $w_i, c^2_{s, i}, \cvisi$ & $w_0=w_1$ &  26 $(8+2\times9)$ \\
\hline
 \varW & $w_i$  & $w_0=w_1$ & 8  \, (9-1)\\
& & $c^2_{s}= \cvis=0$  &  \\
 \hline
\varc & $c^2_{s, i}$, $\cvisi$ & $w=0$ & 18 $(2\times9)$         \\
\hline
 \varweqc & $c^2_{s, i} $ & $w=c^2_{s} = \cvis$ &  9 \,  $(1\times9)$   \\

\hline \hline
\end{tabular}
\end{center}
\caption{List of the  GDM models studied in this work. }
\label{tab:modelnames}
\end{table}
Table~\ref{tab:modelnames} shows a summary of all GDM models considered in this paper.

It is also convenient to define a dimensionless scaled GDM density
\begin{equation} \label{Defomegag}
\omega_g \equiv a^3 \rhob_g\,\frac{8 \pi G}{3\times (100\, \rm{km/s/Mpc})^2},
\end{equation}
in order to facilitate interpreting constraints on $w_i$.
When $w=0$, $\omega_g$ is equal to the conventional (constant) dimensionless CDM density $\omega_c$.
The function  $\omega_g$ is in general time dependent, however, fully determined by the $N+1$ parameters $\omegazero, w_i$.
We use the notation  $\omega_g^{(i)}=\omega_g(a_i)$  and similarly for other functions with subscripts, so that the present day DM abundance
is $\omegazero = \omega_g(a_0)$.  Functions without subscript we write instead as $w_i=w(a_i)$.


\section{Methodology}
\label{sec:methods}

\subsection{Numerical solutions}\label{sec:priors}
In order to perform our analysis we implemented the GDM fluid equations \eqref{GDMconservation} and \eqref{GDMperts}
in the Cosmic  Linear  Anisotropy  Solving  System  (CLASS)  code~\cite{Lesgourgues2011}.
CLASS  numerically  solves  the Boltzmann  equation  for  each  relevant  component  coupled to the Einstein equations and calculates the CMB
and matter power spectra given a set of model parameters.
Our modification of the CLASS code adds an additional GDM component based on the dark energy fluid
(with free equation of state and  sound speed) implemented  by the original authors~\cite{LesgourguesTram2011}, which we further improved to allow for nonzero viscosity.

Our modification of CLASS makes it easy to define as many bins as necessary for all three GDM functions $\{w,c_s^2,\cvis\}$ through the standard CLASS interface
and to set the amplitudes for each of these functions in each bin.
Following this work, our code is made publicly available\footnote{\url{https://github.com/s-ilic/gdm_class_public} } with instructions on how to use it.

We also independently modified a different Boltzmann code~\cite[DASh,][]{KaplighatKnoxSkordis2002}
to include the full GDM parametrization.  We performed a full comparison between the codes in the case of constant GDM parameters,
 including the background evolution, perturbation evolution,
the CMB angular power spectra, matter power spectrum and lensing potential. The numerical difference of the two codes in the case of the GDM model is similar to the corresponding
 difference  in  the  case  of  $\Lambda$CDM,  within $\sim0.1\%$.   This  level
of  agreement  holds  for  all  quantities  in  both  the  synchronous gauge and the conformal Newtonian gauges.

\subsection{Datasets and sampling technique}
Our constraints are obtained using the same datasets as in \cite{KoppSkordisThomasEtal2018}. Specifically we used the Planck 2015 data release \cite{PlanckCollaborationXI2015} of
the CMB anisotropies power spectra, composed of the low-$\ell$ T/E/B likelihood and the
full TT/TE/EE high-$\ell$ likelihood with  its complete set of nuisance parameters.
The combination of these likelihoods is thereafter referred to as Planck power spectra (PPS).
We also selectively added the HST key project prior on $H_0$~\cite{RiessMacriCasertanoEtal2011},
BAO measurements from the 6dF Galaxy Survey~\cite{BeutlerBlakeCollessEtAl2011} and the Baryon Oscillation Spectroscopic Survey Sloan Digital Sky Survey~\cite{AndersonAubourgBaileyEtal2014}, and the
Planck 2015 CMB lensing likelihood (respectively referred to as HST, BAO and Lens). Although more recent cosmological datasets are available \citep[see e.g][]{Aghanim:2018eyx},
keeping to the ones mentioned above allows us to perform a robust comparison with our previous work \cite{KoppSkordisThomasEtal2018} where $c_s^2=\cvis=0$ in order to elucidate
 the effects of the new enlarged parameter space.

Our total cosmological parameter set (not including the Planck likelihood nuisance parameters)
 \begin{equation}
 (\omega_b, \omegazero, H_0,n_s, \tau, \ln 10^{10} A_s, w_i, c^2_{s,i}, \cvisi)
 \end{equation}
consists of 6 $\Lambda$CDM parameters and  8 values $w_i$, 9 values $c^2_{s,i}$ and 9 values $\cvisi$.
We assumed adiabatic initial conditions, described in~\cite{KoppSkordisThomas2016}.

We investigated the constraints on our selection of GDM (sub)models coming from our choice of datasets using
a standard Markov Chain Monte Carlo (MCMC) approach.
For this purpose we used ECLAIR, a publicly available\footnote{\url{https://github.com/s-ilic/ECLAIR}} suite of codes
that uses the numerical output from CLASS, combined with likelihoods of state-of-the-art datasets, and efficient sampling methods.

To sample the parameter space we used the Goodman-Weare affine-invariant ensemble sampling technique~\cite{GoodmanWeare2010} via our ECLAIR framework
which internally uses the technique's  Python implementation \texttt{emcee} \cite{emcee}. The convergence of the MCMC chains was assessed using graphical
and numerical tools included in the ECLAIR code package (see Appendix~\ref{app:ECLAIR} for more details).
The ECLAIR suite was also used to find the point in parameter space corresponding to the maximum likelihood of each model.
The resulting chains were used to determine the marginalized posterior distributions of the parameters using the publicly available code \texttt{getdist} \cite{Lewis2019}.


\subsection{Priors}\label{sec:priors}
We set uniform priors as specified in Table~\ref{tab:priors} unless otherwise stated. We used the same priors on Planck nuisance parameters
 and the same neutrino treatment as in \cite{KoppSkordisThomasEtal2018, ThomasKoppSkordis2016}.
The helium fraction was set to $Y_{\rm He}=0.24667$~\cite{PlanckCollaborationXIII2015}. We have checked that letting $Y_{\rm He}$ be
 an additional free parameter in our MCMC analysis does not affect our results and conclusions.

We set flat priors on standard  cosmological parameters as well as the GDM parameters (see Table~\ref{tab:priors}).\footnote{Throughout $H_{\rm 0}$ is in units of km s${}^{-1}$ Mpc${}^{-1}$, and  $H^{-1}_{\rm eq}$ and $r^{\rm drag}_s$ are in units of Mpc.}
The choice of priors is always a sensitive issue in any type of Bayesian analysis. This is particularly true
in our case, as many of our parameters have a physical lower bound -- namely sound speed and viscosity need to be positive at all times.
These bounds form a (multidimensional) ``corner'' which due to volume effects becomes highly disfavored during the MCMC exploration regardless
of whether the data favor this region of the parameter space or not. This situation is particularly problematic in our case because this corner
 corresponds to the standard CDM paradigm (zero sound speed and viscosity) and could be, thus, erroneously excluded by our MCMC analysis.
Moreover, due to correlations with all other cosmological parameters, the corresponding marginalized posteriors of the set of sound speed and viscosity parameters
might also be affected.
\begin{figure*}[t!]
\begin{center}
\includegraphics[width=0.48 \textwidth]{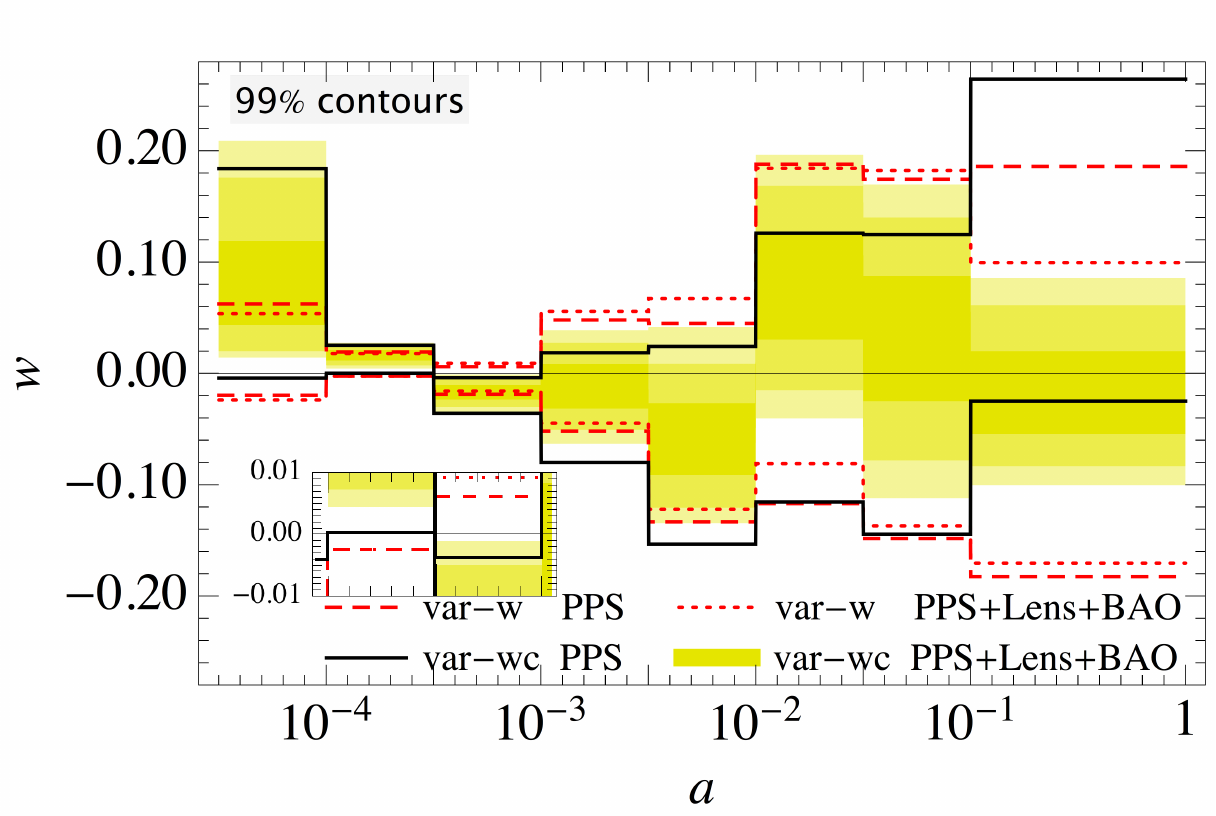}
\includegraphics[width=0.48 \textwidth]{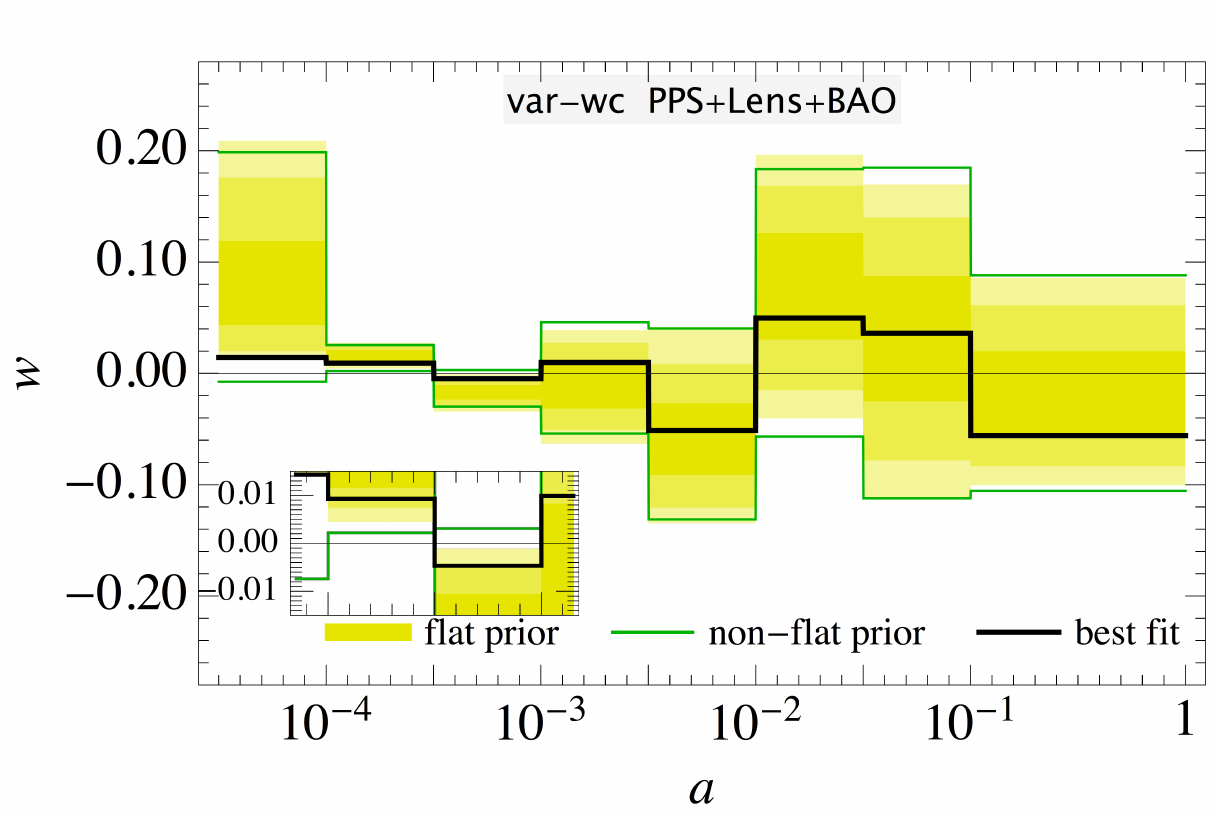}
\end{center}
\vspace{-0.3cm}
\caption{
Shown are the 99\% credible regions on the $w_i$ parameters parametrized the EoS of DM.
The large ticks  on the $a$-axis specify the bin boundaries.
The line styles correspond to different datasets and models specified in the legend.
The insets zoom into the region enclosing $w=0$ and have the same ticks on the $a$-axis. \emph{Left}:
We show the credible regions for the \varwc~and \varW~models when  PPS and PPS+Lens+BAO dataset combinations are used.
 In the \varwc~model, the \Lcdm model (thin black solid line $w=0$) lies outside the 99\% credible region for bins 8, 7 and 6 when PPS+Lens+BAO (yellow shaded regions;
darker shades correspond to $95\%$ and $68\%$) was used while when PPS (black full line) was used only bin 7 is marginally inconsistent with \Lcdm\!\!.
 The \varW~ model (red dashed  and red dotted) is, however, consistent with \Lcdm for all datasets.
\emph{Right}: Comparing flat and nonflat priors in the \varwc~model with the dataset PPS+Lens+BAO combination.
The yellow shaded region corresponds to 99\% (darker shades as on the left) credible regions.
For the nonflat priors (green full line) only bin 7 does not include $w=0$.
The best fit model is shown as thick black line which deviates significantly from the mean
in the early universe. In bin 8 (leftmost bin), the best fit model lies at the lower edge of the lower 99\% credible region of
the flat prior case (lower boundary of the yellow region), while it is well contained for the nonflat prior (full green).
}
\label{varwovertime_Four}
\end{figure*}

In order to alleviate these effects and test the
sensitivity of our constraints on our choice of priors, we also used nonflat priors for $c^2_{s,i}$ and $\cvisi$, keeping the priors on the other parameters unchanged.
For this test, we used flat priors on the combinations
\begin{align} \label{cp2anddDef}
 c^2_{+,i} &\equiv  c^2_{s,i}+\frac{8}{15} \cvisi
\\
 \dpar_{i} &\equiv \frac{15c^2_{s,i} }{15c^2_{s,i}+ 8 \cvisi }
\notag
 \end{align}
which results in nonflat priors for the $c^2_{s,i}$ and $\cvisi$ set of parameters since the measure transforms as
\begin{equation}
dc^2_{+,i} d\dpar_{i}  \propto  \frac{dc^2_{s,i} d\cvisi}{c^{2}_{+,i}}.
\label{cp2dmeasure}
\end{equation}
 We refer to these priors
 as ``nonflat priors'' when discussing the \varwc~and \varc~models.

These priors are physically motivated. During GDM domination the scale below which the gravitational potential decays
is determined by $c_{+} \eta$, where $\eta$ is the conformal time~\cite{ThomasKoppSkordis2016}.
The $\dpar$ parameter interpolates linearly between the two extremes of $100\%$ sound speed or $100\%$ viscosity contribution to $c_{+}^2$.
After the leading-order effect determined by $c_{+}$ sets in, the quantity $\dpar= c_s^2/c_+^2$ results in subleading effects, given a fixed $c_+^2$.
Thus, it seems  natural to assume flat priors on the $\{c^2_{+,i}, \dpar_i\}$ set of parameters
rather than on $c^2_{s,i}$ and $\cvisi$. Flat priors on $c^2_{+}$  and  $\dpar$ translate
then into  priors on $c^2_{s}$ and $\cvis$ which peak at the values $c^2_{s}=0$ and $\cvis =0$, as implied by \eqref{cp2dmeasure}.
 Hence, we call these ``nonflat priors''.
 When viewed  in the $c^2_{s}$-$\cvis$ plane, these nonflat priors give more weight to the CDM ``corner''  $c^2_{s}= \cvis =0$ and thus are expected
to lead to tighter constraints on $c^2_{s}$ and $\cvis$.

\begin{table}
\begin{center}
\begin{tabular}{|l |l |l  |}
\hline
Parameter & Prior & Model\\
\hline \hline
$\omega_b$ & [0., 1.] & all\\
$\omegazero$ & [0., 1.] & all\\
$H_0$ & [45., 90.] & all\\
$\ln(10^{10}A_{s })$ & [2., 4.] & all\\
$n_s$ & [0.8, 1.2] & all\\
$\tau_{\rm reio}$  & [0.01,  0.8] & all\\
\hline
$w_i$  & [-1., 1] & var-w \& var-wc\\
\hline
$c^2_{s, i}$ &  [0., 1.]  & var-c \,\& var-wc\\
$\cvisi$ &  [0., 1.] & var-c \,\& var-wc\\
\hline \hline
\end{tabular}
\end{center}
\caption{List of free cosmological parameters and priors. }
\label{tab:priors}
\end{table}



\section{Results}
\label{sec:results}

\begin{figure*}[t!]
\begin{center}
\includegraphics[width=0.48 \textwidth]{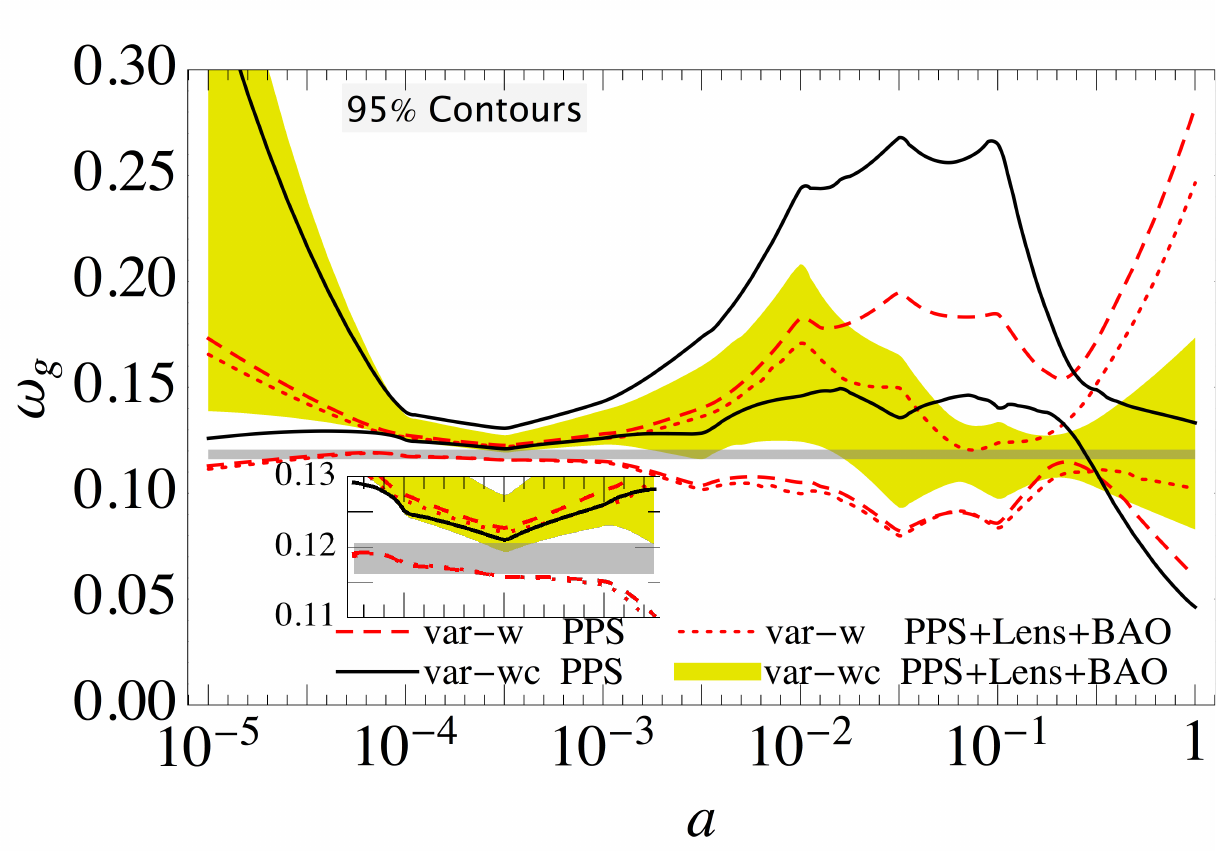}
\includegraphics[width=0.48 \textwidth]{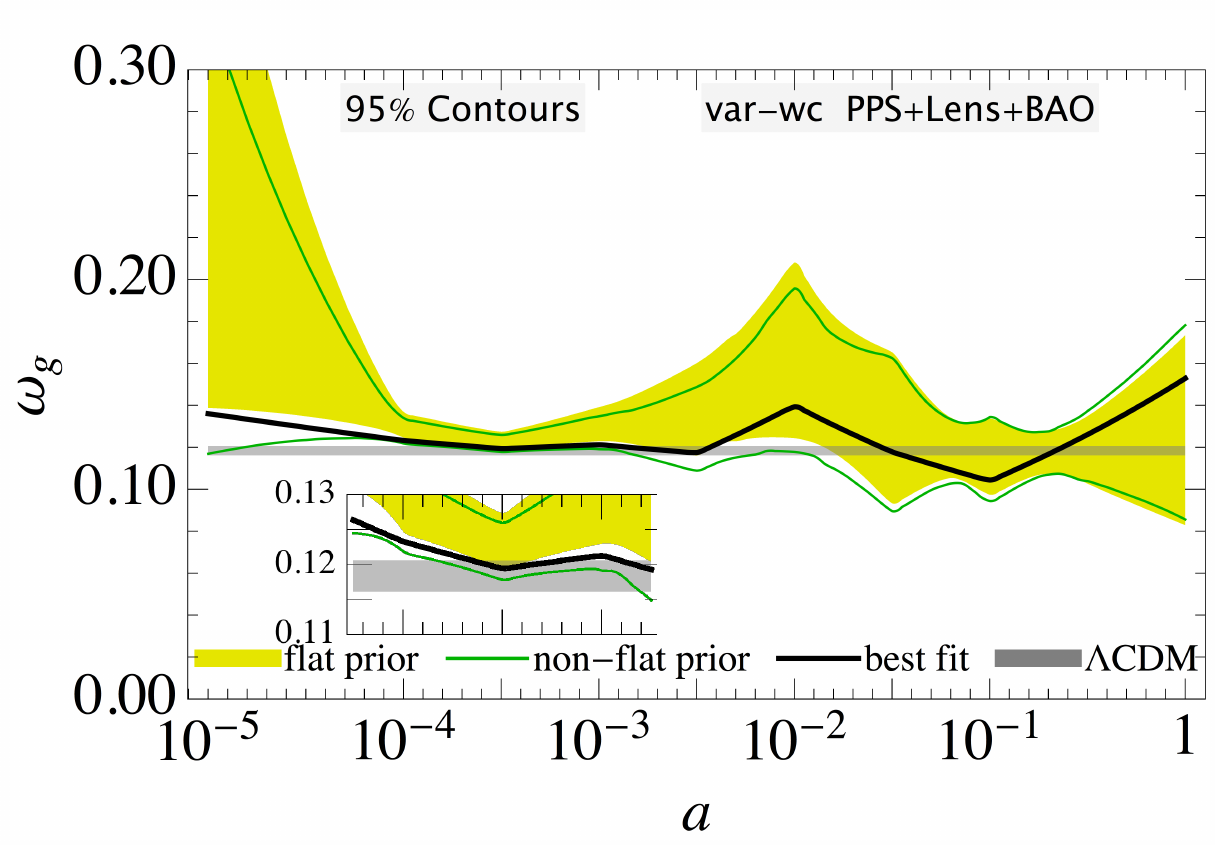}
\end{center}
\vspace{-0.3cm}
\caption{Shown are the 95\% credible regions on the scaled DM abundance $\omega_g(a)$ with same color scheme as in Fig.\,\ref{varwovertime_Four}. The grey band corresponds to the \Lcdm constraint for PPS+Lens+BAO.
}
\label{rhoovertime}
\end{figure*}

The main results of this work are (i) the constraints on the time dependence of the DM EoS $w(a)$ and abundance $\omega_g(a)$ from the \varwc~model shown in Figs.\,\ref{varwovertime_Four}
and  \ref{rhoovertime} and (ii) the constraints on $c^2_s(a)$ and $\cvis(a)$ from \varwc~and \varc~shown in Fig.  \ref{cs2cv2covertime}.
For comparison, we also show the constraints on the \varW~model discussed previously in~\cite{KoppSkordisThomasEtal2018}.

Interestingly, comparing the best (i.e. lowest) $\chi^2$ for the three GDM models to the corresponding $\chi^2$ in $\Lambda$CDM, i.e. $\Delta \chi^2_{\rm GDM} \equiv \chi^2_{\Lambda\mathrm{CDM}} - \chi^2_{\rm GDM}$,
we find  $\Delta \chi^2_{\varwc} \simeq \Delta \chi^2_{\varW} \simeq 8$ and $\Delta \chi^2_{\varc}  \simeq 0$.
To elaborate, adding the 8 new $w_i$ parameters improves the fit only marginally ($\Delta \chi^2 \simeq 8$).
However, adding the 18 new parameters for sound speed and viscosity
 yields virtually no improvement to the fit whether added by themselves (\varc~submodel) or within the full \varwc~model.
We note that since we do not expect our numerous new GDM parameters to be physical, it makes little sense to apply model selection criteria to our GDM models.

The list of the 68\% and 95\% credible regions of the 1\Dim-posteriors as well as best-fit values of the \varwc~and \varc~models for all parameters and datasets
may be found in Appendix~\ref{app:bigtables}. In the following sections we discuss the constraints in detail.


\subsection{Constraints on DM EoS and abundance: \varwc~and \varW}

\subsubsection{Equation of state, $w(a)$}
In Fig.\,\ref{varwovertime_Four} we show constraints on $w$ contrasting several models (\varW,\,\varwc), datasets (PPS, PPS+Lens+BAO)
 and priors (flat and nonflat priors for sound speed and viscosity).  Quite interestingly, we observe on the left panel  and
in the case of the \varwc~model that  \Lcdm lies outside the 99\% credible region in the earliest universe bins ($i=$8,7, and 6) when the PPS+Lens+BAO dataset was used (yellow shaded region).
In contrast, when the same dataset was used to constrain the \varW~model (e.g. with $c^2_s=\cvis=0$) the credible regions of $w_i$
 are consistent with zero, and thus \Lcdm\!\!, in all of the bins (red dotted lines).

Consider now the right panel of Fig.\,\ref{varwovertime_Four} which singles out the \varwc~model constrained with the PPS+Lens+BAO dataset combination
-- the most discrepant with \Lcdm -- on the left panel.
There we display the impact of using different priors on constraining this model: the flat priors on $c_{s,i}^2$ and $\cvisi$ versus the nonflat priors on the same parameters,
the latter corresponding to flat priors on the parameters defined by \eqref{cp2anddDef}. We see that using the nonflat priors
makes the early universe credible regions shift significantly (green lines) so that all bins, except the 7th bin, become consistent with \Lcdm\!\!,
although even the 7th bin's tension with \Lcdm is reduced to $\sim 3\sigma$.
Furthermore, the best fit model lies close to \Lcdm in bins 6, 7 and 8 and so we cannot decisively claim any nonzero detection of $w$.

We consider now the \varwc~versus the \varW~model. Even in the late universe (rightmost) bins where the priors on $c_{s,i}^2$ and $\cvisi$ do not have profound impact
(see right panel of Fig.\,\ref{varwovertime_Four}),  marginalization over $c_s^2$ and $\cvis$ in the \varwc~model shifts the credible regions
significantly (see the left panel of Fig.\,\ref{varwovertime_Four} and compare the red dotted lines with the yellow shaded region).
These differences between the two models at late times are present also when the PPS dataset is used (contrasting red lines versus black dashed line on the left of Fig.\,\ref{varwovertime_Four}).
Clearly then, marginalization over $c_s^2$ and $\cvis$ in the \varwc~model does not lead to the same constraints on  $w$ as simply setting $c_s^2=\cvis=0$ (the \varW~model).
We discuss in more detail the origin of the  differences between these two models in
the late  and  early universe  in Sec\,.\ref{sec:discussion}.

We find strongest constraints on $w$ between $a_6$ and $a_5$ which enclose the matter-radiation equality $a_{\rm eq} \simeq 3\times10^{-4}$.
In other bins the constraints on $w$ weaken significantly.   Adding
the BAO or HST dataset has only a minor effect on \varW~constraints and only
tightens limits in the rightmost bin.
 Contrary to the case of the \constw~model \cite[where $w$ is a constant throughout the evolution of the universe and $c_s^2$ and $\cvis$ are zero,][]{ThomasKoppSkordis2016}, and the \varW~model \cite{KoppSkordisThomasEtal2018},
adding CMB lensing in the \varwc~model significantly shifts the constraints on $w_8$ away from \Lcdm\!\!. This is seen by
comparing the yellow shaded with the solid black region in the left panel of Fig.\,\ref{varwovertime_Four}.
The reasons for this will be discussed in detail in Sec\,.\ref{sec:discussion}.

\subsubsection{Dark matter abundance, $\omega_g(a)$}
The derived parameter $\omega_g(a)$, shown in Fig.\,\ref{rhoovertime}, provides a better intuition on the meaning of the constraints on $w_i$.
The $\omega_g(a)$ parameter is constructed via analytically  integrating \eqref{GDMconservation} given a set of $w_i$ and $\omegazero$.
Figure~\ref{rhoovertime} shows the same model and dataset combinations as Fig.\,\ref{varwovertime_Four}.
Paradoxically, having free sound speed and viscosity in the \varwc~model restricts the posterior of $\omegazero$. This explains why
after marginalizing over $c_s^2$ and $\cvis$, the $w_0$ constraints improve compared to the \varW~submodel. The improvement of the $\omegazero$ posterior in
 \varwc~compared to \varW~is discussed in Sec\,.\ref{sec:discussion}.

The most striking difference with the $w(a)$-constraints is the persistent offset of $\omega_g$ between the \varwc~and \Lcdm models in the early universe, i.e. for
all $a<10^{-2}$.  During the best-constrained period in $a$, around the time of matter-radiation equality $a_{\rm eq} \simeq 3\times10^{-4}$,
we found $\omega^{\rm eq}_g = 0.1236^{+0.0044}_{-0.0041}$ for the $95\%$ credible interval in the case of the \varwc~model when the PPS+Lens+BAO dataset combination was used.
For comparison, we obtain $\omega^{\rm eq}_c = 0.1184^{+0.0021}_{-0.0020}$ in \Lcdm with the same dataset combination.
The same offset is present when only the PPS dataset is used, as
seen in the inset of the left panel of Fig.\,\ref{rhoovertime} (black lines), whereas the \varW~model (red dotted line) leads to virtually the same value for $\omega^{\rm eq}_g$ as $\Lambda$CDM
for both dataset combinations.

\begin{figure*}[t!]
\begin{center}
\includegraphics[width=0.49 \textwidth]{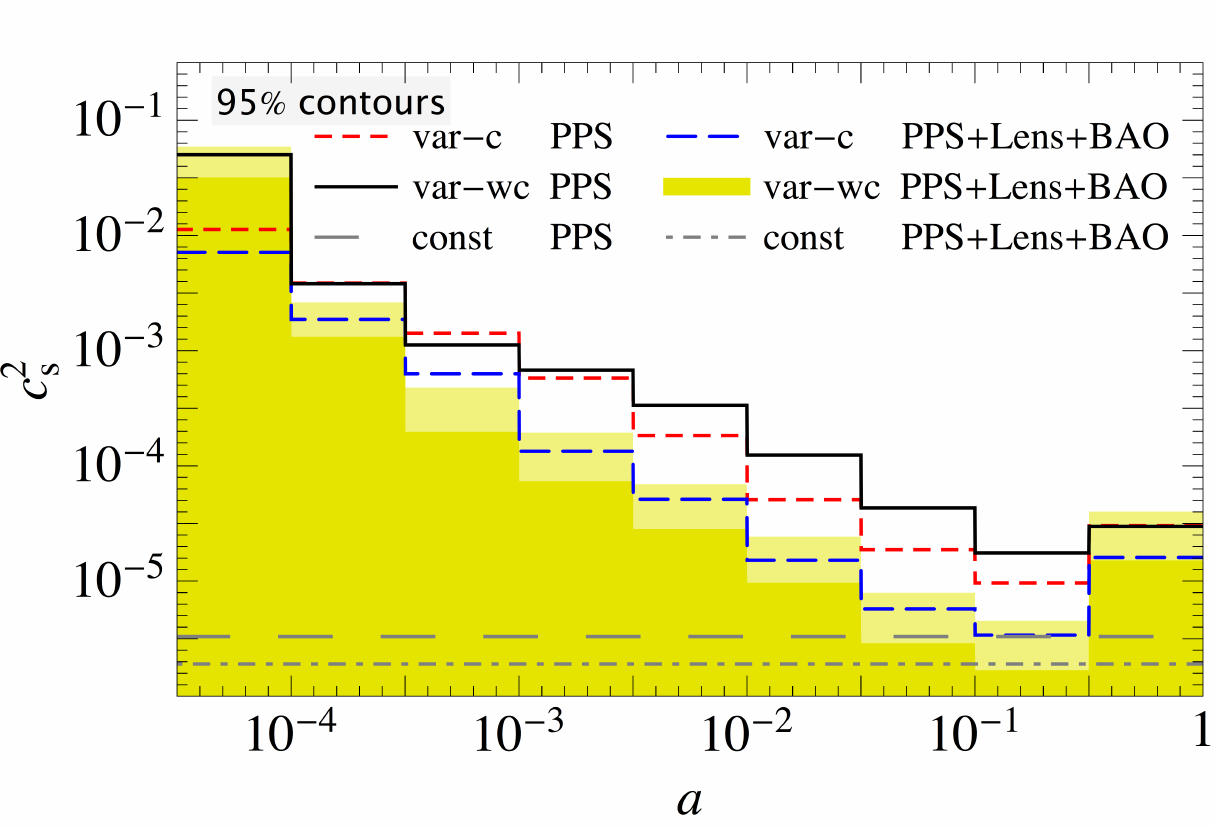}
\includegraphics[width=0.49 \textwidth]{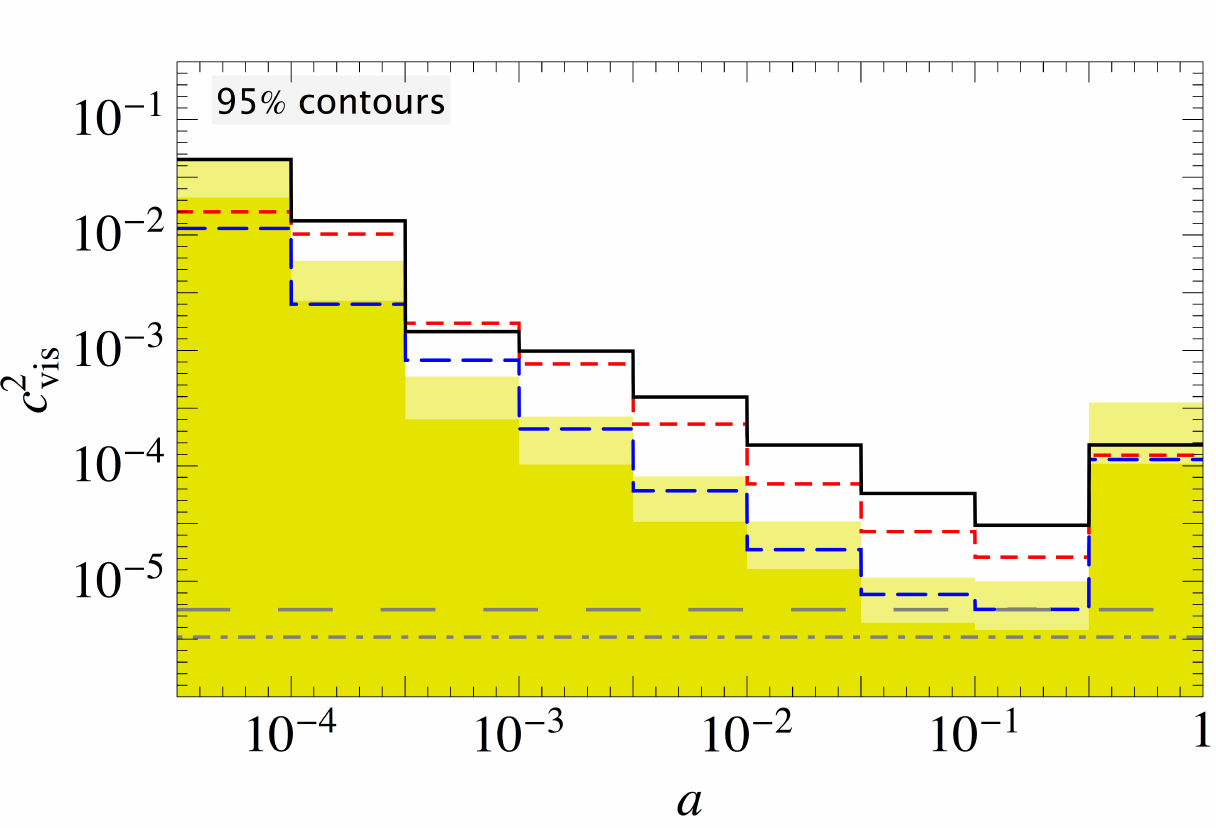}
%
%
\includegraphics[width=0.49 \textwidth]{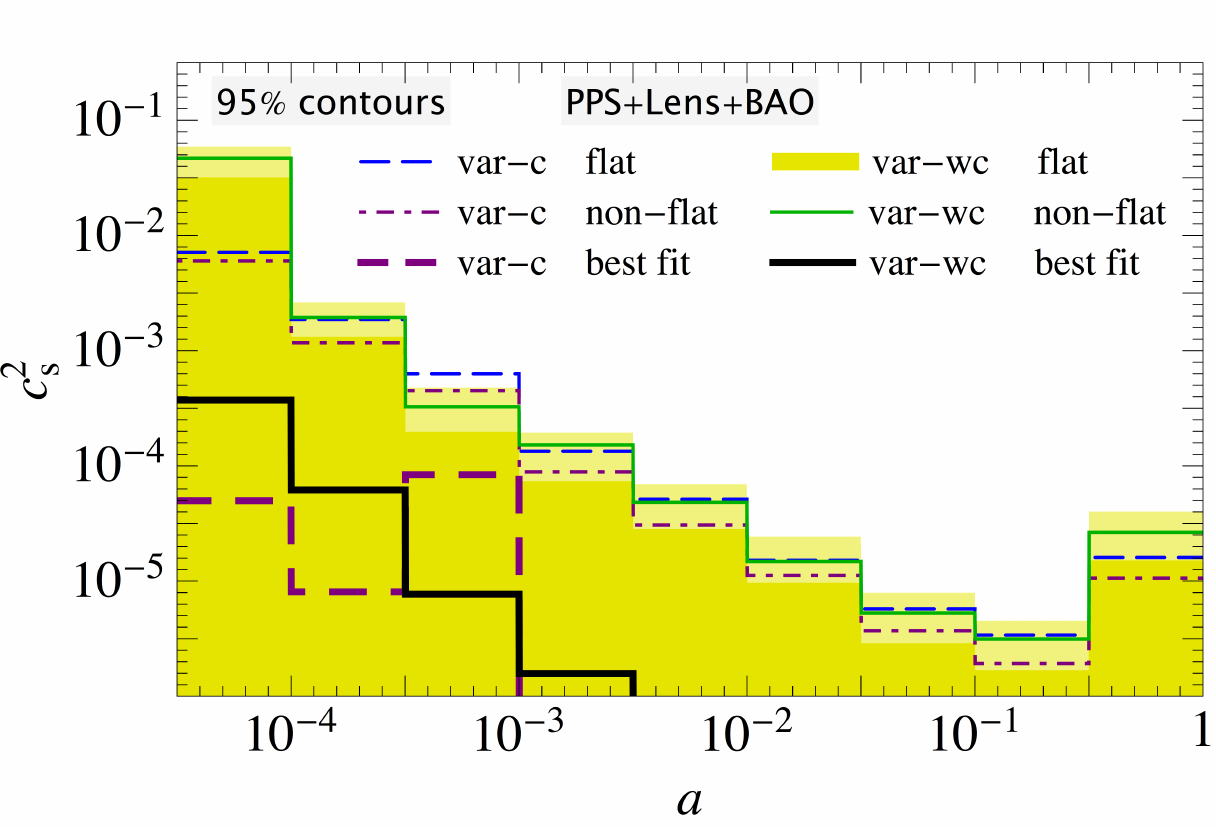}
\includegraphics[width=0.49 \textwidth]{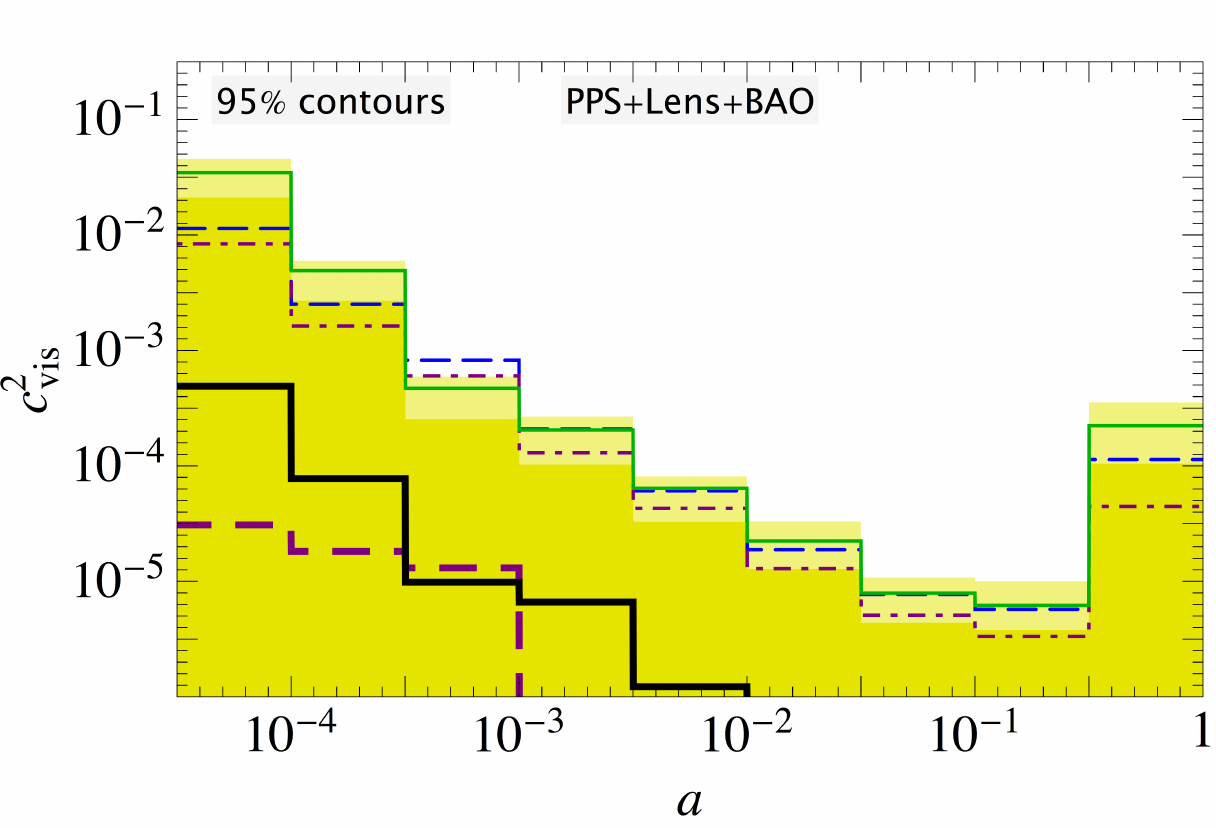}
\end{center}
\vspace{-0.3cm}
\caption{The upper panels display the $95\%$ credible regions of $c_s^2$ (left) and $\cvis$ (right) when PPS alone and the PPS+Lens+BAO dataset combination was used
for constraining the \varc~and \varwc~models. The constraints on constant  $c_s^2$ and $\cvis$ parameters from~\cite{ThomasKoppSkordis2016} are superimposed in grey.
The lower panels display the $95\%$ credible regions of $c_s^2$ (left) and $\cvis$ (right) when the PPS+Lens+BAO dataset combination was used,
 showing the effect of having nonflat priors as well as the best fit model. In all panels does the darker yellow shading display the $68\%$ confidence regions.
Note that the vertical axis is logarithmic so that the constraining power of the later universe data is quite drastic compared to the early universe data.}
\label{cs2cv2covertime}
\label{cs2cv2covertimerepar}
\end{figure*}

The right panel in Fig.\,\ref{rhoovertime} focuses on the impact of priors on $c_s^2$ and $\cvis$. The green lines display the $95\%$ credible regions
 obtained when using nonflat priors on $c_s^2$ and $\cvis$ that give more weight to $\Lambda$CDM, see \eqref{cp2anddDef}.
We see that the flat prior credible region (yellow shaded) is very similar to the nonflat prior region (green lines), suggesting that the offset of $\omega_g$ in the early
universe is not caused by the choice of priors. The black line shows the best fit model obtained through maximization of the log-likelihood.
The best fit model, which does not depend on priors, also stays above the \Lcdm $95\%$ credible region (grey band), favoring  higher values of $\omega_g$ in the pre-recombination era.


\subsection{Constraints on sound speed and viscosity: \varwc~and \varc}

\subsubsection{Constraints on $c_s^2$ and $\cvis$}
We now turn to the constraints on the perturbative GDM parameters $c_{s,i}^2$ and $\cvisi$ with $0 \leq i \leq 9$.
In the upper panel of Fig. \ref{cs2cv2covertime} we compare the \varwc~and \varc~models (for which $w=0$) for each of the dataset combinations, PPS and PPS+Lens+BAO.
We also display the  constraints on constant $c_s^2$ and $\cvis$, labeled  as ``const'', found previously in \cite{ThomasKoppSkordis2016}
using the same respective dataset combinations (dashed grey and  dot-dashed grey lines).
We see from the upper panel of Fig. \ref{cs2cv2covertime}  that  CMB alone (PPS) constrains the $c_{s,i}^2$ and  $\cvisi$ in all redshift bins.
The best constraints, nearly as good as for the constant parameter case (``const''), are achieved in the bin $i=1$ for which $0.1<a<10^{-0.5}$ (redshift $2.2\lesssim z<9$),
where the gravitational lensing of the CMB is most efficient \cite[see e.g.][]{Manzotti2017}.
 This secondary anisotropy, which smoothes the amplitude of the peaks and troughs of the CMB spectra without changing their location,
 most strongly constrains the perturbative GDM parameters.

\begin{figure*}[t!]
\begin{center}
\includegraphics[width=0.39 \textwidth]{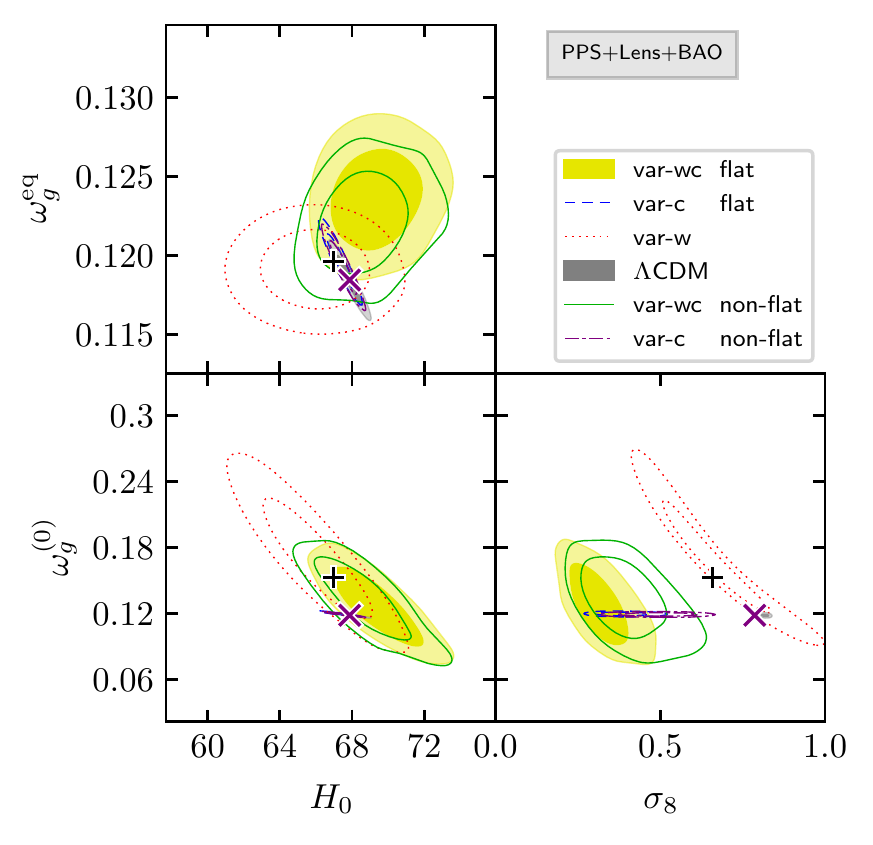}
\includegraphics[width=0.37 \textwidth]{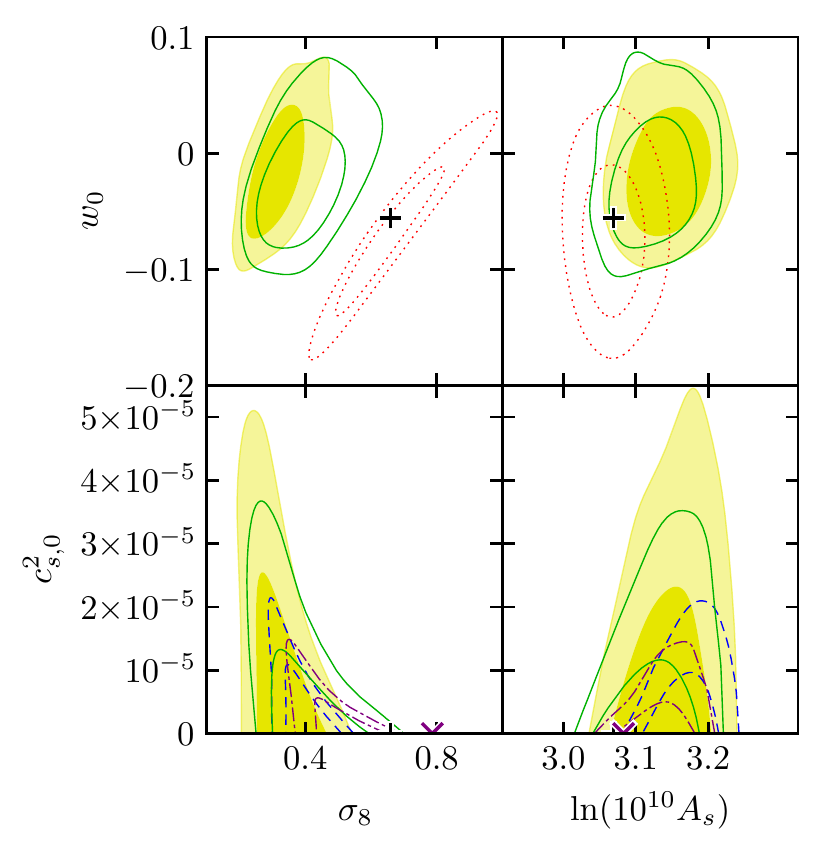}
\raisebox{0.1cm}{\includegraphics[width=0.22 \textwidth]{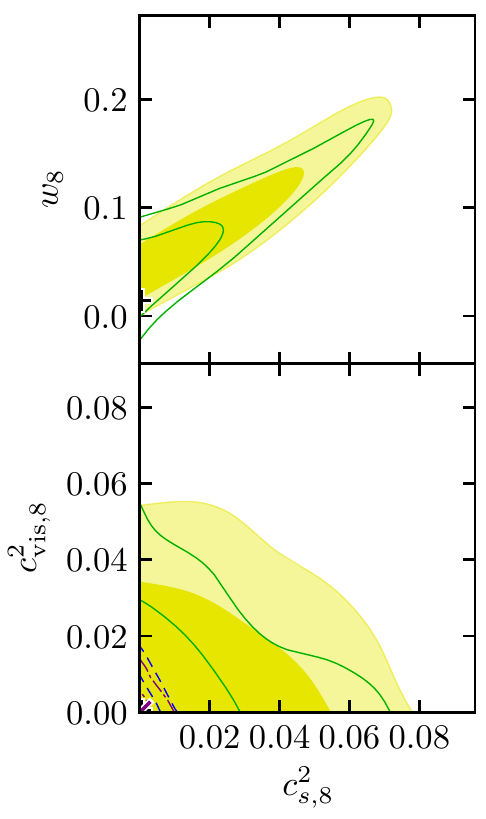}}
\end{center}
\vspace{-0.3cm}
\caption{
The 2\Dim~contours of the $68\%$ and $95\%$ credible regions of various parameter combinations in the set $\{\omega^{(0)}_g,\omega^{{\rm eq}}_g,H_0,\sigma_8,w_0,w_8,A_s,c^2_{s,0},c^2_{s,8},\cvisI{8}\}$ when the PPS+Lens+BAO dataset combination was used,
showing the effect of nonflat priors. Displayed are the \varwc~model with flat priors (yellow shades) and nonflat priors (green lines),
the \varc~model with flat priors (blue dashed lines) and nonflat priors (purple dot-dashed lines),
the \varW~model (red dotted lines) and \Lcdm (grey shades).
The best-fit points are indicated by a black plus for the \varwc~and purple cross for the \varc~model.}
\label{2D_contours}
\end{figure*}

The earliest parameters $c_{s,8}^2$ and $\cvisI{8}$ are mostly constrained, however, through the primary CMB anisotropies.
This may be inferred by comparing the yellow and black regions of the top two panels of Fig.~\ref{cs2cv2covertime},
or the PPS and PPS+Lens columns
(blue numbers)
of Table~\ref{tab:alldatasets1D} (see Appendix~\ref{app:bigtables})
which indicates that CMB lensing (Lens dataset) indeed improves all constraints by a factor 2-4 except for $c_{s,8}^2$ and $\cvisI{8}$.
Note that this is a much bigger improvement than for the constant parameters, where BAO+Lens reduced the upper limits by a factor less than 2~\cite{ThomasKoppSkordis2016}.


The effect of using different priors, flat versus nonflat (see Sec.~\ref{sec:priors}), is depicted in the lower panel of
 Fig.\,\ref{cs2cv2covertimerepar}. There, only the PPS+Lens+BAO dataset combination is chosen. We also show the best-fit \varc~and \varwc~models which are prior-independent.
 Since the nonflat prior favors the $\Lambda$CDM corner in the $c_s^2$-$\cvis$ plane, we expect tighter constraints in the nonflat prior case. This is what is observed
 in the lower panels of
in Fig.\,\ref{cs2cv2covertimerepar}.\footnote{Note that there are only upper limits on the perturbative GDM parameters and their best fit value are always significantly closer to zero than their upper limit.}

\subsubsection{Constraints on the $c_+^2$ and $\dpar$ combinations}
In Table~\ref{tab:constraints_cp2} (see Appendix~\ref{app:bigtables})  we summarize the constraints on $c_{+,i}^2$ in the  nonflat prior case.
The shape of the upper limits on $c_{+,i}^2$ follows those of $c^2_{s,i}$  (and of $c^2_{{\rm vis}, i}$) shown in the lower panels of Fig.\,\ref{cs2cv2covertimerepar},
but are about a factor of 3 larger. This is partly a consequence of error propagation and partly an effect of the nonflat priors.
We remind the reader that ``nonflat'' priors refer to \emph{flat priors} on $c_{+,i}^2$  and $\dpar_i$ as described in detail in Sec.~\ref{sec:priors} and specifically \eqref{cp2anddDef}.
Thus constraints on $c^2_{s,i}$ and $\cvis$ are stronger than those on $c_{+,i}^2$.

Interpreting the constraints on  $c_{+,i}^2$ by imposing flat priors on $c_{s,i}^2$ and $\cvisi$ is difficult since even a uniform 2\Dim-posterior
in the $c_{s,i}^2$-$\cvisi$ quadrant would  lead to a peak at nonzero $c_{+,i}^2$  for the 1\Dim-posterior of $c_{+,i}^2$.
This may be understood through  \eqref{cp2dmeasure} which implies that the 1\Dim-marginalized prior for $c_{+,i}^2$ is proportional to $c_{+,i}^2$.
The parameter $0<\dpar<1$ remains largely unconstrained at present. It is expected, however, that structure formation data which include smaller scales could provide constraints on $\dpar$,
 or put differently, break the degeneracy between $c_s^2$ and $\cvis$ \cite{ThomasKoppMarkovic2019}.

\begin{figure*}[t!]
\begin{center}
\includegraphics[width=0.45 \textwidth]{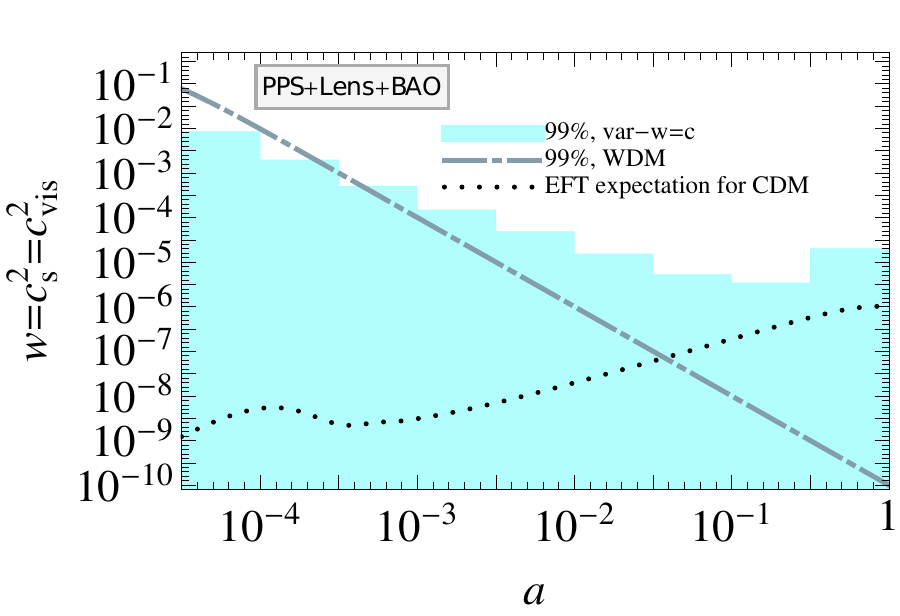}
\includegraphics[width=0.44 \textwidth]{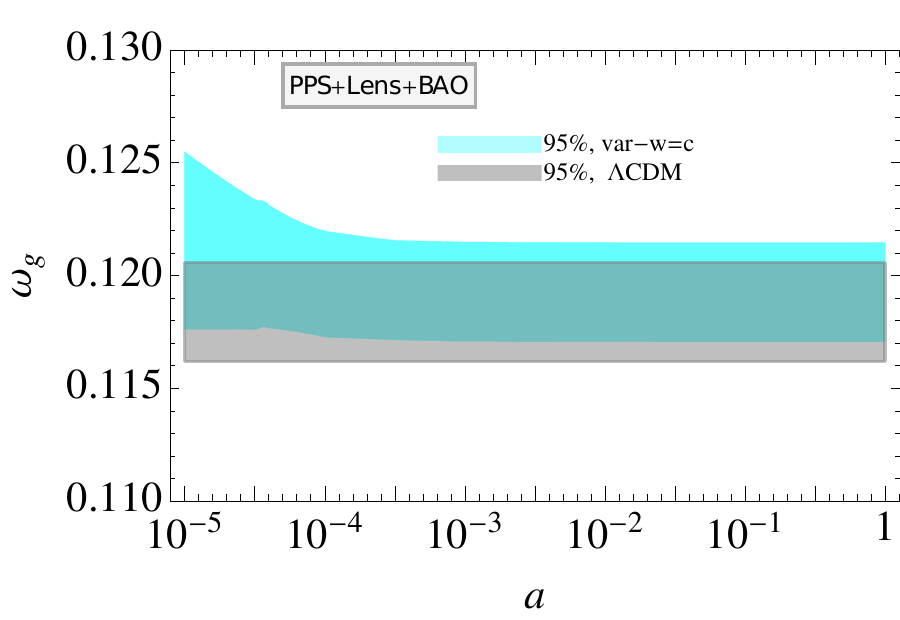}
\end{center}
\vspace{-0.3cm}
\caption{{\it Left:}  Shown are the $99\%$ credible regions (shaded blue) of $w_i=c_{s,i}^2=\cvisi$ model when the dataset combination PPS+Lens+BAO was used.
The blue dot-dashed line is the constraint from \cite{KunzNesserisSawicki2016} while
the dotted line is the rough expectation from the EFTofLSS \cite[see][]{KoppSkordisThomas2016}.
{\it Right:} $95\%$ credible regions  on $\omega_g(a)$ contrasted with those of $\Lambda$CDM in grey.
}
\label{weqcovertime}
\end{figure*}

\subsection{Degeneracies and shifts in the credible regions}
In Fig.\,\ref{2D_contours} we display several 2\Dim-posteriors for the models considered when the PPS+Lens+BAO dataset combination was used.
As evidenced by this figure the best-fit parameters of the models containing $c^2_{s,i}$ and $\cvisi$ (that is \varwc~and \varc)
 lie significantly outside their credible regions.
 This is a consequence of the location of the peak of our likelihood (i.e. the best-fit point) being close to the border of our prior volume, which itself has a nontrivial shape.
The best-fit point of the \varwc~model is marked with a black plus sign and the corresponding one for the varc~submodel with a purple cross.
Choosing nonflat priors reduces, quite generally, the distance between the best-fit points and the credibility contours, confirming our suspicion that the effects
 coming from a choice of priors may be responsible.
We find that the best-fit parameters lie within the credible regions of the corresponding nested models with $c^2_{s,i}=\cvisi=0$,
consistent with the fact that adding the perturbative GDM parameters does not increase the goodness of fit.
Specifically, the best-fit point of \varwc\ lies inside the $68\%$ credible region of \varW,
 and similarly the best-fit point of \varc\ lies inside the $68\%$ credible region of $\Lambda$CDM.

As was already observed when constraints on constant GDM parameters were obtained in an earlier work \cite{ThomasKoppSkordis2016},
there is a degeneracy between the present day values $\omegazero$, $H_0$ and $w_0$.
This degeneracy persists also in the  \varwc~model.

Even more interesting are the various shifts of the credible region contours of the \varwc~model (yellow shades and green lines)
  with respect to either the \varW~(red dotted) or the \Lcdm models.
For some parameter combinations, a shift is also seen between the  \varc~(blue dashed and purple dot-dashed) and the \varW/\Lcdm models, e.g. in the $\{\omega^{(0)}_g,\sigma_8\}$-plane ,
while for others no such shift is observed.
Thus these shifts occur through the interplay between $w$ and $c_s^2$ and $\cvis$.

The most important shift is the one occurring in the $\omega_g^{\rm eq}$-$H_0$ and $\omegazero$-$H_0$ planes, both cases involving $H_0$.
This shift allows the Hubble constant $H_0$ to be pushed to higher values today $H_0 = 69.3^{+3.3}_{-3.0}$ than the $\Lambda$CDM
value $H_0 = 67.89^{+0.93}_{-0.93}$, while keeping $\omegazero$ centered  closer to the $\Lambda$CDM value than in the case of the \varW~model.
The physical mechanism for the increased value of $H_0$ is related to the increased value
of $\omega_g^{\rm eq}$ and is discussed below in Sec.\,\ref{sec:discussion}.

Another shift occurs in the clustering strength at late times, $\sigma_8$,  and at early times, $A_s$
seen in the middle panel of Fig.\,\ref{2D_contours}. While the presence of
non-negative $c_s^2$ and $\cvis$ in the \varc~model is sufficient to shift $\sigma_8$ toward smaller values compared to $\Lambda$CDM,
in combination with letting $w$ free in the \varwc~model, the $\sigma_8$ parameter shifts to even smaller values, while $A_s$ increases, see the middle panel of Fig.\,\ref{2D_contours}.
The  increase in $A_s$ is a consequence of the $w_8$-$c_{s,8}^2$ degeneracy seen in the right panel of Fig.\,\ref{2D_contours}.
As we explain in Sec.\ref{sec:discussion} the $w_8$-$c_{s,8}^2$ degeneracy
is caused by properties of the adiabatic initial conditions in the GDM model.
This degeneracy, in combination with the positivity of $c_{s,8}^2$, is also responsible for the increase in $\omega_g^{\rm eq}$
and  all the other shifts discussed above.

It is also worth pointing out that the usual degeneracy between $c_{s}^2$ and $\cvis$ which keeps the combination $c_{+}^2=c_{s}^2 + \tfrac{8}{15} \cvis$ constant
 is broken in bin 8 for the case of the \varwc~model, as seen on the lower right panel of Fig.\,\ref{2D_contours}.
This is because of the dependence of the adiabatic initial conditions on $w_8,c_{s,8}^2$ and $\cvisI{8}$. The adiabatic initial conditions
may be found in \cite{KoppSkordisThomas2016}, however, \eqref{delta_g_LS} below shows how this effect works.
This degeneracy is restored in the later bins, which is reflected in the red, and thus positive, diagonal in the $c_{s,i}^2$-$\cvisi$
correlation matrix displayed in Fig.\,\ref{corrmatComparison} (see Appendix~\ref{app:bigtables}).


\subsection{Constraints on the submodel \varweqc} \label{varweqc}
The final submodel of \varwc~we study is the case where $w_i=c_{s,i}^2=\cvisi$, denoted by \varweqc. This case is interesting because freely streaming matter satisfies
this condition either exactly (in case of ultrarelativistic radiation), or approximately. One example of the latter
is the warm dark matter (WDM) model. A second example of an approximate $w_i=c_{s,i}^2=\cvisi$ relation is the CDM model on linear scales
with the inclusion of backreaction terms coming from integrating out nonlinear scales as in the effective field theory of large-scale structure \cite[EFTofLSS,][]{BaumannNicolisSenatoreEtal2012,CarrascoHertzbergSenatore2012,CarrollLeichenauerPollack2013,ForemanSenatore2015}.

 In the left panel of Fig.\,\ref{weqcovertime} we show our constraints on the $w_i=c_{s,i}^2=\cvisi$ submodel in shaded blue. We superimpose
 the WDM constraints from \cite{KunzNesserisSawicki2016} (dot-dashed line) where approximately $w=c^2_s=\cvis =(\frac{1}{3} + \frac{c_{s,0}^2}{a^{2}})^{-1}$
and also a rough estimate for the EFTofLSS (black dashed line) discussed in \cite{KoppSkordisThomas2016}.

 In the right panel we show the derived parameter $\omega_g(a)$. Comparing to \varwc~ and other submodels in Fig.\,\ref{rhoovertime}, it is clear that the DM abundance in this submodel
 is much more tightly constrained. This is due to the much tighter constraints on $w_i$ which in the late universe are
driven by the upper limits on $c_{s,i}^2=\cvis$ which cannot be larger than the upper limits on $\{w_i$, $c_{s,i}^2$, $\cvisi\}$
in the \varwc~model, so that they approximately follow those of $c_{+,i}^2$ in Table~\ref{tab:constraints_cp2}.
The constraints on standard cosmological parameters in the \varweqc~model are as tight as in $\Lambda$CDM,
with all 2\Dim-posteriors overlapping except for $\sigma_8$ caused by the diminished DM growth due to the non-negative sound speed and viscosity.
The constraint on the Hubble constant $H_0=67.60^{+0.96}_{-0.93} $ (95\%.C.L.) is virtually the same as corresponding $\Lambda$CDM value of $H_0=67.89{}^{+0.93}_{-0.93}$.

A recurring theme throughout our results is the improvement of the constraints on $c^2_s$ and $c^2_\text{vis}$ due to the CMB lensing spectrum, which is most significant for bin 1 (as shown by Fig.~\ref{cs2cv2covertimerepar}). The lensing spectrum used in this work will be substantially improved by upcoming CMB experiments, such as Simons Observatory \cite[][see Fig.~6 in that work for the expected improvement in the noise compared to the data used in this paper]{Ade:2018sbj}. This improvement from upcoming surveys is particularly significant given the EFTofLSS model line in Fig.~\ref{weqcovertime}. This model has no free amplitude parameter that can be used to shift the model prediction up or down, and the upper limit we have achieved in bin 1 is only a factor of a few above the model. Given that this bin benefits the most by the inclusion of the lensing spectrum, a detection of an EFTofLSS-type GDM signal is likely with the improved lensing spectrum from upcoming experiments.

\begin{figure*}[t!]
    \begin{center}
    \includegraphics[width=0.32 \textwidth]{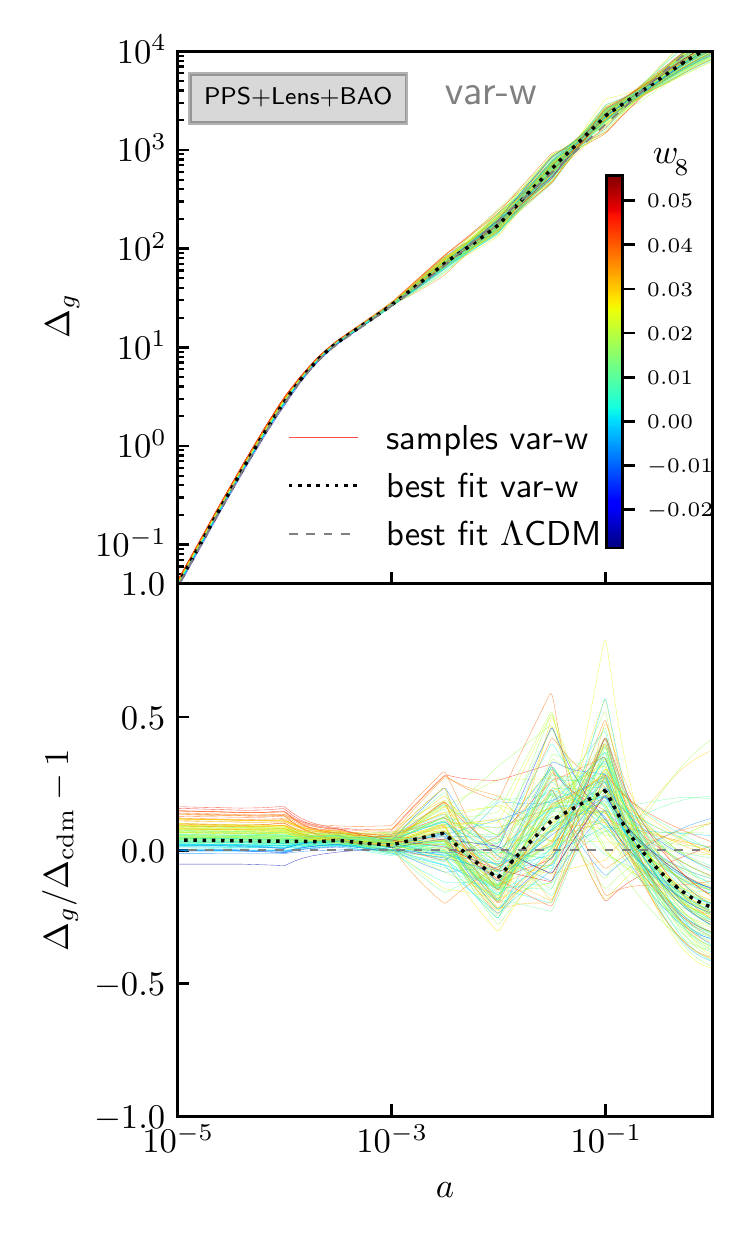}
    \includegraphics[width=0.32 \textwidth]{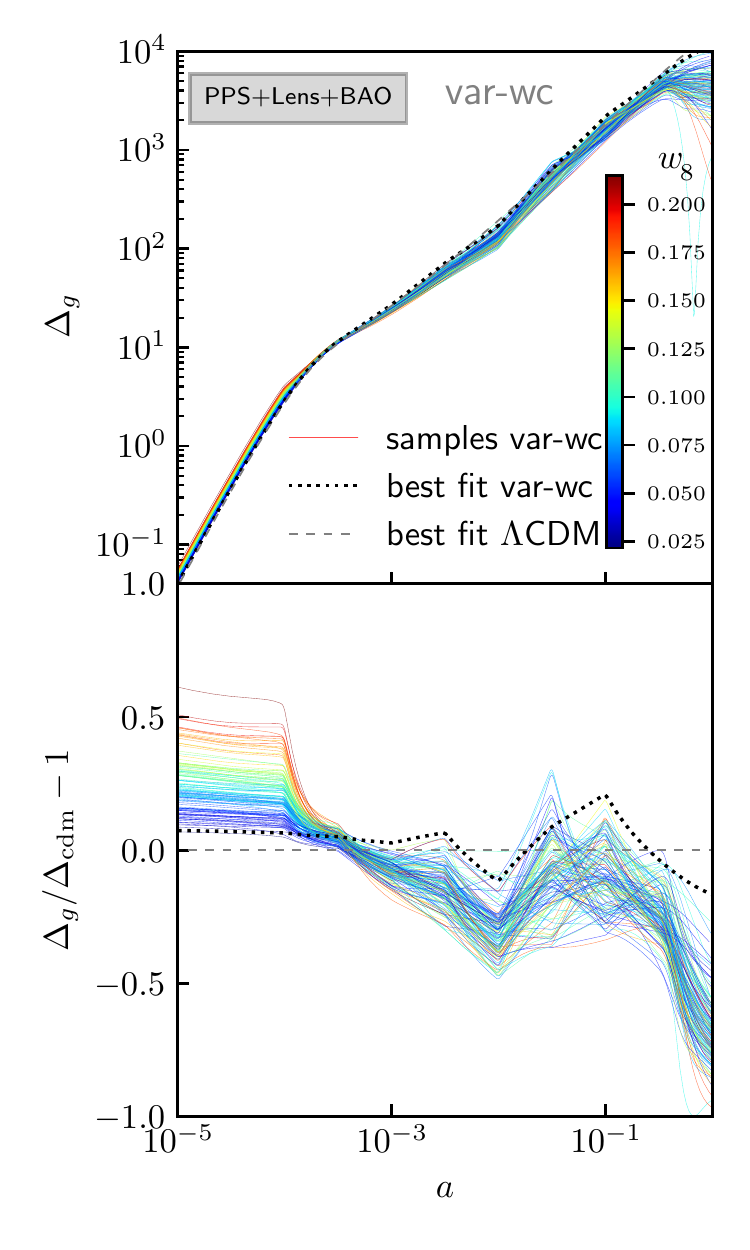}
    \includegraphics[width=0.32 \textwidth]{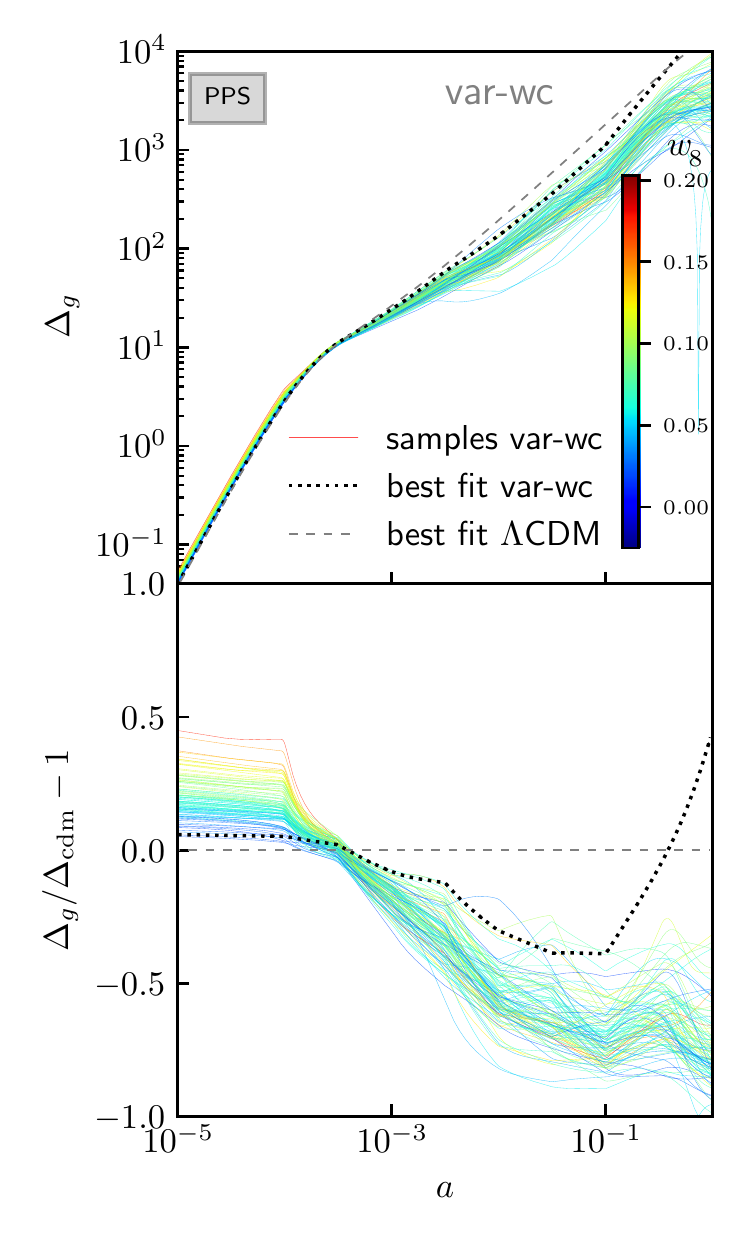}
    \end{center}
    \vspace{-0.7cm}
    \caption{
    Samples from a MCMC chain, showing the evolution of the GDM comoving density perturbation $\DeltaGI_g(a)$ at a fixed scale $k=0.085\,\mathrm{Mpc}^{-1}$.
    In the lower set of panels we show the relative difference compared to the $\Lambda$CDM best-fit in order to make the effects described in the text more visible.
     Just as in the case of  $\omega_g(a)$, $\DeltaGI_g(a)$ also squeezes to a small allowed range around the time of matter-radiation equality $a_{\rm eq}$.
    This is facilitated by a spread at early times due to the $w$-dependence of the time evolution of superhorizon modes during the radiation era.
    On the \emph{left} we show the case of the \varW~model which is to be contrasted with the \varwc~model in the \emph{middle} panel,
    both according to the PPS+Lens+BAO dataset combination.
    The degeneracy between $c^2_{s,8}$ and $w_8$ in the adiabatic mode solution during that time shifts
     the distribution to larger and positive values of $w_8$ as compared to \varW.
    The \emph{right} panel shows the evolution of $\DeltaGI_g(a)$ in \varwc~model according to PPS alone.
    Observe that the best fit (dotted) is not affected by this shift but it remains close to the \Lcdm\!\! case (dashed).
    }
    \label{Deltaovertime}
    \end{figure*}


\section{Discussion and Implications}
\label{sec:discussion}

In this section we discuss the physical mechanism underlying our most interesting results, namely the increase of $\omega_g$ in the early universe,
leading to an increase of $H_0$ for the \varwc~model.

Let us first examine the origin of the increase of $\omega_g^{\rm eq}$, the value of  $\omega_g$ at matter-radiation equality, in the \varwc~model compared to either \varW~or $\Lambda$CDM.
Increasing $\omega_g^{\rm eq}$ is ultimately tied to  \varwc~favoring a higher EoS $w$ in the early universe.
To exemplify, the heights of the CMB peaks severely constrain the evolution of the potential $\Phi$ between radiation and matter eras.
Specifically, $\Phi$ can only evolve within a narrow band of allowed values, otherwise it would cause either too much or too little
 early ISW and acoustic driving in the CMB. These two effects are
controlled by  $a_{{\rm eq}}$  which in turn translates to a small range of allowed values for $w_{\rm eq}$ (leading to the $w_{\rm eq}$-$\omega_g^{\rm eq}$ degeneracy).

Now, during the matter era $\Phi$ is closely linked to the GDM comoving density perturbation $\DeltaGI_g = \delta_g + 3 a H (1+w) \theta_g$,
hence, this narrow band of values for $\Phi$
translates to an equivalent range in $\Delta_g$. However, during the radiation era $\Phi$ is sourced by $\Delta_{{\rm radiation}}$ and thus the link to $\Delta_g$ is broken.
Therefore, the data select trajectories for $\Delta_g$ which may start within a fairly wide range of initial values but
subsequently all squeeze within a narrow range of values around the time of matter-radiation equality.
 This is precisely what is observed in Fig.\,\ref{Deltaovertime}.
There  we plot the evolution of a single $k$-mode ($k= 0.085 {\rm Mpc}^{-1}$) of $\DeltaGI_g$ for a representative sample of our Monte Carlo Markov Chains
 color-coded  by their $w_8$ value. The effect just described is clearly visible in all panels, but less so in the left which displays the $\DeltaGI_g$ evolution in the
\varW~submodel.  This behavior  is similar to the GDM abundance $\omega_g$ which squeezes within a narrow range of  values around $a_{\rm eq}$ in Fig.\,\ref{rhoovertime}.\,
Thus, the CMB constrains both the DM abundance and the amplitude of DM perturbations most strongly around $a_{\rm eq}$.

Although the squeeze in $\DeltaGI_g$ is present in both models, in the \varwc~model it is more pronounced.
The reason is that the evolution of $\DeltaGI_g$ is affected by both $w$ and $c_s^2$. Meanwhile,
during the radiation dominated era, the GDM density perturbation in the synchronous gauge evolves on large scales ($k \eta \ll 1$) as
\begin{equation}
\delta_g = \zeta_{\rm ini} \left(  -\frac{1}{4} + \frac{3c^2_s-5w }{8} \right) (k \eta)^2,
\label{delta_g_LS}
\end{equation}
when adiabatic initial conditions specified by the initial curvature perturbation  $\zeta_{\rm ini}$ are set~\cite{KoppSkordisThomas2016}.
Therefore, the GDM parameters $w_8$ and $c^2_{s,8}$ affect the initial amplitude of the GDM density perturbations and consequently the initial $\DeltaGI_g$. While $w_8$ takes
both positive and negative values, $c^2_{s,8}$ must be non-negative. The two compensate each other only for $w_8>0$, implied by \eqref{delta_g_LS},
leading to the degeneracy shown in the right panel of Fig.\,\ref{2D_contours}. However, the slope of the degeneracy is not what one would expect from
\eqref{delta_g_LS} under the naive assumption $\DeltaGI^{\rm ini}_g = \DeltaGI_{\rm cdm}^{\rm ini}$, because both $w_8$ and $c^2_{s,8}$ affect the subsequent evolution of $\DeltaGI_g$ into the squeezed region around $a_{\rm eq}$.
Nevertheless, this degeneracy combined with the squeezed region around $a_{\rm eq}$ drive the posteriors of both $w_8$ and $c^2_{s,8}$ to more positive values.
Due to the $w-\omega_g$ degeneracy discussed first in~\cite{Hu1998a} and recently demonstrated in~\cite{ThomasKoppSkordis2016},
once $w_8$ shifts to positive values it then leads to an increase of the early universe abundance of DM $\omega^{(8)}_g$ and consequently $\omega_g^{\rm eq}$.

Before discussing the increase of $H_0$ in the \varwc~model, we briefly comment on a few minor points.
First, we showed in Sec.\ref{sec:results} that
 marginalizing over $c_s^2$ and $\cvis$ in the \varwc~model does not exactly lead to the same posteriors as in the \varW~submodel
 where $c_s^2=\cvis=0$. This is a consequence of our discussion of the $w_8$ and $c^2_{s,8}$ degeneracy combined with the squeeze in $\DeltaGI_g$ at equality.
Second, adding the Lens dataset widens the 1\Dim-posteriors of $c^2_{s,8}$, compared to PPS alone, a
somewhat counter-intuitive result also inferred from Table~\ref{tab:alldatasets1D}.  In the right of Fig.\,\ref{Deltaovertime}
we plot the $\DeltaGI_g$ samples using PPS alone, showing a distribution of curves shifted toward smaller $ \DeltaGI_g/ \DeltaGI_{\rm cdm} -1 $ in the late universe.
This is the result of diminished constraining power on DM clustering at that time and
since growth is an integrated effect, the drop of $ \DeltaGI_g/ \DeltaGI_{\rm cdm} -1 $ then requires a more CDM-like growth in the early universe,
thus favoring  smaller values of $c^2_{s,8}$.
Lastly, since $c^2_{s,8}$ and $w_8$ are correlated,
adding the Lens dataset favors larger values of $w_8$ and explains why this dataset affects significantly the constraints on $w_8$ in the \varwc~model.

\begin{figure*}[t!]
    \begin{center}
    \includegraphics[width=0.47 \textwidth]{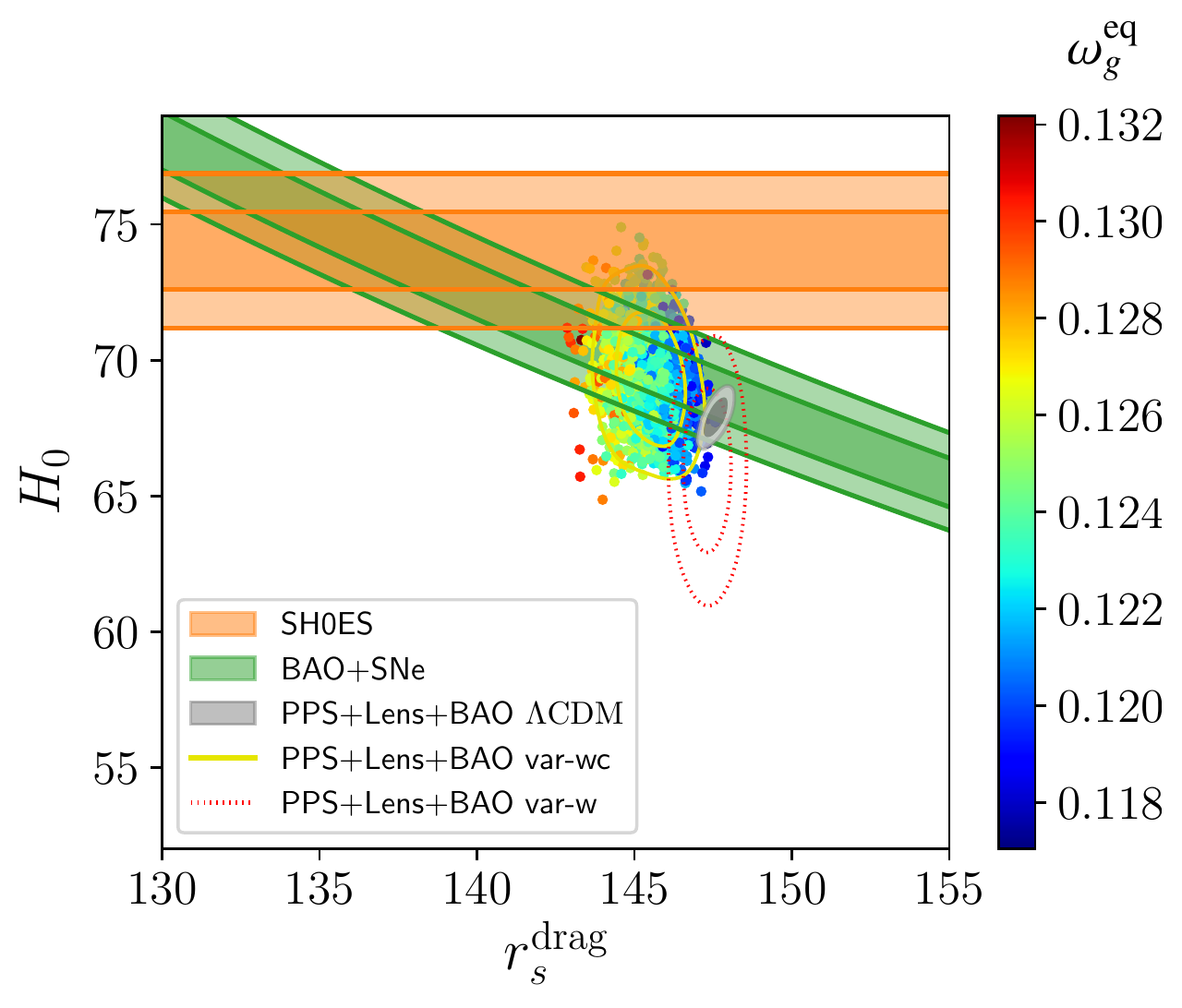}
    \qquad \includegraphics[width=0.37 \textwidth]{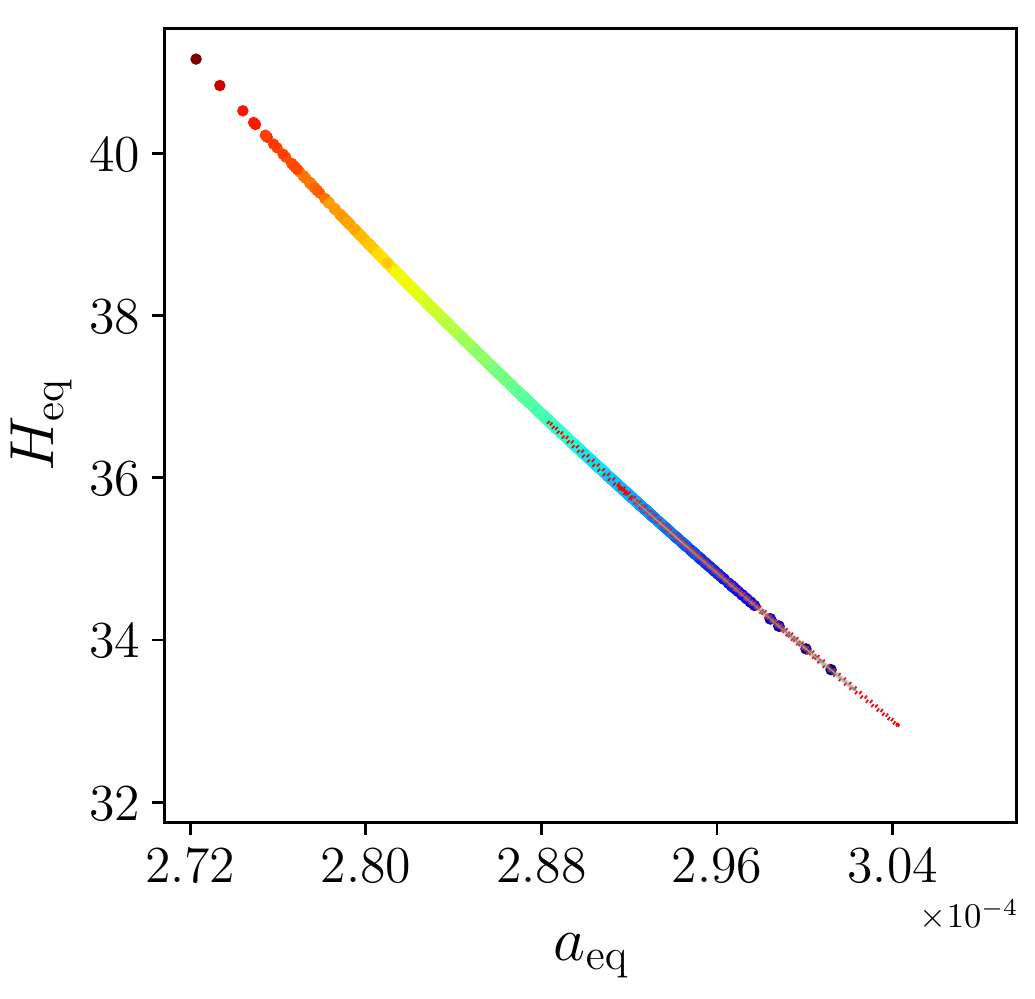}
    \end{center}
    \vspace{-0.3cm}
    \caption{{\it Left: } We show the 68\% and 95\% credible regions for various datasets and models used to constrain $H_0$ and $r_s^{\rm drag}$.
    The SH0ES and the BAO+SNe credible regions are adopted from \cite{KnoxMillea2019}.
    {\it Right: } We show the 68\% and 95\% credible regions of the $H_{\rm eq}$ and $a_{\rm eq}$ plane,
     demonstrating  that $\omega_g^{\rm eq}$ uniquely determines $H_{\rm eq}$ and $a_{\rm eq}$.
    }
    \label{rsdragH0plane}
\end{figure*}

We now discuss the reasons for the increase of the mean of $H_0$ in the \varwc~model, making $H_0$ more consistent with supernovae estimates of its value.
 As was explained in \cite{KoppSkordisThomasEtal2018}, there is strong degeneracy between $\omegazero$, $w_0$ and $H_0$ in the \varW~model,
since these late universe parameters determine a large fraction of the angular diameter distance $d^*_A$ to the last scattering surface.
This degeneracy is still observed to be present in the \varwc~model. However, as can be seen in the left panel of Fig.\,\ref{2D_contours},
the contours in the $\omegazero-H_0$ plane shift to larger values of $H_0$ and smaller values of $\omegazero$ along the degeneracy direction,
compared to the \varW~case. This happens because increasing $w_0$ allows easier structure  growth
 in the late universe (see $w_0$-$\sigma_8$ degeneracy in Fig.\,\ref{2D_contours} and App.\,\ref{app:sigma8w0Degeneracy}).
%
%
In the \varwc~model  the perturbative GDM parameters shift $\sigma_8$ to small values
and the more negative values of $w_0$ become disfavored  by CMB lensing, so that overall $w_0$ tends to be more positive. 
This tighter and more positive distribution of $w_0$ then leads to a tighter distribution of $\omegazero$ with smaller mean (compared to \varW).
Therefore a larger $H_0$ compared to \varW~is required to get the right $d^*_A$.

However, the mean of $\omegazero$ in the \varwc~model is not too different from its values in $\Lambda$CDM and hence, there is more to this story.
The intrinsic size of the sound horizon at the end of recombination, or rather the baryon drag epoch, $r_s^{\rm drag}$, is different in \varwc~case.
As was pointed out in \cite{KnoxMillea2019}, an increase in  $r_s^{\rm drag}$ is one of the most natural ways
for increasing $H_0$ as inferred from early universe data and in turn relaxing the $H_0$-tension.

In Fig.\,\ref{rsdragH0plane} we reproduce Fig.1 of \cite{KnoxMillea2019}, showing two model independent constraints in the $H_0$-$r_s^{\rm drag}$ plane
from distance ladder-calibrated supernovae (SH0ES) and supernovae-calibrated angular diameter distance measurements of $r_s^{\rm drag}$ (BAO+SNe).
We replaced the high-$\ell$ and low-$\ell$ $\Lambda$CDM constraints which were part of the original figure
by several GDM constraints on $H_0$ and $r_s^{\rm drag}$ coming from the PPS+Lens+BAO dataset combination.
We show the $68\%$ and $95\%$ credible regions for the \varwc~model (yellow curves), $\Lambda$CDM (grey) and \varW~(red dotted),
 and additionally a set of samples color-coded according to $\omega_g^{\rm eq}$ using the \varwc~model.
While the original Fig.1 of \cite{KnoxMillea2019} revealed that changing $\omegazero$ cannot reconcile SH0ES and BAO+SNe,
it is clear from our Fig.\,\ref{rsdragH0plane} that an independent increase of $\omega_g^{\rm eq}$ can move the contours to larger $r_s^{\rm drag}$ and larger $H_0$
 toward a region where SH0ES and BAO+SNe overlap.

As is clear from the right panel of Fig.\,\ref{rsdragH0plane},  $\omega_g^{\rm eq}$
is uniquely tied to $H_{\rm eq}$ and $a_{\rm eq}$, such that $r_s^{\rm drag}$ is larger due to an increase
in the prerecombination Hubble parameter and earlier matter-radiation equality (while keeping $\omegazero$ at a smaller value).
Thus, an increased $\omega_g^{\rm eq}$, whose origin we already explained above, is the source of an increased $H_0$ in the \varwc~model.

We note in passing  that although the \varweqc~submodel allows in principle for  $\omega_g^{\rm eq} > \omegazero$, the data does not favor such a shift.
We see in Fig.\,\ref{weqcovertime} that $\omega^{\rm eq}_g$ is very close to $\omegazero$, which in turn is very close to the $\Lambda$CDM value.
This results in a low $H_0$ value which is comparable to that of $\Lambda$CDM, so that a \varweqc~type model cannot resolve the $H_0$ tension.

Our analysis indicates that a designed GDM model with only a couple of additional parameters from $\Lambda$CDM (rather than 26) may be
able to address and further investigate the $H_0$ and $\sigma_8$ tensions.
Our models also further elucidate the mechanism by which decaying dark matter,
see e.g. \cite{Buen-AbadSchmaltzLesgourguesEtal2018,Bringmann2018,Vattis2019} can increase the CMB inference of $H_0$ and decrease $\sigma_8$.
Such a model naturally implements a positive $w_8$ and $c_{s,8}^2$ when decaying DM and dark radiation are collectively modeled as a single GDM fluid.

Finally, we note that in our analysis we assumed that dark energy is a cosmological constant. Letting the dark energy EoS vary in redshift could, in principle, lead to degeneracies with 
$w$ (dark matter EoS). The case of late DE  was investigated in \cite{TutusausEtal2016}, where the degeneracy was solved by using both early- and late-time observables of the perturbations -- for 
instance, CMB data combined with a forecast for galaxy clustering data. It is also possible to have an early dark energy (EDE) component (see e.g. \cite{KarwalKamionkowski2016,
HillMcDonoughToomeyEtal2020}). If the EDE is uncoupled with the DM and has negligible perturbations, or if the EDE is tightly-coupled with the DM, the combined EDE+DM fluid can be well described 
by a single GDM fluid \cite{KoppSkordisThomas2016}, so that one could interpret our constraints on GDM parameters as constraints on such mixture scenarios. An EDE component which is not 
tightly-coupled to the DM is unlikely to be the cause of the early-time marginal effects discussed in this section, since for those effects to appear a varying DM sound speed is necessary, and in 
addition, they do not occur when only the EoS is varied.


\section{Conclusion}
\label{sec:conclusion}
We have presented the most exhaustive parameter search of DM properties to date, using the generalized dark matter model (GDM).
We allowed all three GDM parametric functions to have a fairly general time dependence by binning
$w$ in 8 and $c^2_s$ and $\cvis$ in 9 scale factor bins, that is, $26$ new parameters beyond $\Lambda$CDM in total.
We found no convincing evidence for any of these parameters to be nonzero. We expect that merging some bins will tighten the constraints for each bin; however,
we do not expect that significant non-CDM behavior would emerge as even in the extreme case of constant GDM parameters, this doesn't happen. Thus, our result should be seen as 
depicting the most general time-variation of DM properties allowed by the data.

We analyzed four nested models: \varwc~(all 26 parameters free), \varc~(setting $w_i=0$), \varW~(setting $c^2_s=\cvis=0$; previously studied in \cite{KoppSkordisThomasEtal2018})
and \varW=c where the constraint $w=c^2_s=\cvis$ was imposed.
Our strongest constraints on $w$ were in the early universe around matter-radiation equality, while the strongest constraints on $c^2_s$ and $\cvis$ are
between redshift $2$ and $9$ where the constraining power of the CMB lensing peaks.

Our analysis was performed using flat and nonflat priors for the perturbative GDM parameters,
 in order to ensure robustness of the results. Indeed, while three early universe bins show significant shifts for $w$ away from zero in the \varwc~model when
flat priors were used, these become less significant when nonflat priors were used.

Having a varying $w$ improved the fits marginally while letting $c^2_s$ and $\cvis$ be free led to virtually no improvement.
However, $c^2_s$ and $\cvis$ introduced some rather interesting features.
We observed a number of interesting shifts in the 2\Dim~posteriors between the \varW~and \varwc~models. Specifically, the \varwc~model shifts the DM abundance around equality \omegaeq~
to higher values while today \omegazero decreases and the Hubble constant $H_0$ increases compared to \varW~or to $\Lambda$CDM.
Interestingly, $\sigma_8$ also shifts to lower values in the \varwc~model. These shifts indicate the potential of GDM to alleviate the $H_0$ and $\sigma_8$ tensions driven by
early and late universe data. In particular, an \emph{a-posteriori} constructed GDM model with only a couple of more parameters than $\Lambda$CDM  may be
 favored  over the latter while simultaneously addressing these tensions.
Our present analysis paves the way for further work in this direction using more recent cosmological datasets \cite[e.g][]{Aghanim:2018eyx} as well as investigating possible 
$k^2$-dependencies in the $c_s^2$ and $\cvis$ parametric functions.

Finally, we reiterate our assertion that upcoming CMB experiments, such as Simons Observatory \cite{Ade:2018sbj} will substantially
improve the CMB lensing constraining power. This improvement is expected to have profound impact on further constraining and quite possibly detecting
DM properties in the late universe, specifically  $c^2_s$ and $c^2_\text{vis}$.
We predict that a detection of an EFTofLSS type GDM signal is likely with the improved lensing spectrum from upcoming experiments.



\begin{acknowledgements}
    The research leading to these results
    has received funding from the European Research Council under the European Union's Seventh Framework Programme (FP7/2007-2013) / ERC Grant Agreement no. 617656 ``Theories
     and Models of the Dark Sector: Dark Matter, Dark Energy and Gravity''. The Primary Investigator is C. Skordis.
\end{acknowledgements}

\bibliographystyle{aa}
\bibliography{var-gdm-paper.bib}


\begin{appendix}

\section{The DM growth rate in a $\Lambda$$w$DM universe}
\label{app:sigma8w0Degeneracy}

\subsection{Derivation of the growth rate formula}
Here we derive the DM growth rate in the GDM model with constant $w$ equation of state and zero perturbative parameters, i.e. $c_s^2=\cvis=0$,
assuming that the only components affecting the late Universe is GDM and cosmological constant $\Lambda$. Hence, we have
\begin{align}
\Omega_g+\Omega_\Lambda &=1
\\
\Omega_g &= \frac{   \Omega_{0g} }{ \Omega_{0g}  + (1- \Omega_{0g} )   a^{3(1+w)}    }
\end{align}
where $ \Omega_{0g}$ is the relative GDM density today.

Consider  now the gauge-invariant Bardeen potentials
\begin{align}
\Phi =& -\frac{h + k^2\nu}{6}  + \frac{1}{2} a^2 H \dot{\nu}
\\
\Psi =& -\frac{1}{2} a^2  \ddot{\nu} - a^2 H \dot{\nu}
\end{align}
corresponding to the conformal Newtonian gauge metric perturbations.
Exchanging $\Psi$ with the  gauge-invariant metric variable
\begin{equation}
\Rcal = \Phi + \frac{2}{3}\frac{\dot{\Phi} + H \Psi }{(1+w)H}
\end{equation}
we find that the evolution of  $\Phi$ and $\Rcal$ is dictated by
\begin{align}
H^{-1}\dot{\Phi} &=   \frac{3}{2}(1+w) \Omega_g(\Rcal - \Phi) - \Phi
  \label{PhiGIinTermsOfTime}
\\
H^{-1}\dot{\Rcal} & = 0\,,
\end{align}
which is a special case of the more general case where $c_s^2$ and $\cvis$ are nonzero, as presented in~\cite{KoppSkordisThomas2016}. The solution
is given in terms of the ordinary hypergeometric function ${}_2 F_1$ as
\begin{align}
\Phi &=\Phi_{\rm ini} \left[
  \frac{5+3w}{3 (1+w)}
- \frac{2\, {}_2 F_1\left(\frac{1}{2}, \frac{5 +3 w}{6(1+w)},  \frac{11+9w}{6(1+w)}, -\frac{\Omega_\Lambda}{\Omega_g} \right)}{\sqrt{\Omega_g} (5+ 3 w)} \right]
\\
\Rcal&=\frac{5+3w}{3 (1+w)}  \Phi_{\rm ini}= \mathrm{const} \,.
\label{Rcal_solution}
\end{align}
where $\Phi_{\rm ini}$ is an initial condition.

The above exact solution is not particularly useful. A more
 useful result is obtained by observing that the growth of $\Phi$ is scale independent and thus
we can  define  $\gamma(a)$ via $d \Phi/ d \ln a =  \Phi (\Omega_g^\gamma -1)$ and use $\Omega_g$ as the time variable \cite[see e.g.][]{FerreiraSkordis2010}.
Using  $d\Omega_g/ d \ln a = 3(1+w) (1-\Omega_g ) \Omega_g$
we find that the evolution equation \eqref{PhiGIinTermsOfTime} for $\Phi$    simplifies to
\begin{align}
\Phi \Omega_g^\gamma  &=\frac{3}{2}(1+w) \Omega_g(\Rcal - \Phi) \\
\Rightarrow \quad \Phi &= \frac{5+3 w}{2 \Omega_g^{\gamma-1}+3(1+w)}  \Phi_{\rm ini}
\,,\label{PhiGIinTermsOfGamma}
\end{align}
where \eqref{Rcal_solution} was used and where $\gamma(\Omega_g)$ is at this point still undetermined.

We proceed to find an approximate solution for $\gamma(\Omega_g)$ assuming that $\Omega_\Lambda$ is small and Taylor expand the equation for $\gamma$
(by plugging \eqref{PhiGIinTermsOfGamma}
back into \eqref{PhiGIinTermsOfTime})  around $\Omega_\Lambda=0$ (thus around $w$DM dominated era). This leads to a hierarchy of algebraic equations
for the Taylor coefficients of $\gamma$. The lowest two give
\begin{equation}
\gamma =\frac{6(1+w)}{11+9w}\left[1+\frac{ \Omega_\Lambda (1+3w)(5+3w)}{2(11+9w)(17+15w)}+\Ocal(\Omega_\Lambda^2)\right]\,.
\end{equation}
This approximation is accurate to $1\%$ for $\Omega^{(0)}_g>0.2$ and $a\leq1$. The expression reduces to the standard $\Lambda$CDM result when \mbox{$w=0$}~\cite{FerreiraSkordis2010}.
Beware also that $w$ here is for DM, not dark energy. Hence, the formula for $\gamma$ above, differs from the usual Wang-Steinhardt result
\cite{WangSteinhardt1998,FerreiraSkordis2010}.

 \subsection{Integrated Sachs-Wolfe effect}
Define the growth factor
\begin{equation}
D(a) \equiv \Phi/\Phi_{\rm ini}.
\end{equation}
Then in the case $\Omega_\Lambda=0$ we obtain $D=1$ and thus as explained in \cite{KoppSkordisThomas2016} there is no
integrated Sachs-Wolfe  (ISW) effect in a $w$DM dominated universe.
However, in a $\Lambda$$w$DM universe $w$ can directly affect the late ISW effect.

Evaluating $D(a)$ at $a=1$ we find that for sensible values of $w$ and fixed $\Omega^{(0)}_{g}$  the change to the \Lcdm late ISW effect are small.
A positive (negative) $w$ gives rise to less (more) potential decay, as might be anticipated from the fact that the absolute value of $w_{\rm tot} = \Omega_g (1+w) -1 $
is reduced (increased).
For the case $\Omega^{(0)}_{g}=0.3$ we find a $2\%$ deviation of $D$ between the cases $w_0=0.1$ and $w_0=0$
 and thus given any fixed $\Omega^{(0)}_{g}$ the effect of nonzero $w$  on the ISW is negligible.
Any role played by $w$ to affect the late ISW effect in GDM is thus dominated by a free $w$ to increase the allowed range for $\Omega^{(0)}_\Lambda$ rather than through $w\neq0$ directly. 

\subsection{Density growth}
The GDM comoving density perturbation $ \DeltaGI_g \equiv \delta_g + 3 a H (1+w) \theta_g$
is related to $\Phi$ through the Einstein constraint equation $\grad^2 \Phi = 4 \pi G \rhob^{(0)}_{g} a^{-1-3 w} \DeltaGI_g
 $. The growth rate of $\DeltaGI_g$ is
\begin{equation}
f\equiv  \frac{d \ln (D a^{1+3w}) }{d \ln a} = \Omega_g^{\gamma}+3 w.
\end{equation}
We now focus on  $a=1$ such that the factor $a^{-1-3 w}$ relating potential and density growth is irrelevant.
Since most of the growth happens in the late universe bins, we then expect $\sigma_8$ and $w_0$ to become correlated in the \varW~and \varwc~models.
This is what we see in the middle panel of Fig.\,\ref{2D_contours} and Fig.\,\ref{corrmatComparisonCosmo}.

\section{ECLAIR -- ``Ensemble of Codes for Likelihood Analysis, Inference, and Reporting'': a short description.}
\label{app:ECLAIR}
We present here a brief overview of the characteristics of the ECLAIR (``Ensemble of Codes for Likelihood Analysis, Inference, and Reporting") suite of codes, used to derive all our results in the present work.

ECLAIR is at its core a Monte Carlo code written for extracting constraints on cosmological (and also nuisance) parameters.
The main sampling algorithm used in the current version is the Goodman-Weare affine invariant MCMC ensemble sampling \cite{GoodmanWeare2010},
and most specifically its Python implementation in the \texttt{emcee} module \cite[which needs to be installed before using the ECLAIR suite][]{emcee}.
This method allows for a quick and almost tuning-free exploration of the full parameter space via the use
of ``walkers,'' spreading simultaneously and updating their positions at each MCMC step by using the positions of all the other walkers.
 The implementation of other sampling methods, such as the powerful nested sampling approach through the MultiNest \cite{multinest} algorithm,
 is underway and will appear in future ECLAIR releases.

ECLAIR contains interfaces and likelihood codes for several recent experiments and surveys, including all releases of the Planck likelihood,
with new likelihoods being continuously added through its ongoing development. It allows for a straightforward interface to the CLASS Boltzmann code~\cite{Lesgourgues2011}
for computing and retrieving cosmological quantities and observables used subsequently by the likelihood codes. In particular, it uses the convenient \texttt{classy} python wrapper of CLASS.
 It comes with a bundle of scripts that allow the monitoring of the Markov chain,
including a visual and convenient estimator of its convergence.
A robust maximizer is included for finding the best likelihood of the explored models: it relies on a combination of the simulated annealing technique and ensemble sampling to converge reliably toward the global maximum of the posterior function.
ECLAIR outputs are also easily interfaceable with the popular \texttt{getdist} python module for the computation of 1\Dim\ credible regions and 2\Dim\ contours.

The ECLAIR suite comes with a powerful parser and all configuration is done via a simple \texttt{.ini} file, with a variety of convenient settings.
As an example, the parameter (and prior) redefinition between $\{c_s^2, c_v^2\}$ and $\{c_+^2, d\}$ (described in Sec.~\ref{sec:priors}) takes a simple line of code
which creates on the fly an equivalence between the new MCMC parameters ($c_+^2$ and $d$) with the parameters taken as input by CLASS ($c_s^2$ and $c_v^2$).
Despite all the included capabilities, potential improvements or further case-specific modifications may be easily implemented, as the main code itself is short, well-commented,
and easy to ``hack.'' Developed specifically for the work presented in this paper, ECLAIR is nonetheless versatile and can be applied to a variety of situations;
It is free to use and modify provided the present work and ECLAIR's dedicated explanatory article \cite{ECLAIRexplanatory} are cited and the license conditions fulfilled.

\section{Tables of 1\Dim~constraints of model parameters and correlation matrices}
\label{app:bigtables}
We present here tables of our constraints on model parameters for a number of (sub) models and dataset combinations.
In Table~\ref{tab:alldatasets1D}, we list
the parameter constraints for six dataset combinations when flat priors were used, showing the mean value of each parameter and the
boundaries of the 68\%  and 95\% credible intervals.
In Table~\ref{tab:constraintsPPSvsLensvsPrior} we focus on two datasets, PPS and PPS+Lens+BAO, and contrast \varwc, \varc~and~\varW~together with an alternative choice of priors.
This table also provides the best-fit values, found by maximization
of the likelihood $\Lcal_{\rm max}$  ( $- \ln \Lcal_{\rm max}$ corresponds to $\chi^2/2$ in the case of a Gaussian likelihood).
Table~\ref{tab:constraints_cp2} lists
the 68\% and 95\% credible regions for $c^2_{+,i}$ when the PPS and PPS+Lens+BAO dataset combinations were used
using ``nonflat priors'' (see Sec.~\ref{sec:priors}).

\begin{figure*}[t!]
\begin{center}
\includegraphics[width=\textwidth]{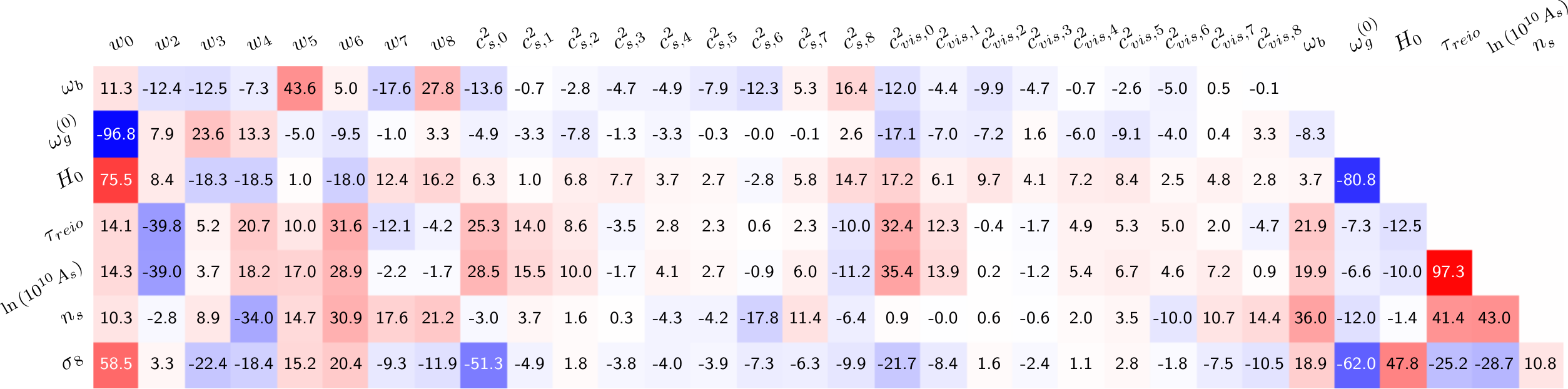}
\end{center}
\caption{
Correlation matrix of standard cosmological parameters with GDM parameters in the \varwc~model when the PPS+Lens+BAO dataset combination was used
with flat priors. The number in each matrix element indicates the value of the corresponding correlation coefficient, multiplied by a factor of $100$. The color of each element also reflects the strength of the correlation, from dark blue ($=-1$) to dark red ($=1$) through white ($=0$).
}
\label{corrmatComparisonCosmo}
\end{figure*}

\begin{figure*}[t!]
\begin{center}
\includegraphics[width=\textwidth]{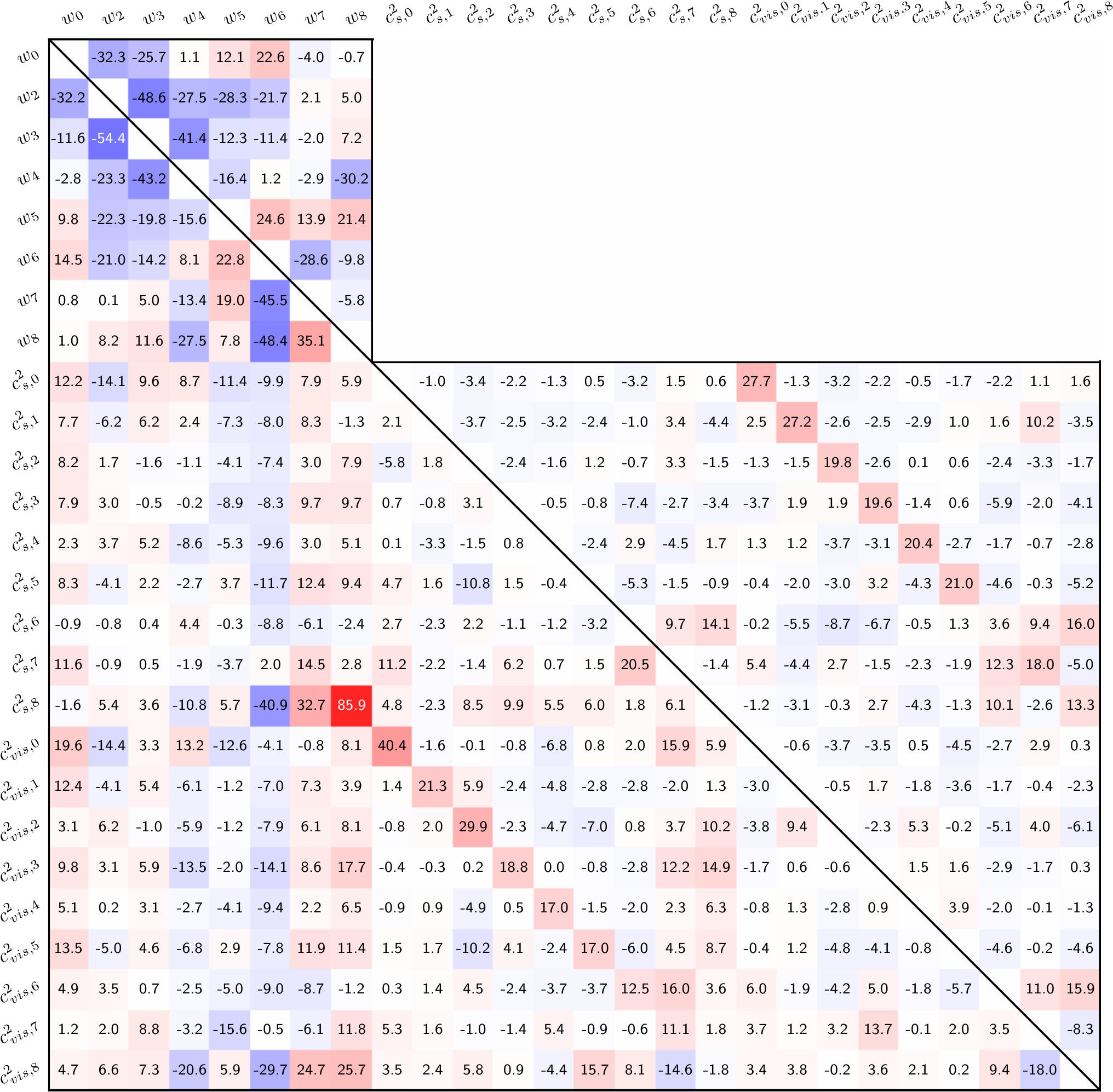}
\end{center}
\caption{
Correlation matrix of GDM parameters for \varwc~(lower triangle), \varW~(upper triangle) and \varc~(right triangle) when the PPS+Lens+BAO dataset combination
 with nonflat priors. The number in each matrix element indicates the value of the corresponding correlation coefficient, multiplied by a factor of $100$. The color of each element also reflects the strength of the correlation, from dark blue ($=-1$) to dark red ($=1$) through white ($=0$).
}
\label{corrmatComparison}
\end{figure*}

Correlation matrices  when the PPS+Lens+BAO dataset combination was used
are displayed in Fig.\ref{corrmatComparisonCosmo} (\varwc~model; cosmological and GDM parameters; flat priors) and Fig.\ref{corrmatComparison}
(GDM parameters only; nonflat priors) respectively.

 \renewcommand{\arraystretch}{1}
\begin{table*}
\resizebox{\textwidth}{!}{%
\begin{tabular} { |l| l| l| l| l| l| l| }  \hline \hline
  \backslashbox[22mm]{\!Parameter}{Data} &  PPS     & PPS+Lens        & PPS+BAO    & PPS+HST  & PPS+Lens+BAO & PPS+Lens+HST      \\

\hline
$100 \omega_{\rm b }$ &
\colorvarc{$2.187^{+0.018}_{-0.018}$  $ {}^{+0.035}_{-0.034}$ }&
\colorvarc{$2.201^{+0.016}_{-0.017}$  $ {}^{+0.032}_{-0.033}$ }&
\colorvarc{$2.207^{+0.016}_{-0.016}$  $ {}^{+0.032}_{-0.031}$}  &
\colorvarc{$2.199^{+0.017}_{-0.017}$ $ {}^{+0.034}_{-0.033}$}&
\colorvarc{ $2.218^{+0.015}_{-0.015}$ $ {}^{+0.030}_{-0.030}$}&
    \colorvarc{$2.210^{+0.017}_{-0.016}$ $ {}^{+0.032}_{-0.033}$}\\[1ex]

& \colorvarwc{ $2.179^{+0.022}_{-0.022}$  $ {}^{+0.043}_{-0.044}$}    &
  \colorvarwc{$2.188^{+0.022}_{-0.022}$  $ {}^{+0.042}_{-0.043}$} &
  \colorvarwc{$2.182^{+0.023}_{-0.023}$  $ {}^{+0.045}_{-0.044}$} &
\colorvarwc{$2.185^{+0.022}_{-0.022}$ $ {}^{+0.044}_{-0.043}$}&
  \colorvarwc{$2.194^{+0.021}_{-0.021}$ $ {}^{+0.041}_{-0.041}$}&
    \colorvarwc{$2.196^{+0.020}_{-0.020}$ $ {}^{+0.039}_{-0.040}$}\\[1ex]
 \hline

 $\omegazero$  &
 \colorvarc{ $0.1245^{+0.0020}_{-0.0023}$  $ {}^{+0.0045}_{-0.0040}$} &
 \colorvarc{$0.1229^{+0.0018}_{-0.0018}$  $ {}^{+0.0035}_{-0.0033}$} &
\colorvarc{$0.1200^{+0.0012}_{-0.0012}$  $ {}^{+0.0024}_{-0.0024}$  }&
\colorvarc{$0.1223^{+0.0018}_{-0.0020}$ $ {}^{+0.0038}_{-0.0035}$}&
\colorvarc{ $0.1197^{+0.0011}_{-0.0011}$ $ {}^{+0.0022}_{-0.0022}$}&
    \colorvarc{$0.1213^{+0.0016}_{-0.0016}$ $ {}^{+0.0032}_{-0.0031}$}\\[1ex]

 & \colorvarwc{$0.084^{+0.012}_{-0.028}$  $ {}^{+0.049}_{-0.038}$}   &
 \colorvarwc{$0.091^{+0.015}_{-0.026}$  $ {}^{+0.045}_{-0.036}$}   &
 \colorvarwc{$0.121^{+0.025}_{-0.032}$  $ {}^{+0.053}_{-0.049}$}  &
\colorvarwc{$0.073^{+0.013}_{-0.023}$ $ {}^{+0.041}_{-0.034}$}&
  \colorvarwc{$0.127^{+0.024}_{-0.024}$ $ {}^{+0.046}_{-0.044}$}&
    \colorvarwc{$0.082^{+0.015}_{-0.025}$ $ {}^{+0.046}_{-0.036}$}\\[1ex]
\hline

  $H_{\rm 0}$ &
\colorvarc{ $65.3^{+0.85}_{-0.84}$  $ {}^{+1.6}_{-1.7}$ }&
\colorvarc{$66.0^{+0.74}_{-0.73}$  $ {}^{+1.4}_{-1.4}$  }&
\colorvarc{$67.1^{+0.51}_{-0.51}$  $ {}^{+1.0}_{-1.0}$} &
\colorvarc{$66.2^{+0.75}_{-0.74}$ $ {}^{+1.4}_{-1.5}$}&
\colorvarc{ $67.33^{+0.49}_{-0.50}$ $ {}^{+0.99}_{-0.95}$}&
    \colorvarc{$66.7^{+0.67}_{-0.67}$ $ {}^{+1.3}_{-1.4}$}\\[1ex]

&  \colorvarwc{$< 49.5$  $< 55.6         $}  &
 \colorvarwc{$< 54.6$  $< 63.3         $} &
 \colorvarwc{$69.6^{+1.9}_{-1.9}$  $ {}^{+3.6}_{-3.6}$}   &
\colorvarwc{$70.9^{+2.3}_{-2.3}$ $ {}^{+4.6}_{-4.6}$}&
  \colorvarwc{$69.3^{+1.6}_{-1.7}$ $ {}^{+3.3}_{-3.0}$}&
    \colorvarwc{$72.0^{+2.4}_{-2.3}$ $ {}^{+4.7}_{-4.6}$}\\[1ex]

 \hline

$\tau_{\rm reio } $  &
\colorvarc{ $0.108^{+0.018}_{-0.018}$  $ {}^{+0.036}_{-0.035}$ }  &
\colorvarc{ $0.106^{+0.015}_{-0.015}$  $ {}^{+0.032}_{-0.030}$ }&
\colorvarc{ $0.121^{+0.017}_{-0.017}$  $ {}^{+0.034}_{-0.034}$ }   &
\colorvarc{$0.114^{+0.018}_{-0.018}$ $ {}^{+0.036}_{-0.037}$}&
\colorvarc{ $0.118^{+0.014}_{-0.016}$ $ {}^{+0.030}_{-0.028}$}&
    \colorvarc{$0.113^{+0.016}_{-0.016}$ $ {}^{+0.032}_{-0.031}$}\\[1ex]

& \colorvarwc{$0.106^{+0.018}_{-0.018}$  $ {}^{+0.034}_{-0.034}$}   &
 \colorvarwc{$0.101^{+0.019}_{-0.020}$  $ {}^{+0.037}_{-0.039}$} &
  \colorvarwc{$0.102^{+0.022}_{-0.022}$  $ {}^{+0.042}_{-0.041}$}  &
\colorvarwc{$0.105^{+0.021}_{-0.019}$ $ {}^{+0.039}_{-0.041}$}&
  \colorvarwc{$0.095^{+0.019}_{-0.020}$ $ {}^{+0.039}_{-0.037}$}&
    \colorvarwc{$0.098^{+0.020}_{-0.020}$ $ {}^{+0.038}_{-0.037}$}\\[1ex]

\hline

$\ln(10^{10}A_{s })$ &
\colorvarc{$3.159^{+0.035}_{-0.035}$  $ {}^{+0.070}_{-0.068}$ }   &
\colorvarc{$3.151^{+0.029}_{-0.029}$  $ {}^{+0.060}_{-0.057}$ }&
\colorvarc{ $3.174^{+0.034}_{-0.035}$  $ {}^{+0.067}_{-0.067}$  } &
\colorvarc{$3.164^{+0.036}_{-0.035}$ $ {}^{+0.071}_{-0.071}$}&
\colorvarc{ $3.170^{+0.029}_{-0.029}$ $ {}^{+0.059}_{-0.055}$}&
    \colorvarc{$3.163^{+0.030}_{-0.030}$ $ {}^{+0.061}_{-0.059}$}\\[1ex]

&  \colorvarwc{$3.164^{+0.037}_{-0.036}$  $ {}^{+0.070}_{-0.068}$}   &
 \colorvarwc{ $3.153^{+0.039}_{-0.039}$  $ {}^{+0.075}_{-0.077}$} &
   \colorvarwc{$3.158^{+0.044}_{-0.044}$  $ {}^{+0.084}_{-0.082}$}  &
\colorvarwc{$3.164^{+0.043}_{-0.039}$ $ {}^{+0.079}_{-0.083}$}&
  \colorvarwc{$3.145^{+0.038}_{-0.039}$ $ {}^{+0.076}_{-0.073}$}&
    \colorvarwc{$3.149^{+0.038}_{-0.039}$ $ {}^{+0.074}_{-0.074}$}\\[1ex]
\hline

$n_{s } $ &
\colorvarc{$0.956^{+0.0057}_{-0.0057}$  $ {}^{+0.011}_{-0.011}$}  &
\colorvarc{ $0.958^{+0.0053}_{-0.0054}$  $ {}^{+0.010}_{-0.011}$ }   &
\colorvarc{  $0.9640^{+0.0050}_{-0.0050}$  $ {}^{+0.0097}_{-0.010}$}  &
\colorvarc{$0.960^{+0.0055}_{-0.0055}$ $ {}^{+0.011}_{-0.011}$}&
\colorvarc{ $0.9653^{+0.0044}_{-0.0044}$ $ {}^{+0.0086}_{-0.0091}$}&
    \colorvarc{$0.962^{+0.0050}_{-0.0050}$ $ {}^{+0.010}_{-0.010}$}\\[1ex]

&  \colorvarwc{$0.963^{+0.012}_{-0.012}$  $ {}^{+0.023}_{-0.022}$}  &
 \colorvarwc{$0.971^{+0.011}_{-0.011}$  $ {}^{+0.022}_{-0.021}$}  &
   \colorvarwc{$0.966^{+0.011}_{-0.011}$  $ {}^{+0.022}_{-0.022}$}  &
\colorvarwc{$0.969^{+0.011}_{-0.012}$ $ {}^{+0.023}_{-0.021}$}&
  \colorvarwc{$0.973^{+0.010}_{-0.011}$ $ {}^{+0.022}_{-0.020}$}&
    \colorvarwc{$0.976^{+0.011}_{-0.011}$ $ {}^{+0.021}_{-0.020}$}\\[1ex]

\hline \hline

$w_0 $

&  \colorvarwc{$0.13^{+0.070}_{-0.047}$  $ {}^{+0.11}_{-0.13}$}  &
 \colorvarwc{$0.0998^{+0.058}_{-0.049}$  $ {}^{+0.094}_{-0.099}$}  &
   \colorvarwc{$0.000^{+0.043}_{-0.051}$  $ {}^{+0.086}_{-0.082}$}  &
\colorvarwc{$0.103^{+0.060}_{-0.049}$ $ {}^{+0.098}_{-0.11}$}&
  \colorvarwc{$-0.014^{+0.034}_{-0.041}$ $ {}^{+0.075}_{-0.070}$}&
    \colorvarwc{$0.073^{+0.058}_{-0.048}$ $ {}^{+0.094}_{-0.11}$}\\[1ex]

\hline

$w_2 $

&  \colorvarwc{$-0.01^{+0.054}_{-0.053}$  $ {}^{+0.11}_{-0.10}$}  &
 \colorvarwc{$0.01^{+0.056}_{-0.055}$  $ {}^{+0.11}_{-0.11}$}  &
   \colorvarwc{$0.02^{+0.054}_{-0.054}$  $ {}^{+0.11}_{-0.11}$}  &
\colorvarwc{$0.01^{+0.051}_{-0.051}$ $ {}^{+0.10}_{-0.10}$}&
  \colorvarwc{$0.03^{+0.057}_{-0.055}$ $ {}^{+0.11}_{-0.11}$}&
    \colorvarwc{$0.03^{+0.057}_{-0.058}$ $ {}^{+0.11}_{-0.11}$}\\[1ex]

\hline

$w_3 $

&  \colorvarwc{$-0.004^{+0.042}_{-0.050}$  $ {}^{+0.094}_{-0.085}$}  &
 \colorvarwc{$0.003^{+0.048}_{-0.055}$  $ {}^{+0.11}_{-0.094}$}  &
   \colorvarwc{$0.078^{+0.049}_{-0.050}$  $ {}^{+0.093}_{-0.094}$}  &
\colorvarwc{$0.044^{+0.044}_{-0.045}$ $ {}^{+0.087}_{-0.090}$}&
  \colorvarwc{$0.078^{+0.048}_{-0.047}$ $ {}^{+0.091}_{-0.093}$}&
    \colorvarwc{$0.050^{+0.048}_{-0.048}$ $ {}^{+0.093}_{-0.096}$}\\[1ex]

\hline

$w_4 $

&  \colorvarwc{$-0.075^{+0.029}_{-0.036}$  $ {}^{+0.068}_{-0.065}$}  &
 \colorvarwc{$-0.073^{+0.035}_{-0.035}$  $ {}^{+0.069}_{-0.067}$}  &
   \colorvarwc{$-0.045^{+0.031}_{-0.036}$  $ {}^{+0.071}_{-0.064}$}  &
\colorvarwc{$-0.061^{+0.034}_{-0.033}$ $ {}^{+0.065}_{-0.063}$}&
  \colorvarwc{$-0.057^{+0.030}_{-0.034}$ $ {}^{+0.065}_{-0.064}$}&
    \colorvarwc{$-0.065^{+0.032}_{-0.037}$ $ {}^{+0.070}_{-0.067}$}\\[1ex]

\hline

$w_5 $

&  \colorvarwc{ $-0.030^{+0.019}_{-0.019}$  $ {}^{+0.036}_{-0.037}$}  &
 \colorvarwc{$-0.024^{+0.020}_{-0.020}$  $ {}^{+0.038}_{-0.040}$}  &
   \colorvarwc{$-0.017^{+0.020}_{-0.019}$  $ {}^{+0.037}_{-0.039}$}  &
\colorvarwc{$-0.021^{+0.019}_{-0.019}$ $ {}^{+0.039}_{-0.037}$}&
  \colorvarwc{$-0.012^{+0.020}_{-0.020}$ $ {}^{+0.039}_{-0.039}$}&
    \colorvarwc{$-0.015^{+0.021}_{-0.019}$ $ {}^{+0.038}_{-0.040}$}\\[1ex]

\hline

$w_6 $

&  \colorvarwc{ $-0.020^{+0.0060}_{-0.0060}$  $ {}^{+0.012}_{-0.012}$}  &
 \colorvarwc{$-0.020^{+0.0062}_{-0.0062}$  $ {}^{+0.012}_{-0.012}$}  &
   \colorvarwc{$-0.016^{+0.0061}_{-0.0060}$  $ {}^{+0.012}_{-0.012}$}  &
\colorvarwc{$-0.017^{+0.0062}_{-0.0061}$ $ {}^{+0.012}_{-0.012}$}&
  \colorvarwc{$-0.017^{+0.0067}_{-0.0062}$ $ {}^{+0.012}_{-0.013}$}&
    \colorvarwc{$-0.017^{+0.0059}_{-0.0059}$ $ {}^{+0.011}_{-0.012}$}\\[1ex]

\hline

$w_7 $

&  \colorvarwc{  $0.0131^{+0.0049}_{-0.0048}$  $ {}^{+0.0094}_{-0.0095}$}  &
 \colorvarwc{ $0.0152^{+0.0045}_{-0.0046}$  $ {}^{+0.0089}_{-0.0092}$}  &
   \colorvarwc{ $0.0141^{+0.0045}_{-0.0046}$  $ {}^{+0.0094}_{-0.0094}$}  &
\colorvarwc{$0.0137^{+0.0047}_{-0.0046}$ $ {}^{+0.0091}_{-0.0093}$}&
  \colorvarwc{$0.0162^{+0.0046}_{-0.0046}$ $ {}^{+0.0087}_{-0.0088}$}&
    \colorvarwc{$0.0153^{+0.0046}_{-0.0042}$ $ {}^{+0.0083}_{-0.0088}$}\\[1ex]

\hline

$w_8 $

&  \colorvarwc{  $0.070^{+0.026}_{-0.041}$  $ {}^{+0.076}_{-0.066}$}  &
 \colorvarwc{$0.089^{+0.029}_{-0.046}$  $ {}^{+0.084}_{-0.070}$}  &
   \colorvarwc{$0.069^{+0.026}_{-0.040}$  $ {}^{+0.077}_{-0.066}$}  &
\colorvarwc{$0.075^{+0.027}_{-0.042}$ $ {}^{+0.077}_{-0.065}$}&
  \colorvarwc{$0.090^{+0.029}_{-0.047}$ $ {}^{+0.086}_{-0.070}$}&
    \colorvarwc{$0.091^{+0.028}_{-0.044}$ $ {}^{+0.082}_{-0.067}$}\\[1ex]

\hline \hline

$c^2_{s,0} $ &
\colorvarc{$< 1.08\cdot 10^{-5}$  $< 3.04\cdot 10^{-5}$}  &
\colorvarc{$< 7.46\cdot 10^{-6}$  $< 1.97\cdot 10^{-5}$}   &
\colorvarc{ $< 1.02\cdot 10^{-5}$  $< 2.71\cdot 10^{-5}$}  &
\colorvarc{$< 9.75\cdot 10^{-6}$ $< 2.81\cdot 10^{-5}$}&
\colorvarc{ $< 6.23\cdot 10^{-6}$ $< 1.61\cdot 10^{-5}$}&
    \colorvarc{$< 7.46\cdot 10^{-6}$ $< 2.07\cdot 10^{-5}$}\\[1ex]

&  \colorvarwc{  $< 1.10\cdot 10^{-5}$  $< 2.99\cdot 10^{-5}$}  &
 \colorvarwc{$< 2.29\cdot 10^{-5}$  $< 6.11\cdot 10^{-5}$}  &
   \colorvarwc{$< 1.42\cdot 10^{-5}$  $< 4.04\cdot 10^{-5}$}  &
\colorvarwc{$< 1.35\cdot 10^{-5}$ $< 4.06\cdot 10^{-5}$}&
  \colorvarwc{$< 1.51\cdot 10^{-5}$ $< 3.99\cdot 10^{-5}$}&
    \colorvarwc{$< 2.15\cdot 10^{-5}$ $< 5.47\cdot 10^{-5}$}\\[1ex]

\hline

$c^2_{s,1} $ &
\colorvarc{ $< 3.76\cdot 10^{-6}$  $< 9.65\cdot 10^{-6}$ }  &
\colorvarc{$< 1.38\cdot 10^{-6}$  $< 3.75\cdot 10^{-6}$}   &
\colorvarc{   $< 3.35\cdot 10^{-6}$  $< 8.92\cdot 10^{-6}$}  &
\colorvarc{$< 3.50\cdot 10^{-6}$ $< 9.21\cdot 10^{-6}$}&
\colorvarc{ $< 1.24\cdot 10^{-6}$ $< 3.41\cdot 10^{-6}$}&
    \colorvarc{$< 1.25\cdot 10^{-6}$ $< 3.34\cdot 10^{-6}$}\\[1ex]

&  \colorvarwc{ $< 6.61\cdot 10^{-6}$  $< 1.76\cdot 10^{-5}$}  &
 \colorvarwc{$< 2.19\cdot 10^{-6}$  $< 5.85\cdot 10^{-6}$}  &
   \colorvarwc{$< 4.38\cdot 10^{-6}$  $< 1.12\cdot 10^{-5}$}  &
\colorvarwc{$< 4.53\cdot 10^{-6}$ $< 1.22\cdot 10^{-5}$}&
  \colorvarwc{$< 1.68\cdot 10^{-6}$ $< 4.51\cdot 10^{-6}$}&
    \colorvarwc{$< 1.95\cdot 10^{-6}$ $< 5.74\cdot 10^{-6}$}\\[1ex]

\hline

$c^2_{s,2} $ &
\colorvarc{$< 7.04\cdot 10^{-6}$  $< 1.87\cdot 10^{-5}$}  &
\colorvarc{ $< 2.32\cdot 10^{-6}$  $< 6.22\cdot 10^{-6}$}   &
\colorvarc{  $< 6.35\cdot 10^{-6}$  $< 1.63\cdot 10^{-5}$}  &
\colorvarc{$< 6.55\cdot 10^{-6}$ $< 1.74\cdot 10^{-5}$}&
\colorvarc{ $< 2.21\cdot 10^{-6}$ $< 5.74\cdot 10^{-6}$}&
    \colorvarc{$< 2.19\cdot 10^{-6}$ $< 5.56\cdot 10^{-6}$}\\[1ex]

&  \colorvarwc{  $< 1.67\cdot 10^{-5}$  $< 4.32\cdot 10^{-5}$}  &
 \colorvarwc{$< 4.48\cdot 10^{-6}$  $< 1.08\cdot 10^{-5}$}  &
   \colorvarwc{$< 8.58\cdot 10^{-6}$  $< 2.13\cdot 10^{-5}$}  &
\colorvarwc{$< 9.95\cdot 10^{-6}$ $< 2.63\cdot 10^{-5}$}&
  \colorvarwc{$< 2.88\cdot 10^{-6}$ $< 7.90\cdot 10^{-6}$}&
    \colorvarwc{$< 3.60\cdot 10^{-6}$ $< 8.97\cdot 10^{-6}$}\\[1ex]

\hline

$c^2_{s,3} $ &
\colorvarc{$< 2.01\cdot 10^{-5}$  $< 5.09\cdot 10^{-5}$}  &
\colorvarc{ $< 6.35\cdot 10^{-6}$  $< 1.61\cdot 10^{-5}$ }   &
\colorvarc{$< 1.98\cdot 10^{-5}$  $< 5.31\cdot 10^{-5}$}  &
\colorvarc{$< 2.03\cdot 10^{-5}$ $< 5.42\cdot 10^{-5}$}&
\colorvarc{ $< 6.15\cdot 10^{-6}$ $< 1.52\cdot 10^{-5}$}&
    \colorvarc{$< 6.28\cdot 10^{-6}$ $< 1.62\cdot 10^{-5}$}\\[1ex]

&  \colorvarwc{  $< 4.90\cdot 10^{-5}$  $< 0.000124$}  &
 \colorvarwc{$< 1.35\cdot 10^{-5}$  $< 3.52\cdot 10^{-5}$}  &
   \colorvarwc{$< 2.93\cdot 10^{-5}$  $< 8.54\cdot 10^{-5}$}  &
\colorvarwc{$< 3.24\cdot 10^{-5}$ $< 9.80\cdot 10^{-5}$}&
  \colorvarwc{$< 9.65\cdot 10^{-6}$ $< 2.43\cdot 10^{-5}$}&
    \colorvarwc{$< 1.07\cdot 10^{-5}$ $< 3.12\cdot 10^{-5}$}\\[1ex]

\hline

$c^2_{s,4} $ &
\colorvarc{ $< 6.75\cdot 10^{-5}$  $< 0.000184$ }  &
\colorvarc{$< 2.09\cdot 10^{-5}$  $< 5.50\cdot 10^{-5}$}   &
\colorvarc{  $< 5.68\cdot 10^{-5}$  $< 0.000147$}  &
\colorvarc{$< 6.06\cdot 10^{-5}$ $< 0.000169$}&
\colorvarc{ $< 1.94\cdot 10^{-5}$ $< 5.13\cdot 10^{-5}$}&
    \colorvarc{$< 1.90\cdot 10^{-5}$ $< 4.96\cdot 10^{-5}$}\\[1ex]

&  \colorvarwc{ $< 0.000131$  $< 0.000336 $}  &
 \colorvarwc{$< 3.21\cdot 10^{-5}$  $< 8.15\cdot 10^{-5}$}  &
   \colorvarwc{$< 8.37\cdot 10^{-5}$  $< 0.000219$}  &
\colorvarwc{$< 0.000102$ $< 0.000254 $}&
  \colorvarwc{$< 2.83\cdot 10^{-5}$ $< 6.92\cdot 10^{-5}$}&
    \colorvarwc{$< 3.04\cdot 10^{-5}$ $< 8.22\cdot 10^{-5}$}\\[1ex]

\hline

$c^2_{s,5} $ &
\colorvarc{$< 0.000215$  $< 0.000580 $}  &
\colorvarc{ $< 6.23\cdot 10^{-5}$  $< 0.000161$}   &
\colorvarc{  $< 0.000199$  $< 0.000541 $}  &
\colorvarc{$< 0.000209$ $< 0.000548 $}&
\colorvarc{ $< 5.31\cdot 10^{-5}$ $< 0.000134$}&
    \colorvarc{$< 5.90\cdot 10^{-5}$ $< 0.000145$}\\[1ex]

&  \colorvarwc{  $< 0.000264$  $< 0.000679 $}  &
 \colorvarwc{$< 9.52\cdot 10^{-5}$  $< 0.000238$}  &
   \colorvarwc{$< 0.000244$  $< 0.000685 $}  &
\colorvarwc{$< 0.000248$ $< 0.000700 $}&
  \colorvarwc{$< 7.36\cdot 10^{-5}$ $< 0.000194$}&
    \colorvarwc{$< 8.06\cdot 10^{-5}$ $< 0.000210$}\\[1ex]

\hline

$c^2_{s,6} $ &
\colorvarc{$< 0.000952$  $< 0.00142  $}  &
\colorvarc{ $< 0.000350$  $< 0.000735 $ }   &
\colorvarc{  $< 0.000663$  $< 0.00119  $}  &
\colorvarc{$< 0.000783$ $< 0.00132  $}&
\colorvarc{ $< 0.000317$ $< 0.000631 $}&
    \colorvarc{$< 0.000326$ $< 0.000672 $}\\[1ex]

&  \colorvarwc{  $< 0.000567$  $< 0.00112  $}  &
 \colorvarwc{$< 0.000239$  $< 0.000534 $}  &
   \colorvarwc{$< 0.000456$  $< 0.000927 $}  &
\colorvarwc{$< 0.000481$ $< 0.000992 $}&
  \colorvarwc{$< 0.000197$ $< 0.000476 $}&
    \colorvarwc{$< 0.000199$ $< 0.000511 $}\\[1ex]

\hline

$c^2_{s,7} $ &
\colorvarc{ $< 0.00223$  $< 0.00387   $ }  &
\colorvarc{ $< 0.00131$  $< 0.00256   $}   &
\colorvarc{  $< 0.00174$  $< 0.00325   $}  &
\colorvarc{$< 0.00212$ $< 0.00383   $}&
\colorvarc{ $< 0.000883$ $< 0.00187  $}&
    \colorvarc{$< 0.00114$ $< 0.00233   $}\\[1ex]

&  \colorvarwc{  $< 0.00200$  $< 0.00382   $}  &
 \colorvarwc{$< 0.00133$  $< 0.00276   $}  &
   \colorvarwc{$< 0.00193$  $< 0.00363   $}  &
\colorvarwc{$< 0.00198$ $< 0.00378   $}&
  \colorvarwc{$< 0.00132$ $< 0.00261   $}&
    \colorvarwc{$< 0.00139$ $< 0.00287   $}\\[1ex]

\hline

$c^2_{s,8} $ &
\colorvarc{$< 0.00501$  $< 0.0113    $}  &
\colorvarc{ $< 0.00380$  $< 0.00896   $}   &
\colorvarc{  $< 0.00469$  $< 0.0110    $}  &
\colorvarc{$< 0.00524$ $< 0.0120    $}&
\colorvarc{ $< 0.00304$ $< 0.00714   $}&
    \colorvarc{$< 0.00348$ $< 0.00841   $}\\[1ex]

&  \colorvarwc{  $< 0.0247$  $< 0.0503     $}  &
 \colorvarwc{$< 0.0311$  $< 0.0587     $}  &
   \colorvarwc{$< 0.0243$  $< 0.0503     $}  &
\colorvarwc{$< 0.0250$ $< 0.0497     $}&
  \colorvarwc{$< 0.0318$ $< 0.0589     $}&
    \colorvarwc{$< 0.0304$ $< 0.0580     $}\\[1ex]

\hline
\hline

$\cvisI{0} $ &
\colorvarc{$< 4.10\cdot 10^{-5}$  $< 0.000124$}  &
\colorvarc{ $< 4.16\cdot 10^{-5}$  $< 0.000128$ }   &
\colorvarc{  $< 3.72\cdot 10^{-5}$  $< 0.000113$}  &
\colorvarc{$< 4.07\cdot 10^{-5}$ $< 0.000127$}&
\colorvarc{ $< 3.26\cdot 10^{-5}$ $< 0.000114$}&
    \colorvarc{$< 3.90\cdot 10^{-5}$ $< 0.000137$}\\[1ex]

&  \colorvarwc{   $< 4.64\cdot 10^{-5}$  $< 0.000152$}  &
 \colorvarwc{$< 0.000321$  $< 0.000701 $}  &
   \colorvarwc{ $< 5.53\cdot 10^{-5}$  $< 0.000198$}  &
\colorvarwc{$< 7.70\cdot 10^{-5}$ $< 0.000253$}&
  \colorvarwc{$< 0.000104$ $< 0.000355 $}&
    \colorvarwc{$< 0.000180$ $< 0.000474 $}\\[1ex]

\hline

$\cvisI{1} $ &
\colorvarc{ $< 6.00\cdot 10^{-6}$  $< 1.62\cdot 10^{-5}$ }  &
\colorvarc{$< 2.34\cdot 10^{-6}$  $< 6.40\cdot 10^{-6}$}   &
\colorvarc{  $< 5.57\cdot 10^{-6}$  $< 1.49\cdot 10^{-5}$}  &
\colorvarc{$< 5.81\cdot 10^{-6}$ $< 1.46\cdot 10^{-5}$}&
\colorvarc{ $< 2.22\cdot 10^{-6}$ $< 5.73\cdot 10^{-6}$}&
    \colorvarc{$< 2.50\cdot 10^{-6}$ $< 7.20\cdot 10^{-6}$}\\[1ex]

&  \colorvarwc{ $< 1.18\cdot 10^{-5}$  $< 3.06\cdot 10^{-5}$}  &
 \colorvarwc{$< 4.92\cdot 10^{-6}$  $< 1.34\cdot 10^{-5}$}  &
   \colorvarwc{$< 7.27\cdot 10^{-6}$  $< 2.14\cdot 10^{-5}$}  &
\colorvarwc{$< 9.12\cdot 10^{-6}$ $< 2.41\cdot 10^{-5}$}&
  \colorvarwc{$< 3.78\cdot 10^{-6}$ $< 1.00\cdot 10^{-5}$}&
    \colorvarwc{$< 4.11\cdot 10^{-6}$ $< 1.22\cdot 10^{-5}$}\\[1ex]

\hline

$\cvisI{2} $ &
\colorvarc{$< 9.11\cdot 10^{-6}$  $< 2.70\cdot 10^{-5}$}  &
\colorvarc{ $< 3.06\cdot 10^{-6}$  $< 8.05\cdot 10^{-6}$}   &
\colorvarc{  $< 7.97\cdot 10^{-6}$  $< 2.13\cdot 10^{-5}$}  &
\colorvarc{$< 7.84\cdot 10^{-6}$ $< 1.98\cdot 10^{-5}$}&
\colorvarc{ $< 2.88\cdot 10^{-6}$ $< 7.73\cdot 10^{-6}$}&
    \colorvarc{$< 2.98\cdot 10^{-6}$ $< 7.75\cdot 10^{-6}$}\\[1ex]

&  \colorvarwc{  $< 2.18\cdot 10^{-5}$  $< 5.78\cdot 10^{-5}$}  &
 \colorvarwc{$< 5.95\cdot 10^{-6}$  $< 1.57\cdot 10^{-5}$}  &
   \colorvarwc{$< 1.11\cdot 10^{-5}$  $< 3.01\cdot 10^{-5}$}  &
\colorvarwc{$< 1.34\cdot 10^{-5}$ $< 3.51\cdot 10^{-5}$}&
  \colorvarwc{$< 4.36\cdot 10^{-6}$ $< 1.08\cdot 10^{-5}$}&
    \colorvarwc{$< 4.55\cdot 10^{-6}$ $< 1.24\cdot 10^{-5}$}\\[1ex]

\hline

$\cvisI{3} $ &
\colorvarc{$< 2.51\cdot 10^{-5}$  $< 6.99\cdot 10^{-5}$}  &
\colorvarc{ $< 8.09\cdot 10^{-6}$  $< 2.19\cdot 10^{-5}$ }   &
\colorvarc{  $< 2.20\cdot 10^{-5}$ $< 6.26\cdot 10^{-5}$}  &
\colorvarc{$< 2.43\cdot 10^{-5}$ $< 6.20\cdot 10^{-5}$}&
\colorvarc{ $< 7.28\cdot 10^{-6}$ $< 1.88\cdot 10^{-5}$}&
    \colorvarc{$< 8.09\cdot 10^{-6}$ $< 2.13\cdot 10^{-5}$}\\[1ex]

&  \colorvarwc{ $< 5.88\cdot 10^{-5}$  $< 0.000152$}  &
 \colorvarwc{$< 1.37\cdot 10^{-5}$  $< 3.57\cdot 10^{-5}$}  &
   \colorvarwc{$< 3.58\cdot 10^{-5}$  $< 9.26\cdot 10^{-5}$}  &
\colorvarwc{$< 4.16\cdot 10^{-5}$ $< 0.000105$}&
  \colorvarwc{$< 1.28\cdot 10^{-5}$ $< 3.30\cdot 10^{-5}$}&
    \colorvarwc{$< 1.25\cdot 10^{-5}$ $< 3.33\cdot 10^{-5}$}\\[1ex]

\hline

$\cvisI{4} $ &
\colorvarc{ $< 8.17\cdot 10^{-5}$  $< 0.000231$ }  &
\colorvarc{$< 2.43\cdot 10^{-5}$  $< 6.58\cdot 10^{-5}$}   &
\colorvarc{  $< 6.84\cdot 10^{-5}$  $< 0.000177$}  &
\colorvarc{$< 7.61\cdot 10^{-5}$ $< 0.000208$}&
\colorvarc{ $< 2.39\cdot 10^{-5}$ $< 6.10\cdot 10^{-5}$}&
    \colorvarc{$< 2.40\cdot 10^{-5}$ $< 6.19\cdot 10^{-5}$}\\[1ex]

&  \colorvarwc{ $< 0.000152$  $< 0.000395 $}  &
 \colorvarwc{$< 3.79\cdot 10^{-5}$  $< 0.000101$}  &
   \colorvarwc{$< 0.000100$  $< 0.000276 $}  &
\colorvarwc{$< 0.000115$ $< 0.000332 $}&
  \colorvarwc{$< 3.27\cdot 10^{-5}$ $< 8.09\cdot 10^{-5}$}&
    \colorvarwc{$< 3.49\cdot 10^{-5}$ $< 9.30\cdot 10^{-5}$}\\[1ex]

\hline

$\cvisI{5} $ &
\colorvarc{$< 0.000273$  $< 0.000765 $}  &
\colorvarc{ $< 8.37\cdot 10^{-5}$  $< 0.000223$}   &
\colorvarc{$< 0.000263$  $< 0.000722 $}  &
\colorvarc{$< 0.000273$ $< 0.000744 $}&
\colorvarc{ $< 8.01\cdot 10^{-5}$ $< 0.000209$}&
    \colorvarc{$< 7.94\cdot 10^{-5}$ $< 0.000215$}\\[1ex]

&  \colorvarwc{  $< 0.000377$  $< 0.000988 $}  &
 \colorvarwc{$< 0.000101$  $< 0.000278 $}  &
   \colorvarwc{$< 0.000292$  $< 0.000821 $}  &
\colorvarwc{$< 0.000333$ $< 0.000952 $}&
  \colorvarwc{$< 0.000102$ $< 0.000267 $}&
    \colorvarwc{$< 0.000104$ $< 0.000269 $}\\[1ex]

\hline

$\cvisI{6} $ &
\colorvarc{$< 0.000786$  $< 0.00172  $}  &
\colorvarc{ $< 0.000364$  $< 0.000811 $ }   &
\colorvarc{   $< 0.000798$  $< 0.00167  $}  &
\colorvarc{$< 0.000786$ $< 0.00166  $}&
\colorvarc{ $< 0.000356$ $< 0.000823 $}&
    \colorvarc{$< 0.000361$ $< 0.000853 $}\\[1ex]

&  \colorvarwc{  $< 0.000622$  $< 0.00146  $}  &
 \colorvarwc{ $< 0.000292$  $< 0.000640 $}  &
   \colorvarwc{$< 0.000514$  $< 0.00124  $}  &
\colorvarwc{$< 0.000549$ $< 0.00128  $}&
  \colorvarwc{$< 0.000253$ $< 0.000591 $}&
    \colorvarwc{$< 0.000255$ $< 0.000637 $}\\[1ex]

\hline

$\cvisI{7} $ &
\colorvarc{ $< 0.00660$  $< 0.0102    $ }  &
\colorvarc{$< 0.00174$  $< 0.00393   $}   &
\colorvarc{  $< 0.00258$  $< 0.00526   $}  &
\colorvarc{$< 0.00406$ $< 0.00772   $}&
\colorvarc{ $< 0.00107$ $< 0.00252   $}&
    \colorvarc{$< 0.00135$ $< 0.00322   $}\\[1ex]

&  \colorvarwc{  $< 0.00764$  $< 0.0133    $}  &
 \colorvarwc{ $< 0.00298$  $< 0.00677   $}  &
   \colorvarwc{ $< 0.00621$  $< 0.0119    $}  &
\colorvarwc{$< 0.00693$ $< 0.0126    $}&
  \colorvarwc{$< 0.00268$ $< 0.00596   $}&
    \colorvarwc{$< 0.00281$ $< 0.00617   $}\\[1ex]

\hline

$\cvisI{8} $ &
\colorvarc{$< 0.00681$  $< 0.0159    $}  &
\colorvarc{$< 0.00517$  $< 0.0121    $-}   &
\colorvarc{$< 0.00747$  $< 0.0173    $}  &
\colorvarc{$< 0.00723$ $< 0.0172    $}&
\colorvarc{ $< 0.00478$ $< 0.0114    $}&
    \colorvarc{$< 0.00473$ $< 0.0112    $}\\[1ex]

&  \colorvarwc{  $< 0.0212$  $< 0.0452     $}  &
 \colorvarwc{$< 0.0209$  $< 0.0444     $}  &
   \colorvarwc{$< 0.0208$  $< 0.0431     $}  &
\colorvarwc{$< 0.0210$ $< 0.0460     $}&
  \colorvarwc{$< 0.0211$ $< 0.0455     $}&
    \colorvarwc{$< 0.0201$ $< 0.0428     $}\\[1ex]

\hline
 \hline
  $\sigma_{8}$ &
  \colorvarc{  $0.278^{+0.041}_{-0.047}$  $ {}^{+0.091}_{-0.081}$} &
 \colorvarc{  $0.366^{+0.045}_{-0.061}$  $ {}^{+0.11}_{-0.096}$ }&
  \colorvarc{   $0.290^{+0.041}_{-0.049}$  $ {}^{+0.094}_{-0.083}$} &
\colorvarc{$ 0.285^{+0.042}_{-0.049}$ $ {}^{+0.089}_{-0.085}$}&
\colorvarc{  $0.386^{+0.049}_{-0.061}$ $ {}^{+0.11}_{-0.099}$}&
    \colorvarc{$0.37^{+0.050}_{-0.064}$ $ {}^{+0.11}_{-0.10}$}\\[1ex]

  & \colorvarwc{$0.41^{+0.084}_{-0.10}$  $ {}^{+0.18}_{-0.18}$}  &
   \colorvarwc{$0.38^{+0.071}_{-0.084}$  $ {}^{+0.15}_{-0.14}$} &
    \colorvarwc{$0.28^{+0.050}_{-0.073}$  $ {}^{+0.13}_{-0.12}$}  &
\colorvarwc{$0.40^{+0.079}_{-0.10}$ $ {}^{+0.18}_{-0.17}$}&
  \colorvarwc{ $0.32^{+0.051}_{-0.070}$ $ {}^{+0.13}_{-0.12}$}&
    \colorvarwc{$0.42^{+0.081}_{-0.098}$ $ {}^{+0.19}_{-0.17}$}\\[1ex]
  \hline  \hline
\end{tabular}}
\caption{\label{tab:alldatasets1D} {
Parameter constraints for six dataset combinations when flat priors are used, showing the mean value of each parameter and the boundaries of the 68\%  and 95\% credible intervals.
\colorvarc{Black}/\colorvarwc{blue} font shows constraints for \colorvarc{var-c}/\colorvarwc{var-wc}.}}
\end{table*}

\renewcommand{\arraystretch}{0.5}
\begin{table*}
\resizebox{\textwidth}{!}{%
\begin{tabular} { |l| l| l| l|  |  l| l| l| }  \hline
 \hline
 \backslashbox[25mm]{\!Parameter}{Data} & \multicolumn{3}{c||}{PPS} &  \multicolumn{3}{c|}{PPS+Lens+BAO} \\

 \hline
 $-\ln(\Lcal_{\rm max}) $ &
  \multicolumn{3}{c||}{\colorvarc{var-c}:\colorvarc{ 6467.722}, \colorvarwc{var-wc}:\colorvarwc{6462.443}, \colorvarw{var-w}:\colorvarw{6462.040}, \colorLCDM{$\Lambda\!$CDM}:\colorLCDM{6467.585}} &
    \multicolumn{3}{c|}{    \colorvarc{6475.377},  \colorvarwc{6471.250}, \colorvarw{6470.89}, \colorLCDM{6475.238}
} \\
 \hline
&  flat prior on $c^2_{s,i},\cvisi$    &  nonflat prior on $c^2_{s,i}, \cvisi$    &   best fit    & flat prior  on $c^2_{s,i},\cvisi$   &  nonflat prior  on $c^2_{s,i}, \cvisi$  &   best fit     \\
\hline
$w_0 $

&  \colorvarwc{$0.13^{+0.070}_{-0.047}$  $ {}^{+0.11}_{-0.13}$}  &
\colorvarwc{$0.133^{+0.052}_{-0.052}$ $ {}^{+0.12}_{-0.090}$}&
   \colorvarwc{0.1139}  &
\colorvarwc{$-0.014^{+0.034}_{-0.041}$  $ {}^{+0.075}_{-0.070}$}&
\colorvarwc{$-0.022^{+0.028}_{-0.048}$ $ {}^{+0.088}_{-0.065}$}&
    \colorvarwc{-0.0558}\\[1ex]

&
\colorvarw{$-0.01^{+0.080}_{-0.084}$ $ {}^{+0.15}_{-0.15}$}&
 \colorvarw{-}  &
   \colorvarw{-0.007}  &
\colorvarw{$-0.070^{+0.039}_{-0.047}$  $ {}^{+0.088}_{-0.087}$}&
  \colorvarw{-}&
    \colorvarw{-0.072}\\[1ex]

\hline

$w_2 $

&  \colorvarwc{$-0.01^{+0.054}_{-0.053}$  $ {}^{+0.11}_{-0.10}$}  &
\colorvarwc{$-0.02^{+0.047}_{-0.055}$ $ {}^{+0.10}_{-0.10}$}&
   \colorvarwc{-0.0092}  &
\colorvarwc{$0.03^{+0.057}_{-0.056}$  $ {}^{+0.11}_{-0.11}$}&
\colorvarwc{$0.03^{+0.056}_{-0.066}$ $ {}^{+0.12}_{-0.11}$}&
    \colorvarwc{0.0361}\\[1ex]

&
\colorvarw{$0.00^{+0.060}_{-0.065}$ $ {}^{+0.13}_{-0.12}$}&
 \colorvarw{-}  &
   \colorvarw{0.0209}  &
\colorvarw{$0.03^{+0.065}_{-0.065}$  $ {}^{+0.12}_{-0.12}$}&
  \colorvarw{-}&
    \colorvarw{0.0342}\\[1ex]

\hline

$w_3 $

&  \colorvarwc{$-0.004^{+0.042}_{-0.050}$  $ {}^{+0.094}_{-0.085}$}  &
\colorvarwc{$-0.015^{+0.041}_{-0.052}$ $ {}^{+0.10}_{-0.085}$}&
   \colorvarwc{-0.0450}  &
\colorvarwc{$0.078^{+0.048}_{-0.047}$  $ {}^{+0.091}_{-0.093}$}&
\colorvarwc{$0.064^{+0.050}_{-0.051}$ $ {}^{+0.095}_{-0.096}$}&
    \colorvarwc{0.0498}\\[1ex]

&
\colorvarw{$0.02^{+0.057}_{-0.068}$ $ {}^{+0.12}_{-0.11}$}&
 \colorvarwc{-}  &
   \colorvarw{-0.002}  &
\colorvarw{$0.05^{+0.054}_{-0.054}$  $ {}^{+0.10}_{-0.10}$}&
  \colorvarw{-}&
    \colorvarw{0.0521}\\[1ex]

\hline

$w_4 $

&  \colorvarwc{$-0.075^{+0.029}_{-0.036}$  $ {}^{+0.068}_{-0.065}$}  &
\colorvarwc{$-0.068^{+0.030}_{-0.035}$ $ {}^{+0.069}_{-0.066}$}&
   \colorvarwc{-0.0643}  &
\colorvarwc{$-0.057^{+0.030}_{-0.034}$  $ {}^{+0.065}_{-0.064}$}&
\colorvarwc{$-0.051^{+0.030}_{-0.036}$ $ {}^{+0.070}_{-0.063}$}&
    \colorvarwc{-0.0512}\\[1ex]

&
\colorvarw{$-0.047^{+0.034}_{-0.034}$ $ {}^{+0.068}_{-0.066}$}&
 \colorvarw{-}  &
   \colorvarw{-0.0533}  &
\colorvarw{$-0.037^{+0.033}_{-0.039}$  $ {}^{+0.073}_{-0.070}$}&
  \colorvarw{-}&
    \colorvarw{-0.0467}\\[1ex]

\hline

$w_5 $

&  \colorvarwc{ $-0.030^{+0.019}_{-0.019}$  $ {}^{+0.036}_{-0.037}$}  &
\colorvarwc{$-0.024^{+0.020}_{-0.019}$ $ {}^{+0.037}_{-0.038}$}&
   \colorvarwc{-0.00691}  &
\colorvarwc{$-0.012^{+0.020}_{-0.020}$  $ {}^{+0.039}_{-0.039}$}&
\colorvarwc{$-0.003^{+0.019}_{-0.019}$ $ {}^{+0.038}_{-0.038}$}&
    \colorvarwc{0.00988}\\[1ex]

&
\colorvarw{$0.002^{+0.020}_{-0.018}$ $ {}^{+0.036}_{-0.040}$}&
 \colorvarw{-}  &
   \colorvarw{0.0041}  &
\colorvarw{$0.007^{+0.019}_{-0.019}$  $ {}^{+0.038}_{-0.039}$}&
\colorvarw{-}&
    \colorvarw{0.0110}\\[1ex]

\hline

$w_6 $

&  \colorvarwc{ $-0.020^{+0.0060}_{-0.0060}$  $ {}^{+0.012}_{-0.012}$}  &
\colorvarwc{$-0.017^{+0.0067}_{-0.0057}$ $ {}^{+0.012}_{-0.012}$}&
   \colorvarwc{-0.00887}  &
\colorvarwc{$-0.017^{+0.0067}_{-0.0062}$  $ {}^{+0.012}_{-0.013}$}&
\colorvarwc{$-0.012^{+0.0063}_{-0.0062}$ $ {}^{+0.012}_{-0.013}$}&
    \colorvarwc{-0.00466}\\[1ex]

&
\colorvarw{$-0.0066^{+0.0051}_{-0.0052}$ $ {}^{+0.0098}_{-0.0095}$}&
 \colorvarw{-}  &
   \colorvarw{-0.00788}  &
\colorvarw{$-0.0034^{+0.0049}_{-0.0048}$  $ {}^{+0.0096}_{-0.0094}$}&
  \colorvarw{-}&
    \colorvarw{-0.00391}\\[1ex]

\hline

$w_7 $

&  \colorvarwc{  $0.0131^{+0.0049}_{-0.0048}$  $ {}^{+0.0094}_{-0.0095}$}  &
\colorvarwc{$0.0123^{+0.0046}_{-0.0047}$ $ {}^{+0.0091}_{-0.0095}$}&
   \colorvarwc{0.00803}  &
\colorvarwc{$0.0162^{+0.0047}_{-0.0046}$  $ {}^{+0.0087}_{-0.0088}$}&
\colorvarwc{$0.0137^{+0.0047}_{-0.0042}$ $ {}^{+0.0089}_{-0.0086}$}&
    \colorvarwc{0.00930}\\[1ex]

&
\colorvarw{$0.0084^{+0.0040}_{-0.0040}$ $ {}^{+0.0080}_{-0.0082}$}&
 \colorvarw{ -}  &
   \colorvarw{-0.00788}  &
\colorvarw{$0.0078^{+0.0039}_{-0.0039}$  $ {}^{+0.0077}_{-0.0078}$}&
  \colorvarw{-}&
    \colorvarw{0.00852}\\[1ex]

\hline

$w_8 $

&  \colorvarwc{  $0.070^{+0.026}_{-0.041}$  $ {}^{+0.076}_{-0.066}$}  &
\colorvarwc{$0.059^{+0.022}_{-0.037}$ $ {}^{+0.068}_{-0.059}$}&
   \colorvarwc{0.01863}  &
\colorvarwc{$0.090^{+0.029}_{-0.047}$  $ {}^{+0.086}_{-0.070}$}&
\colorvarwc{$0.062^{+0.019}_{-0.038}$ $ {}^{+0.082}_{-0.061}$}&
    \colorvarwc{0.0143}\\[1ex]

&
\colorvarw{$0.020^{+0.016}_{-0.016}$ $ {}^{+0.033}_{-0.031}$}&
 \colorvarw{-}  &
   \colorvarw{0.0113}  &
\colorvarw{$0.016^{+0.015}_{-0.015}$  $ {}^{+0.029}_{-0.030}$}&
  \colorvarw{-}&
    \colorvarw{0.0123}\\[1ex]

\hline \hline

$c^2_{s,0} $ &
\colorvarc{$< 1.08\cdot 10^{-5}$  $< 3.04\cdot 10^{-5}$}  &
\colorvarc{$< 4.36\cdot 10^{-6}$ $< 1.63\cdot 10^{-5}$}&
\colorvarc{ $< 4.43\cdot 10^{-8}$}  &
\colorvarc{$< 6.23\cdot 10^{-6}$  $< 1.61\cdot 10^{-5}$}&
\colorvarc{$< 2.94\cdot 10^{-6}$ $< 1.06\cdot 10^{-5}$}&
    \colorvarc{$< 3.03\cdot 10^{-8}$}\\[1ex]

&  \colorvarwc{  $< 1.10\cdot 10^{-5}$  $< 2.99\cdot 10^{-5}$}  &
\colorvarwc{$< 6.46\cdot 10^{-6}$ $< 2.23\cdot 10^{-5}$}&
   \colorvarwc{$< 1.36\cdot 10^{-7}$}  &
\colorvarwc{$< 1.51\cdot 10^{-5}$  $< 4.00\cdot 10^{-5}$}&
\colorvarwc{$< 7.09\cdot 10^{-6}$ $< 2.65\cdot 10^{-5}$}&
    \colorvarwc{$< 9.68\cdot 10^{-8}$}\\[1ex]

\hline

$c^2_{s,1} $ &
\colorvarc{ $< 3.76\cdot 10^{-6}$  $< 9.65\cdot 10^{-6}$ }  &
\colorvarc{$< 1.34\cdot 10^{-6}$ $< 5.14\cdot 10^{-6}$}&
\colorvarc{  $< 1.71\cdot 10^{-8}$}  &
\colorvarc{$< 1.24\cdot 10^{-6}$  $< 3.41\cdot 10^{-6}$}&
\colorvarc{$< 5.28\cdot 10^{-7}$ $< 1.93\cdot 10^{-6}$}&
    \colorvarc{$< 1.71\cdot 10^{-8}$}\\[1ex]

&  \colorvarwc{ $< 6.61\cdot 10^{-6}$  $< 1.76\cdot 10^{-5}$}  &
\colorvarwc{$< 3.61\cdot 10^{-6}$ $< 1.24\cdot 10^{-5}$}&
   \colorvarwc{$< 1.35\cdot 10^{-7}$}  &
\colorvarwc{$< 1.68\cdot 10^{-6}$  $< 4.52\cdot 10^{-6}$}&
\colorvarwc{$< 8.41\cdot 10^{-7}$ $< 3.15\cdot 10^{-6}$}&
    \colorvarwc{$< 2.83\cdot 10^{-8}$ }\\[1ex]

\hline

$c^2_{s,2} $ &
\colorvarc{$< 7.04\cdot 10^{-6}$  $< 1.87\cdot 10^{-5}$}  &
\colorvarc{$< 2.93\cdot 10^{-6}$ $< 1.11\cdot 10^{-5}$}&
\colorvarc{  $3.4\cdot 10^{-8}$}  &
\colorvarc{$< 2.21\cdot 10^{-6}$  $< 5.74\cdot 10^{-6}$}&
\colorvarc{$< 1.00\cdot 10^{-6}$ $< 3.71\cdot 10^{-6}$}&
    \colorvarc{$4.6\cdot 10^{-8}$}\\[1ex]

&  \colorvarwc{  $< 1.67\cdot 10^{-5}$  $< 4.32\cdot 10^{-5}$}  &
\colorvarwc{$< 7.98\cdot 10^{-6}$ $< 3.48\cdot 10^{-5}$}&
   \colorvarwc{$4.8\cdot 10^{-7}$}  &
\colorvarwc{$< 2.88\cdot 10^{-6}$  $< 7.90\cdot 10^{-6}$}&
\colorvarwc{$< 1.41\cdot 10^{-6}$ $< 5.29\cdot 10^{-6}$}&
    \colorvarwc{$< 1.19\cdot 10^{-7}$}\\[1ex]

\hline

$c^2_{s,3} $ &
\colorvarc{$< 2.01\cdot 10^{-5}$  $< 5.09\cdot 10^{-5}$}  &
\colorvarc{$< 8.00\cdot 10^{-6}$ $< 2.89\cdot 10^{-5}$}&
\colorvarc{$1.2\cdot 10^{-7}$}  &
\colorvarc{$< 6.15\cdot 10^{-6}$  $< 1.52\cdot 10^{-5}$}&
\colorvarc{$< 3.07\cdot 10^{-6}$ $< 1.12\cdot 10^{-5}$}&
    \colorvarc{$< 1.00\cdot 10^{-7}$}\\[1ex]

&  \colorvarwc{  $< 4.90\cdot 10^{-5}$  $< 0.000124$}  &
\colorvarwc{$< 2.48\cdot 10^{-5}$ $< 8.89\cdot 10^{-5}$}&
   \colorvarwc{$< 1.19\cdot 10^{-6}$ }  &
\colorvarwc{$< 9.63\cdot 10^{-6}$  $< 2.43\cdot 10^{-5}$}&
\colorvarwc{$< 4.61\cdot 10^{-6}$ $< 1.48\cdot 10^{-5}$}&
    \colorvarwc{$2.2\cdot 10^{-7}$}\\[1ex]

\hline

$c^2_{s,4} $ &
\colorvarc{ $< 6.75\cdot 10^{-5}$  $< 0.000184$ }  &
\colorvarc{$< 2.50\cdot 10^{-5}$ $< 9.70\cdot 10^{-5}$}&
\colorvarc{ $0.52\cdot 10^{-7}$}  &
\colorvarc{$< 1.94\cdot 10^{-5}$  $< 5.13\cdot 10^{-5}$}&
\colorvarc{$< 8.51\cdot 10^{-6}$ $< 3.07\cdot 10^{-5}$}&
    \colorvarc{$< 3.38\cdot 10^{-7}$}\\[1ex]

&  \colorvarwc{ $< 0.000131$  $< 0.000336 $}  &
\colorvarwc{$< 6.34\cdot 10^{-5}$ $< 0.000257$}&
   \colorvarwc{$3.1\cdot 10^{-6}$}  &
\colorvarwc{$< 2.83\cdot 10^{-5}$  $< 6.92\cdot 10^{-5}$}&
\colorvarwc{$< 1.39\cdot 10^{-5}$ $< 4.84\cdot 10^{-5}$}&
    \colorvarwc{$< 7.92\cdot 10^{-7}$}\\[1ex]

\hline

$c^2_{s,5} $ &
\colorvarc{$< 0.000215$  $< 0.000580 $}  &
\colorvarc{$< 7.23\cdot 10^{-5}$ $< 0.000271$}&
\colorvarc{   $1.6\cdot 10^{-6}$}  &
\colorvarc{$< 5.31\cdot 10^{-5}$  $< 0.000134$}&
\colorvarc{$< 2.55\cdot 10^{-5}$ $< 8.89\cdot 10^{-5}$}&
    \colorvarc{$< 7.43\cdot 10^{-7}$ }\\[1ex]

&  \colorvarwc{  $< 0.000264$  $< 0.000679 $}  &
\colorvarwc{$< 0.000157$ $< 0.000554 $}&
   \colorvarwc{$< 4.09\cdot 10^{-6}$}  &
\colorvarwc{$< 7.36\cdot 10^{-5}$  $< 0.000193$}&
\colorvarwc{$< 4.03\cdot 10^{-5}$ $< 0.000152$}&
    \colorvarwc{$< 1.79\cdot 10^{-6}$}\\[1ex]

\hline

$c^2_{s,6} $ &
\colorvarc{$< 0.000952$  $< 0.00142  $}  &
\colorvarc{$< 0.000498$ $< 0.00100  $}&
\colorvarc{  0.000142}  &
\colorvarc{$< 0.000317$  $< 0.000631 $}&
\colorvarc{$< 0.000164$ $< 0.000450 $}&
    \colorvarc{$8.4\cdot 10^{-5}$}\\[1ex]

&  \colorvarwc{  $< 0.000567$  $< 0.00112  $}  &
\colorvarwc{$< 0.000334$ $< 0.000889 $}&
   \colorvarwc{0.000367}  &
\colorvarwc{ $< 0.000197$  $< 0.000477 $}&
\colorvarwc{$< 9.69\cdot 10^{-5}$ $< 0.000326$}&
    \colorvarwc{$< 1.08\cdot 10^{-5}$}\\[1ex]

\hline

$c^2_{s,7} $ &
\colorvarc{ $< 0.00223$  $< 0.00387   $ }  &
\colorvarc{$< 0.00137$ $< 0.00304   $}&
\colorvarc{  $4.0\cdot 10^{-5}$}  &
\colorvarc{$< 0.000883$  $< 0.00187  $}&
\colorvarc{$< 0.000358$ $< 0.00117  $}&
    \colorvarc{$0.81\cdot 10^{-5}$}\\[1ex]

&  \colorvarwc{  $< 0.00200$  $< 0.00382   $}  &
\colorvarwc{$< 0.00150$ $< 0.00307   $}&
   \colorvarwc{0.000920}  &
\colorvarwc{$< 0.00132$  $< 0.00261   $}&
\colorvarwc{$< 0.000725$ $< 0.00195  $}&
    \colorvarwc{$6.18 \cdot 10^{-5}$}\\[1ex]

\hline

$c^2_{s,8} $ &
\colorvarc{$< 0.00501$  $< 0.0113    $}  &
\colorvarc{$< 0.00246$ $< 0.00725   $}&
\colorvarc{ $6.9 \cdot 10^{-5}$}  &
\colorvarc{$< 0.00304$  $< 0.00714   $}&
\colorvarc{$< 0.00168$ $< 0.00602   $}&
    \colorvarc{ $< 6.93\cdot 10^{-5}$}\\[1ex]

&  \colorvarwc{  $< 0.0247$  $< 0.0503     $}  &
\colorvarwc{$< 0.0168$ $< 0.0424     $}&
   \colorvarwc{$0.00026$}  &
\colorvarwc{$< 0.0318$  $< 0.0589     $}&
\colorvarwc{$< 0.0156$ $< 0.0469     $}&
    \colorvarwc{$< 0.000576$}\\[1ex]

\hline
\hline

$\cvisI{0} $ &
\colorvarc{$< 4.10\cdot 10^{-5}$  $< 0.000124$}  &
\colorvarc{$< 1.21\cdot 10^{-5}$ $< 5.82\cdot 10^{-5}$}&
\colorvarc{ $< 9.68\cdot 10^{-8}$}  &
\colorvarc{ $< 3.26\cdot 10^{-5}$  $< 0.000114$}&
\colorvarc{$< 8.95\cdot 10^{-6}$ $< 4.46\cdot 10^{-5}$}&
    \colorvarc{$< 3.13\cdot 10^{-8}$}\\[1ex]

&  \colorvarwc{   $< 4.64\cdot 10^{-5}$  $< 0.000152$}  &
\colorvarwc{$< 2.09\cdot 10^{-5}$ $< 0.000101$}&
   \colorvarwc{ $< 7.50\cdot 10^{-7}$}  &
\colorvarwc{$< 0.000104$  $< 0.000355 $}&
\colorvarwc{$< 3.34\cdot 10^{-5}$ $< 0.000223$}&
    \colorvarwc{$< 3.75\cdot 10^{-7}$}\\[1ex]

\hline

$\cvisI{1} $ &
\colorvarc{ $< 6.00\cdot 10^{-6}$  $< 1.62\cdot 10^{-5}$ }  &
\colorvarc{$< 2.57\cdot 10^{-6}$ $< 8.89\cdot 10^{-6}$}&
\colorvarc{  $< 3.12\cdot 10^{-8}$}  &
\colorvarc{$< 2.22\cdot 10^{-6}$  $< 5.73\cdot 10^{-6}$}&
\colorvarc{$< 9.74\cdot 10^{-7}$ $< 3.34\cdot 10^{-6}$}&
    \colorvarc{$1.9\cdot 10^{-8}$}\\[1ex]

&  \colorvarwc{ $< 1.18\cdot 10^{-5}$  $< 3.06\cdot 10^{-5}$}  &
\colorvarwc{$< 6.30\cdot 10^{-6}$ $< 2.28\cdot 10^{-5}$}&
   \colorvarwc{$2.3 \cdot 10^{-7}$}  &
\colorvarwc{$< 3.78\cdot 10^{-6}$  $< 1.00\cdot 10^{-5}$}&
\colorvarwc{$< 1.74\cdot 10^{-6}$ $< 6.20\cdot 10^{-6}$}&
    \colorvarwc{ $< 5.07\cdot 10^{-8}$ }\\[1ex]

\hline

$\cvisI{2} $ &
\colorvarc{$< 9.11\cdot 10^{-6}$  $< 2.70\cdot 10^{-5}$}  &
\colorvarc{$< 4.12\cdot 10^{-6}$ $< 1.45\cdot 10^{-5}$}&
\colorvarc{  $0.67\cdot 10^{-8}$}  &
\colorvarc{$< 2.88\cdot 10^{-6}$  $< 7.73\cdot 10^{-6}$}&
\colorvarc{$< 1.51\cdot 10^{-6}$ $< 5.09\cdot 10^{-6}$}&
    \colorvarc{$< 3.26\cdot 10^{-8}$}\\[1ex]

&  \colorvarwc{  $< 2.18\cdot 10^{-5}$  $< 5.78\cdot 10^{-5}$}  &
\colorvarwc{$< 1.11\cdot 10^{-5}$ $< 4.21\cdot 10^{-5}$}&
   \colorvarwc{$0.80\cdot 10^{-6}$}  &
\colorvarwc{$< 4.35\cdot 10^{-6}$  $< 1.08\cdot 10^{-5}$}&
\colorvarwc{$< 2.19\cdot 10^{-6}$ $< 7.92\cdot 10^{-6}$}&
    \colorvarwc{$< 5.95\cdot 10^{-8}$}\\[1ex]

\hline

$\cvisI{3} $ &
\colorvarc{$< 2.51\cdot 10^{-5}$  $< 6.99\cdot 10^{-5}$}  &
\colorvarc{$< 1.13\cdot 10^{-5}$ $< 4.11\cdot 10^{-5}$}&
\colorvarc{  $1.7\cdot 10^{-7}$ }  &
\colorvarc{$< 7.28\cdot 10^{-6}$  $< 1.88\cdot 10^{-5}$}&
\colorvarc{$< 4.14\cdot 10^{-6}$ $< 1.29\cdot 10^{-5}$}&
    \colorvarc{$< 1.60\cdot 10^{-7}$}\\[1ex]

&  \colorvarwc{ $< 5.88\cdot 10^{-5}$  $< 0.000152$}  &
\colorvarwc{$< 3.33\cdot 10^{-5}$ $< 0.000104$}&
   \colorvarwc{$< 1.78\cdot 10^{-6}$}  &
\colorvarwc{$< 1.28\cdot 10^{-5}$  $< 3.30\cdot 10^{-5}$}&
\colorvarwc{$< 7.01\cdot 10^{-6}$ $< 2.24\cdot 10^{-5}$}&
    \colorvarwc{$5.6\cdot 10^{-7}$}\\[1ex]

\hline

$\cvisI{4} $ &
\colorvarc{ $< 8.17\cdot 10^{-5}$  $< 0.000231$ }  &
\colorvarc{$< 3.81\cdot 10^{-5}$ $< 0.000128$}&
\colorvarc{  $< 3.86\cdot 10^{-7}$}  &
\colorvarc{$< 2.39\cdot 10^{-5}$  $< 6.10\cdot 10^{-5}$}&
\colorvarc{$< 1.27\cdot 10^{-5}$ $< 4.30\cdot 10^{-5}$}&
    \colorvarc{$< 2.18\cdot 10^{-7}$}\\[1ex]

&  \colorvarwc{ $< 0.000152$  $< 0.000395 $}  &
\colorvarwc{$< 8.02\cdot 10^{-5}$ $< 0.000280$}&
   \colorvarwc{$4.8 \cdot 10^{-6}$}  &
\colorvarwc{ $< 3.27\cdot 10^{-5}$  $< 8.09\cdot 10^{-5}$}&
\colorvarwc{$< 1.88\cdot 10^{-5}$ $< 6.41\cdot 10^{-5}$}&
    \colorvarwc{$1.2\cdot 10^{-6}$}\\[1ex]

\hline

$\cvisI{5} $ &
\colorvarc{$< 0.000273$  $< 0.000765 $}  &
\colorvarc{$< 0.000119$ $< 0.000406 $}&
\colorvarc{$< 2.09\cdot 10^{-6}$}  &
\colorvarc{$< 8.01\cdot 10^{-5}$  $< 0.000209$}&
\colorvarc{$< 3.99\cdot 10^{-5}$ $< 0.000130$}&
    \colorvarc{$< 1.30\cdot 10^{-6}$}\\[1ex]

&  \colorvarwc{  $< 0.000377$  $< 0.000988 $}  &
\colorvarwc{$< 0.000252$ $< 0.000760 $}&
   \colorvarwc{$< 6.69\cdot 10^{-6}$ }  &
\colorvarwc{$< 0.000102$  $< 0.000267 $}&
\colorvarwc{$< 6.29\cdot 10^{-5}$ $< 0.000205$}&
    \colorvarwc{$0.66\cdot 10^{-6}$}\\[1ex]

\hline

$\cvisI{6} $ &
\colorvarc{$< 0.000786$  $< 0.00172  $}  &
\colorvarc{$< 0.000511$ $< 0.00122  $}&
\colorvarc{ $1.5 \cdot 10^{-5}$}  &
\colorvarc{$< 0.000356$  $< 0.000823 $}&
\colorvarc{$< 0.000222$ $< 0.000604 $}&
    \colorvarc{$< 1.74\cdot 10^{-5}$}\\[1ex]

&  \colorvarwc{  $< 0.000622$  $< 0.00146  $}  &
\colorvarwc{$< 0.000412$ $< 0.00115  $}&
   \colorvarwc{$< 3.52\cdot 10^{-5}$}  &
\colorvarwc{$< 0.000253$  $< 0.000590 $}&
\colorvarwc{$< 0.000148$ $< 0.000470 $}&
    \colorvarwc{$0.98\cdot 10^{-5}$}\\[1ex]

\hline

$\cvisI{7} $ &
\colorvarc{ $< 0.00660$  $< 0.0102    $ }  &
\colorvarc{$< 0.00331$ $< 0.00717   $}&
\colorvarc{   $< 5.62\cdot 10^{-5}$}  &
\colorvarc{$< 0.00107$  $< 0.00252   $}&
\colorvarc{$< 0.000549$ $< 0.00163  $}&
    \colorvarc{$< 2.49\cdot 10^{-5}$}\\[1ex]

&  \colorvarwc{  $< 0.00764$  $< 0.0133    $}  &
\colorvarwc{$< 0.00539$ $< 0.0115    $}&
   \colorvarwc{ 0.00319}  &
\colorvarwc{$< 0.00267$  $< 0.00596   $}&
\colorvarwc{$< 0.00161$ $< 0.00491   $}&
    \colorvarwc{$7.8 \cdot 10^{-5}$}\\[1ex]

\hline

$\cvisI{8} $ &
\colorvarc{$< 0.00681$  $< 0.0159    $}  &
\colorvarc{$< 0.00382$ $< 0.0118    $}&
\colorvarc{0.000101}  &
\colorvarc{$< 0.00478$  $< 0.0114    $}&
\colorvarc{$< 0.00253$ $< 0.00839   $}&
    \colorvarc{$< 3.61\cdot 10^{-5}$}\\[1ex]

&  \colorvarwc{  $< 0.0212$  $< 0.0452     $}  &
\colorvarwc{$< 0.0154$ $< 0.0390     $}&
   \colorvarwc{$< 0.000659$ }  &
\colorvarwc{$< 0.0211$  $< 0.0455     $}&
\colorvarwc{$< 0.0143$ $< 0.0347     $}&
    \colorvarwc{$0.00049$}\\[1ex]

\hline
 \hline
  $\sigma_{8}$ &
  \colorvarc{  $0.278^{+0.041}_{-0.047}$  $ {}^{+0.091}_{-0.081}$} &
 \colorvarc{  $0.39^{+0.066}_{-0.067}$  $ {}^{+0.13}_{-0.12}$ }&
  \colorvarc{   0.778} &
\colorvarc{$0.386^{+0.049}_{-0.061}$  $ {}^{+0.11}_{-0.099}$}&
\colorvarc{$0.49^{+0.090}_{-0.071}$ $ {}^{+0.14}_{-0.15}$}&
    \colorvarc{0.7868}\\[1ex]

  & \colorvarwc{$0.41^{+0.084}_{-0.10}$  $ {}^{+0.18}_{-0.18}$}  &
   \colorvarwc{$0.50^{+0.098}_{-0.12}$  $ {}^{+0.22}_{-0.21}$} &
    \colorvarwc{0.829}  &
\colorvarwc{$0.32^{+0.051}_{-0.070}$  $ {}^{+0.13}_{-0.12}$}&
\colorvarwc{$0.39^{+0.065}_{-0.11}$ $ {}^{+0.18}_{-0.16}$}&
    \colorvarwc{0.659}\\[1ex]

    & \colorvarw{$0.76^{+0.18}_{-0.29}$  $ {}^{+0.50}_{-0.41}$}  &
   \colorvarw{-} &
    \colorvarw{0.71}  &
\colorvarw{$0.68^{+0.084}_{-0.15}$ $ {}^{+0.30}_{-0.25}$}&
  \colorvarw{ -}&
    \colorvarw{0.656}\\[1ex]

  \hline
 \hline
\end{tabular}}
\caption{\label{tab:constraintsPPSvsLensvsPrior} {Constraints
 on the \colorvarc{var-c}/\colorvarwc{var-wc}/\colorvarw{var-w} (\colorvarc{Black}/\colorvarwc{blue}/\colorvarw{green} respectively) parameters
when two dataset combinations (PPS and PPS+lens+BAO)  and both types of priors were used.
The upper and lower limits of the 68\% and 95\% credible regions around the mean are shown.
 The maximum likelihood $-\ln\Lcal_{\rm max}$
is stated in the first row; $\Lambda$CDM  value in \colorLCDM{grey}.
}}
\end{table*}

\renewcommand{\arraystretch}{1.}
\begin{table}
\resizebox{0.49\textwidth}{!}{%
\begin{tabular} { |l| l| l|  |  l| l| }  \hline
 \hline
 \backslashbox[22mm]{\!Parameter}{Data} & \multicolumn{2}{c||}{PPS} &  \multicolumn{2}{c|}{PPS+Lens+BAO} \\

   \hline
&   68\% and 95\% C.L.    &   best fit    &  68\% and 95\% C.L.     &   best fit     \\

\hline \hline

$c^2_{+,0} $ &
\colorvarc{$< 1.19\cdot 10^{-5}$  $< 4.14\cdot 10^{-5}$}  &
\colorvarc{ $0.71 \cdot 10^{-7}$}  &
\colorvarc{$< 8.53\cdot 10^{-6}$  $< 3.03\cdot 10^{-5}$}&
    \colorvarc{$< 5.48\cdot 10^{-8}$}\\[1ex]

&  \colorvarwc{  $< 1.94\cdot 10^{-5}$  $< 6.58\cdot 10^{-5}$}  &
   \colorvarwc{$3.8\cdot 10^{-7}$}  &
\colorvarwc{$< 2.65\cdot 10^{-5}$  $< 0.000139$}&
    \colorvarwc{$< 3.14\cdot 10^{-7}$}\\[1ex]

\hline
$c^2_{+,1} $ &
\colorvarc{$< 3.03\cdot 10^{-6}$  $< 8.19\cdot 10^{-6}$}  &
\colorvarc{ $2.5\cdot 10^{-8}$}  &
\colorvarc{$< 1.14\cdot 10^{-6}$  $< 3.11\cdot 10^{-6}$}&
    \colorvarc{$2.3\cdot 10^{-8}$}\\[1ex]

&  \colorvarwc{  $< 7.80\cdot 10^{-6}$  $< 1.99\cdot 10^{-5}$}  &
   \colorvarwc{$2.3\cdot 10^{-7}$}  &
\colorvarwc{$< 2.01\cdot 10^{-6}$  $< 5.44\cdot 10^{-6}$}&
    \colorvarwc{$4.0\cdot 10^{-8}$ }\\[1ex]

   \hline
 $c^2_{+,2} $ &
\colorvarc{ $< 5.72\cdot 10^{-6}$  $< 1.61\cdot 10^{-5}$}  &
\colorvarc{ $0.70\cdot 10^{-8}$}  &
\colorvarc{$< 2.04\cdot 10^{-6}$  $< 5.44\cdot 10^{-6}$}&
    \colorvarc{$0.60\cdot 10^{-8}$}\\[1ex]

&  \colorvarwc{  $< 1.53\cdot 10^{-5}$  $< 4.90\cdot 10^{-5}$}  &
   \colorvarwc{ $0.91\cdot 10^{-6}$}  &
\colorvarwc{$< 2.84\cdot 10^{-6}$  $< 8.06\cdot 10^{-6}$}&
    \colorvarwc{$1.09\cdot 10^{-7}$}\\[1ex]

   \hline
 $c^2_{+,3} $ &
\colorvarc{ $< 1.57\cdot 10^{-5}$  $< 4.29\cdot 10^{-5}$}  &
\colorvarc{ $2.1\cdot 10^{-7}$}  &
\colorvarc{$< 5.90\cdot 10^{-6}$  $< 1.53\cdot 10^{-5}$}&
    \colorvarc{$1.4\cdot 10^{-7}$}\\[1ex]

&  \colorvarwc{  $< 4.81\cdot 10^{-5}$  $< 0.000120$}  &
   \colorvarwc{$1.6\cdot 10^{-6}$}  &
\colorvarwc{$< 9.09\cdot 10^{-6}$  $< 2.39\cdot 10^{-5}$}&
    \colorvarwc{$5.2\cdot 10^{-7}$}\\[1ex]

       \hline
 $c^2_{+,4} $ &
\colorvarc{ $< 5.06\cdot 10^{-5}$  $< 0.000141$}  &
\colorvarc{ $0.68\cdot 10^{-6}$}  &
\colorvarc{$< 1.70\cdot 10^{-5}$  $< 4.60\cdot 10^{-5}$}&
    \colorvarc{$3.6\cdot 10^{-7}$}\\[1ex]

&  \colorvarwc{  $< 0.000119$  $< 0.000348 $}  &
   \colorvarwc{$5.6 \cdot 10^{-6}$}  &
\colorvarwc{$< 2.63\cdot 10^{-5}$  $< 6.81\cdot 10^{-5}$}&
    \colorvarwc{ $1.26\cdot 10^{-6}$}\\[1ex]

           \hline
 $c^2_{+,5} $ &
\colorvarc{ $< 0.000151$  $< 0.000411 $}  &
\colorvarc{ $2.4\cdot 10^{-6}$}  &
\colorvarc{$< 5.23\cdot 10^{-5}$  $< 0.000132$}&
    \colorvarc{$1.13\cdot 10^{-6}$ }\\[1ex]

&  \colorvarwc{  $< 0.000321$  $< 0.000829 $}  &
   \colorvarwc{$0.57\cdot 10^{-6}$}  &
\colorvarwc{$< 8.20\cdot 10^{-5}$  $< 0.000223$}&
    \colorvarwc{$5.1\cdot 10^{-6}$}\\[1ex]

         \hline
 $c^2_{+,6} $ &
\colorvarc{ $< 0.000883$  $< 0.00129  $}  &
\colorvarc{ $0.000150$ }  &
\colorvarc{ $< 0.000299$  $< 0.000602 $}&
    \colorvarc{$0.000090$}\\[1ex]

&  \colorvarwc{  $< 0.000588$  $< 0.00118  $}  &
   \colorvarwc{$0.000381$}  &
\colorvarwc{$< 0.000195$  $< 0.000478 $}&
    \colorvarwc{$1.30\cdot 10^{-5}$}\\[1ex]

         \hline
 $c^2_{+,7} $ &
\colorvarc{ $< 0.00338$  $< 0.00545   $}  &
\colorvarc{ $0.000063$}  &
\colorvarc{ $< 0.000708$  $< 0.00168  $}&
    \colorvarc{$1.8\cdot 10^{-5}$ }\\[1ex]

&  \colorvarwc{  $< 0.00474$  $< 0.00737   $}  &
   \colorvarwc{$0.002619$}  &
\colorvarwc{$< 0.00169$  $< 0.00369   $}&
    \colorvarwc{$0.000103$}\\[1ex]

   \hline
 $c^2_{+,8} $ &
\colorvarc{ $< 0.00497$  $< 0.0110    $}  &
\colorvarc{ $0.000123$}  &
\colorvarc{ $< 0.00338$  $< 0.00867   $}&
    \colorvarc{$< 9.08\cdot 10^{-5}$}\\[1ex]

&  \colorvarwc{  $< 0.0262$  $< 0.0505     $}  &
   \colorvarwc{$0.00051$}  &
\colorvarwc{$< 0.0242$  $< 0.0534     $}&
    \colorvarwc{$0.00064$}\\[1ex]

 \hline
\end{tabular}}
\caption{\label{tab:constraints_cp2}{The 68\% and 95\% credible regions for $c^2_{+,i}$ when the PPS and PPS+Lens+BAO dataset combinations were used
using ``nonflat priors'' (see Sec.~\ref{sec:priors}).
\colorvarc{Black}/\colorvarwc{blue} fonts show constraints for \colorvarc{var-c}/\colorvarwc{var-wc}, respectively.
}}
\end{table}

\end{appendix}




\end{document}